\documentclass[iop]{emulateapj}
\usepackage{amsmath}
\usepackage{amssymb}
\usepackage{mathrsfs}
\usepackage{graphicx}

\shorttitle{Evaporation Fronts}
\shortauthors{Estrada, Cuzzi and Morgan}

\begin{document}
\title{Global Modeling of Nebulae with Particle Growth, Drift and Evaporation Fronts. I: Methodology
and Typical Results}

\author{Paul R. Estrada}
\affil{Carl Sagan Center, SETI Institute, 189 N. Bernardo Ave. \# 100, Mountain View, CA 94043}

\author{Jeffrey N. Cuzzi}
\affil{Ames Research Center, NASA; Mail Stop 245-3, Moffett Field, CA 94035}

\and

\author{Demitri A. Morgan}
\affil{USRA, NASA Ames Research Center, Mail Stop 245-3, Moffett Field, CA 94035}
 
\begin{abstract}
We model particle growth in a turbulent, viscously evolving protoplanetary 
nebula, incorporating sticking, bouncing, fragmentation, and mass transfer 
at high speeds. We treat small particles using a moments method and large 
particles using a traditional histogram binning, including a probability 
distribution function of collisional velocities. The fragmentation strength 
of the particles depends on their composition (icy aggregates are stronger 
than silicate aggregates). The particle opacity, which controls the nebula 
thermal structure, evolves as particles grow and mass redistributes. While
growing, particles drift radially due to nebula headwind drag. Particles
of different compositions evaporate at ``evaporation fronts'' (EFs) where
the midplane temperature exceeds their respective evaporation temperatures. 
We track the vapor and solid phases of each component, accounting for 
advection and radial and vertical diffusion. We present characteristic 
results in evolutions lasting $2 \times 10^5$ years. In general, (a) mass 
is transferred from the outer to inner nebula in significant amounts, 
creating radial concentrations of solids at EFs; (b) particle sizes 
are limited by a combination of fragmentation, bouncing, and drift; (c)
``lucky'' large particles never represent a significant amount of mass;
and (d) restricted radial zones just outside each EF become compositionally
enriched in the associated volatiles. We point out implications  for
mm-submm SEDs and inference of nebula mass, radial banding, the role of 
opacity on new mechanisms for generating turbulence, enrichment of 
meteorites in heavy oxygen isotopes, variable and nonsolar redox conditions,
primary accretion of silicate and icy planetesimals, and the makeup of 
Jupiter's core.
\end{abstract}


\section{Introduction}\label{sec:intro}

The very early evolution of solids, as they first decouple from cosmic gas in the protoplanetary nebula and grow into planetesimals such as we see today (asteroids and TNOs) remains poorly understood. We call this stage of growth {\it primary accretion}. The physics of subsequent or {\it secondary} accretion by mutual collisions (including gravitational effects) and sweepup of smaller objects is better understood and the process is more well characterized 
\citep[for a recent review, see, e.g.,][]{ray14}.
In this stage, gravitational focusing by larger objects (so-called runaway accretion) has traditionally played a dominant role, and nebula gas a lesser role, but recently it has been shown that gas can play an important role even during this  secondary stage \citep{ok10,lj12}. Either way, studies of secondary accretion generally make arbitrary assumptions regarding the sizes of the primary bodies that provide their initial conditions. 

Pioneering studies by \citet{wei80,wei84,wei88}, \citet{md88}, \citet{wei89}, \citet{don90}, \citet{bm93}, 
\citet{dt97}, \citet{bw00}, \citet{blu00} and others since (see Testi et al. 2014, and Johansen et al. 2014, 
2015 for reviews including more recent references) have illustrated the importance of the messy physics of sticking and bouncing of grainy, porous, irregular particles. \citet{wei84} recognized the important role of gas turbulence and the great difficulty of forming large objects by ``incremental growth"\footnote{We define incremental growth as slow and steady growth by particles of a given size to larger sizes, by simple physical collision and sticking of same-and-smaller-size particles.}  in the presence of turbulence, due to a combination of fragmentation, slow growth, and rapid radial drift. Consequently, nearly all Weidenschilling's subsequent models 
\citep{wei97,wei00,wei04,wei11} have emphasized conditions where global nebula gas turbulence is vanishingly small, allowing a dense layer of particles to settle to the midplane; this layer itself is stirred by a small amount of self-generated turbulence \citep{wei80,cuz93,cuz94,wei95,bar05,joh06,bs10}, which is however insufficient to suppress incremental growth by sticking. 

The possibility of such a nonturbulent environment has also motivated a completely different approach to primary accretion through various kinds of midplane particle layer instabilities (Safronov 1972, Goldreich \& Ward 1973, Sekiya 1983 {\it et seq.}, Garaud \& Lin 2004, Goodman \& Pindor 2000, Youdin \& Goodman 2005, and subsequent workers). In these models, the uncertain properties of sticking are ultimately made irrelevant by the dominant role of local gravitational and/or drag instabilities which lead directly to km-and-larger planetesimals (see Chiang \& Youdin 2010 for a review that emphasizes work of this type). Midplane instability models all require significant {\it initial} enhancements of the midplane solids-to-gas ratio before they can operate. Because settling of particles into dense midplane layers is frustrated by turbulent vertical diffusion, this requirement in turn places restrictive conditions on the level of global turbulence and/or global mass
enhancement, especially for the mm-size particles that dominate primitive asteroids \citep[for a discussion, see][]{cw06}.  

The level of global nebula turbulence remains very much a matter of active debate. While the predictions of recent MHD models have been trending towards very low intensity near-midplane turbulence \citep{bs13,bai14}, purely hydrodynamical turbulence has been making a comeback \citep[][see \citet{tur14} for a review]{nel13,mar13,mar14,sk14}. Moreover, a number of meteoritic and cometary properties point to an extended period of asteroid formation \citep{con12,ku12,kit13} and an extended radial range of mixing of mineral constituents \citep{zol06,joz12}, which are both difficult to reconcile with a globally nonturbulent nebula in the region of planetesimal formation. It seems appropriate to face the implications of moderate turbulence squarely, including the various barriers it presents to incremental growth. 

Once particles grow into the cm-m size range, depending on location, they drift radially at a rapid pace because of headwind drag with the more slowly rotating, pressure-supported gas \citep{wei77,cw06,wei06,joh14}. This drift can potentially transport particles great distances while they grow only slowly, because turbulence stirs their ``feedstock" and leads to collision velocities which are erosive or destructive. Because particles found at any distance from the sun may have originated at greater distances, proper models should ideally treat the entire radial extent of the nebula at once. 

Early global models with growth and drift were presented by \citet{sv96,sv97} and \citet{kor04}, but these were simplified in several ways. \citet{cc06} studied the implications of drift with simplified growth
models, and \citet{gar07} developed a more comprehensive model including an evolving nebula with several
volatile species, but still with a simplified growth model.
\citet{bra08} were the first to present a complete global (1D) model of particle growth and drift with a detailed model of sticking, which  was enabled by an elaborate implicit numerical solution to the cumbersome coagulation equation. They demonstrated, described, and motivated the various barriers to growth (both fragmentation and drift), while assuming a steady-state, non-evolving gas disk. \citet{ha10,ha12} presented models with evolving gas disks, but a very simple sticking algorithm emphasizing the possible differences
between viscosity $\nu$ and the diffusivity $D$.
\citet{bir10,bir11,bir12a} extended the Brauer results to evolving disks, using their implicit numerical approach, and added several new aspects, including composition-dependent sticking (ice {\it vs.} silicate); their particle collision velocities remained single-valued as a function of particle size, following \citet{oc07}. Their model opacity did not account for the evolving particle size distribution, and the only evaporation they included was the silicate evaporation front at 1500K where all the solids evaporated. There was no explicit treatment of the behavior of vapor of different species.   

In the last decade or so, a large and still growing body of experimental work has clarified and refined sticking coefficients for granular particles of various composition, porosity, and relative velocity \citep{gut10}. The most detailed coagulation models which employ this detailed sticking physics encounter a ``bouncing barrier" for mm-cm size particles in moderate turbulence \citep{zso10}. This means that, while small particles can grow by sticking because their collision velocities are very small, the very act of growing leads them to acquire relative velocities in turbulence \citep{vol80,car08,car10,pp10,pp13,pp14,hub12,hub13,pan14a,pan14b} that exceed the sticking threshold at fairly small sizes - in the mm-cm range at 2.5AU. This ``bouncing barrier" is in intriguing agreement with the sizes of ubiquitous ``chondrules" dominating primitive meteorites, but it poses a problem for actually forming the asteroidal parents of those very meteorites. Sophisticated variants of this model showed that weak dm-size agglomerates of dust-rimmed ``chondrules" could be formed, but only in very low levels of nebula turbulence 
\citep{orm08}. \citet{win12a,win12b}, \citet{gar13}, and \citet{dra13,dra14} refined collision models to include a probability distribution function (PDF) for collision velocity, adding a growth path involving fragmentation of a smaller projectile along with mass transfer to a larger target \citep{wur05,bei11}. They all found that some very few ``lucky" large particles could 
form, although the quantitative details were found to be highly sensitive to the numerics of mass histogram binning. However, these studies either neglected the radial headwind drift or assumed arbitrary radial ``pressure bumps" to prevent it, thus giving only an upper limit on the abundance of large ``lucky" particles. Our models include not only bouncing and partial or total fragmentation effects, mass- and velocity-dependent sticking coefficients, velocity PDFs and ``lucky" particles, but also include full radial drift and a viscously evolving background gas nebula (section \ref{sec:model}).

Sticking properties also depend on particle composition. From laboratory studies and material properties,
it is becoming clear that where temperatures are low enough for water ice or methanol to form, sticking is significantly more effective \citep{bri96,wad09}, so larger and stronger particles can grow more robustly \citep{oku12}. Thus, accretion might operate differently inside and outside of the ``snowline", strengthening the need for global models. Growth might be faster in the cold outer regions, even while subsequent 
\clearpage
\begin{deluxetable*}{ c c c}
\tablewidth{0pt}
\tabletypesize{\scriptsize}
\tablecaption{Selected Symbols and Parameters used in this paper}
\tablehead{
\colhead{{\bf Symbol}}	&	\colhead{{\bf Definition}}	&	\colhead{{\bf Section}}}
\startdata
$c$ & gas sound speed &  sec. 2.1  \\
$D_{{\rm v,d}}$ & (vapor, ``dust") diffusion coefficient&  Sec. 2.2.2  \\
$h_{\rm{d}}$ & effective vertical scale height of ``dust" subdisk & Sec. 2.2.3 \\
$i,\, j, \, k$ & indices for species, radial bin, and mass &  Sec. 2\\
$L_\star$ & time variable stellar luminosity & Sec. 2.3.1 \\
$m, \, m^{\prime}, \, m_k $ & particle mass &  Sec. 2.4\\
$m_*, \, r_*$ & fragmentation mass and radius & Sec. 2.4.2\\
$\dot{M}, \, \dot{M}_i^{{\rm v,d}}$ & mass accretion rate, general and for species $i$ vapor and dust  &  Sec. 2.1, b.2 \\
$M_{\rm{D}}$ & initial disk mass &   sec. 2.1 \\
$M^{\rm{m}}_k $ & total mass of ``migrators" $m_k$ in a radial bin &  Sec. 2.4.4 \\
$Q_{\rm{T}}, \,Q_{\rm{d}}$ & Toomre parameters for gas and particle subdisk &   Sec. 2 \\
$Q_*, \,C_0$ & specific energy for fragmentation and bouncing &   Sec. 2.4.2\\
$Q_{\rm{a}},\,Q_{\rm{s}},\,Q_{\rm{e}},\,Q_{\rm{e}}^{\prime}$ & radiation transfer efficiencies of particles &  Appendix A.3\\
$R, \, R_0$ & radial coordinate, and initial disk radius &   Sec. 1, 2.1 \\
$r, \, r_k$  & particle radii, general or for mass $m_k$ &   Sec. 2.1, 2.4 \\
$r_{\rm{M}}$  & radius of particle containing most mass in radial bin &  Sec. 2.4.6 \\
$r_{\rm{L}}$  & radius of largest particle in radial bin &  Sec. 2.4.1, 3.1.1\\
St,St$_{\rm{L}}$,St$_{\rm{M}}$ & particle Stokes number = $t_{\rm{s}} \Omega$ &   Sec. 2.2.1, 3.2 \\
St$_*$,St$_{\rm{b}}$,St$_{\rm{d}}$ & fragmentation, bouncing and drift Stokes numbers &  Sec. 3.2, 4.2 \\
$S_{{\rm{b}},*}(m,m^{\prime});\,S_{k,l} $ & sticking coefficients for bouncing or fragmentation &   Sec. 2.4.2, 2.4.4\\
$T,\,T_{{\rm ph}}$ & disk midplane and photospheric temperature &  Sec. 2.3.1\\
$t_{\rm{s}}$ & particle stopping time &   Sec. 2.2.1\\
$u, \, v$ & gas radial and azimuthal velocity & Sec. 2.4.3\\
$U_k, \, V_k$ & radial and azimuthal velocity for mass $m_k$ & Sec. 2.4.3\\
$V_k^{{\rm rad}}$ &  radial drift velocity of $m_k$ & Sec. 2.2.2, 2.4.5, Eq. 51\\
$V_{\rm{g}}$ &  viscous accretion velocity of nebula gas & Sec. 2.1, Eqs. 2, 6\\
$V_{{\rm v,d}}$ &  radial drift velocity (vapor, mass-weighted solid) &  Sec. 2.2.2\\
$\alpha_{\rm{t}}$ & turbulence parameter &  Sec. 2.1 \\
$\alpha^{{\rm v,d}}_i$ & (vapor, solid) mass fraction in species $i$ &  Sec. 2.2\\
$\Delta t_{{\rm min}},\, \Delta t_{{\rm sync}}$ & minimum code timestep and synchronization timestep & Sec. 2\\
$\Delta V_{{\rm pg}}$ & particle-gas relative velocity & Sec. 2.4.2, 2.4.3\\
$\Delta V_{{\rm pp}},\, \Delta V^{{\rm pp}}_{k,l}$ & particle-particle relative velocity & Sec. 2.4.2, 2.4.3, 2.4.4\\
$\kappa$ & Rosseland mean opacity & sec. 2.3.1, 2.3.2, Appendix A.3\\
$\lambda_{{\rm mfp}}$ & molecular mean free path & Sec. 2.2.1\\
$\nu,\,\nu_{\rm{m}}$ & turbulent and molecular viscosity &  Sec. 2.1 \\
$\Omega$ & Keplerian orbit frequency &  Sec. 1, 2 \\
$\rho$ & gas mass density &  Sec. 2.1\\
$\rho^{\rm{d}}_k $ & ``dust" particle  mass density for mass $m_k$ &  Sec. 2.2.2, 2.4.3 \\
$\rho^{\rm{m}}_k $ & ``migrator" particle mass density for mass $m_k$ &  Sec. 2.4.4 \\
$\rho_{\rm{p}} $ & particle material (internal) mass density  &   Sec. 2.4.1 \\
$\rho_{\rm{d}}$ & volume mass density of solids in ``dust" population &  Sec. 2.2.2 \\
$\sigma, \sigma(m,m^{\prime})$ & collision cross section &   Sec. 2.4.2, 2.4.4 \\
$\Sigma$ & gas surface mass density &   Sec. 2 \\
$\Sigma_i^{{\rm v,d}}$ & (vapor, solid) surface mass density in species $i$ &   Sec. 2.2.2 \\
$\tau$ & optical depth at thermal wavelengths = $\kappa \Sigma$ &  Sec. 2.3.1 \\
\enddata
\end{deluxetable*}
\noindent
radial drift might carry this material to the inner solar system where the now ice-free refractory material mightgrow less robustly. Our models include these material-dependent sticking properties. 

Of course, the nebula is warmer at smaller radial distances from the sun, due to a combination of solar heating and viscous dissipation. Drifting particles do not just get ``lost to the sun" but evaporate their volatiles along the way \citep{mv84,hg03,cz04,kor04,cc06,gar07}. Each volatile, including ices and silicates, has its own {\it Evaporation Front} (EF) and our model treats them all. Such EFs have a number of important implications, creating  deviations from some uniform ``cosmic" abundance of the different materials. Amongst the meteoritical implications of such deviations are the enrichment of nebula vapor in H$_2$O, with chemical and mineralogical implications \citep{fg06}, and the transport of O-isotopes formed and frozen out in the outer nebula to the inner nebula where they can become incorporated into meteorites \citep{yur07}. Each of these processes has a timescale associated with it (now poorly known), that can in principle be modeled and tied to meteoritical data. In studies of other materials besides H$_2$O, \citet{pas05} and \citet{cie15} have modeled variable sulfur chemistry, and \citet{yc12} have modeled temporal and radial variations of the nebula D/H ratio due to transport, but in neither case were radial drift of large particles included. It is a goal of our models to treat these processes. 

In calculating the thermal evolution of the nebula, our model incorporates two other significant advances. First, the nebula opacity is calculated self-consistently as particles grow (Sec. \ref{subsubsec:opac}). This is very important because once a particle exceeds the typical wavelength of thermal emission, its opacity decreases linearly with its radius. In all other models to date, either ISM (tiny grain) opacities are assumed even while we know growth must be occurring, or some other arbitrary ``representative" value of opacity is assumed. Secondly, we model the plausible luminosity evolution of the early sun.  In the first $10^5$-$10^6$ years of its life, the sun's luminosity is $10\times$ to $3\times$ larger than its main sequence value \citep{kus70,dm94}. Naturally this will have implications for the locations of the ``snowline" and other evaporation fronts. 

In addition to the bouncing, fragmentation, and radial drift barriers (which are driven by aerodynamical forcing of particle velocities by eddies on a range of scales), turbulent nebulae present one additional barrier to incremental growth. In global turbulence, small gas {\it density} fluctuations associated with pressure fluctuations amongst large scale eddies gravitationally excite random velocities of objects in the km- and larger size range, much like giant molecular clouds scatter stars in our galaxy. These relative velocities actually increase with planetesimal size because of the slower damping for larger objects \citep{ng10,gre12}. \citet{ida08} showed that these velocities are strongly disruptive for $1-10$ km objects for nominal levels of nebula turbulence (objects of 100km size and larger are stabilized by their own self gravity). This serial gauntlet of barriers to growth in turbulence was discussed further by \citet{oo13} and \citet{joh14}. 

Recent years have seen the emergence of ``leapfrog" models that circumvent {\it all} these barriers, producing 10-100km size primitive bodies directly, in one stage, from mm-dm size objects. A discussion of these models
\citep{cuz01,cuz08,cuz10,cuz14b,ch12,joh07,joh09,joh11,cha10,car15} is beyond the scope of this paper 
\citep[for a recent review, see][]{joh15} but they share a preference for local enhancement of solids over the cosmic abundance ratio by factors of 10 or more. It is possible that radial drift can scavenge the outer reaches of the nebula, and contribute to such an enhancement of solids in the inner nebula; even the region of TNO formation might, in principle, be augmented in solids by strong radial drifts scavenging the rarefied outer nebula at hundreds of AU from the sun \citep{sv96,sv97,ha10,ha12}. It may even be that such radial redistribution might lead to the ``stubby" radial distribution of solids that \citet{des07} noted is the implication of pre-migration planetary distributions. \citet{cc06} found only small degrees of enhancement for the inner solar system, but they made several simplifying assumptions. In this paper we discuss the possibilities for large-scale rearrangement of the solids-to-gas ratio in the nebula, as a function of time. 

Given the wide range in timescales associated with modeling the nebula over its entire radial extent,
which could extend to as much as $\sim 1000$ AU, coupled with the computationally expensive implementation 
of the collisional coagulation equation \citep{smo16} has previously made such modeling efforts 
prohibitive. This has typically restricted global models to, for example, either studying dust growth and
redistribution alone \citep[e.g.,][]{bra08,bir10}, or studying compositional 
enhancements at evaporation fronts (EFs) due to simplified assumptions about growth \citep[e.g.,][]{cz04, 
cc06,gar07} or ones that treat growth more carefully, but neglect radial drift \citep{win12a,win12b,gar13}. 
Taken together these processes are likely to be critical to the evolutionary history of solids and condensibles.

Our model incorporates what may be an even faster numerical approach than that of \citet{bra08} (the method of moments, see section \ref{subsec:growth}) which avoids direct calculations of the detailed size distribution for small particles, while still continuing to treat the larger particles explicitly with detailed size, strength, and relative velocity distributions, and our opacity uses the local, evolving size distribution. It is the large particles that are of most interest for radial transport of solids, and it is the details of {\it their} sizes that may ultimately distinguish between various proposed ``leapfrog" models of planetesimal formation, so they receive a higher-fidelity treatment using the full coagulation equation.

\vspace{0.25in}
\section{Nebula Model}
\label{sec:model}

The simulations presented in this paper are done using a new $1+1$D radial nebula code
that is capable of simultaneously treating particle growth and radial migration
over many decades of particle size, while simulating the dynamical and thermal evolution of the
circumstellar gas disk. Our model includes self-consistent growth and radial drift of particles of
all sizes, accounts for the vertical diffusion and settling of smaller grains, radial diffusion and advection 
of dust and vapor phases of multiple species, and a self-consistent calculation of opacity and disk
temperature which allows us to track the evaporation and condensation of the various species as they
are transported throughout the disk. Our code is parallelized in radial bins which is a natural
step in attacking problems of this magnitude. The development of further innovations
and techniques to treat some of the more computationally expensive processes helps to make the problem
even more tractable up to and including additional parallelization in mass bins. 

In our code, several indices will be used. In general, we reserve the index $j$ to refer to quantities
that are functions of semimajor axis, $k$ will be used to refer to the histogram
of particle masses, whereas the index $i$ will exclusively be used for compositional species. As the need
arises for additional indices, we assign the index $l$ for dummy variables. 
For our model variables, a lack of subscript generally refers to a nebular quantity such as the gas surface
density $\Sigma$, or the pressure scale height $H$. The subscripts or superscripts $ d $, $ v $ or $ m $ refer to properties
of the dust, vapor or migrators as described in the following sections. We specifically draw attention
to the fact that there are many different velocities that will be discussed in this paper; the ones most widely used are referenced in {\bf Table 1}.
Finally, quantities that can be generally referred to will be done so using a subscript, such as
the total volume density of solid material $\rho_{\rm{d}}$, whereas the same quantity that is
specific to a radial bin will have the subscript transposed to a superscript with the index used as
the subscript, e.g., $\rho^{\rm{d}}_j$. Most of the relevant parameters used in our code 
are summarized in Table 1.

We define the minimum time step $\Delta t_{\rm{min}}$ in our code to be a fraction of the innermost
radial bin's orbital period $P = 2\pi/\Omega$, where $\Omega = \sqrt{GM_\star/R^3}$ is the orbital
frequency, $M_\star$ is the stellar mass, $G$ is the
gravitational constant, and $R$ is the semi-major axis. This
presents a dilemma in that the dynamical times in the innermost portions of the disk where evolution
generally happens more quickly can be orders of magnitude shorter than those in the outer regions of 
the disk. Using the same time step globally is thus inefficient. 

In order to somewhat circumvent this
problem, we employ an asynchronous time stepping scheme which works as follows. Each radial
location in the disk $R$ has its own time step $\Delta t$ associated with it that is the same
fraction of its orbital period as is $\Delta t_{\rm{min}}$. We compare the time ``elapsed'' in a radial
bin with the total simulation time $t_{\rm{sim}} = I\Delta t_{\rm{min}}$, where $I$ is the number
of iterations the code has currently executed. If a bin has executed $N$ steps, then
its next execution will occur if $(N+1)\Delta t \leq I\Delta t_{\rm{min}}$. 
When the condition is satisfied, the counter $N$ is incremented by 1 and a time step
is executed for that bin. In this fashion the global ``time step'' for the code is the fraction of the 
orbital period we choose, usually $\leq P/4$. The innermost radial bin is called every iteration, but we 
avoid unnecessary calls to other radial bins.

Having each radial location effectively at a different evolutionary time poses a problem for the transport
of material across radial boundaries, either through diffusion or radial drift. We thus introduce the
concept of the ``synchronization time'', a predefined number of steps $N_{\rm{sync}}$ in which we 
periodically bring the simulation to the same time globally. When ${\rm{MOD}}(I,N_{\rm{sync}})=0$, all radial bins 
are executed with the time step $\Delta t_{\rm{sync}} = N_{\rm{sync}}\Delta t_{\rm{min}} - N
\Delta t$. In this paper we choose values $N_{\rm{sync}} \leq 100$. During the synchronization step 
we execute global calculations such as the nebula
gas evolution (Sec. \ref{subsec:gasevol}), solve the diffusion-advection equation (Sec. 
\ref{subsec:diffadv}), determine the disk temperature (Sec. \ref{subsubsec:temp}) and write out simulation 
data. On the other hand, radial drift of material can be on time scales which are quite fast
so that waiting for a synchronization step is not practical. Instead, we have developed a method to account
for radial drift that is called at all $\Delta t$ (Sec. \ref{subsubsec:drift}).
Once a synchronization step is completed, the
time counters ($I,N$) are set back to zero. In full global simulations, this synchronization
step will still happen much sooner than an orbit period in the outermost portions of the disk, but the savings in time can be significant. 
  
\subsection{Gas Evolution}
\label{subsec:gasevol}
%

For this work, we use a radial 1-D model for the time-dependent evolution of the gas surface density 
$\Sigma(R,t)$ and radial velocity $V_{\rm{g}}(R,t)$. In the case that the gravitational potential is due to a central point mass $M_\star$, the equations
for the radial evolution and velocity of the nebula gas under viscosity $\nu$ can be derived from the continuity
and angular momentum conservation equations, and are given by \citep{prin81}

\begin{equation}
\label{equ:sgas}
\frac{\partial \Sigma}{\partial t} = \frac{3}{R}\frac{\partial}{\partial R}\left\{R^{1/2}
\frac{\partial}{\partial R}(R^{1/2}\nu\Sigma)\right\},
\end{equation}

\begin{equation}
\label{equ:vgas}
V_{\rm{g}} = -\frac{3}{R^{1/2}\Sigma}\frac{\partial}{\partial R}\left(R^{1/2}\nu\Sigma\right).
\end{equation}

\noindent
The corresponding disk mass accretion rate is then
$\dot{M} = - 2\pi R\Sigma V_{\rm{g}}$ where the sign is chosen such that a positive value of $\dot{M}$
indicates accretion onto the central star. The {\it total} viscosity $\nu$ can be related to 
``$\alpha-$models'' in which $\nu = \alpha_{\rm{t}} c H = \alpha_{\rm{t}} c^2/\Omega$, where 
$\alpha_{\rm{t}}$ parametrizes 
the turbulent intensity, $H$ is the nebula gas pressure scale height, 
and the gas sound speed  can be expressed in terms of temperature $T$ as

\begin{equation}
\label{equ:sound}
c = \left(\frac{\gamma k_{\rm{B}} T}{\mu_{\rm{H}}}\right)^{1/2}.
\end{equation}

\noindent
In Eq. (\ref{equ:sound}) above, $k_{\rm{B}}$ is the Boltzmann constant, the adiabatic index $\gamma=1.4$
for a diatomic gas and $\mu_{\rm{H}} = 3.9\times 10^{-24}$ g is the mean mass per molecule in a mixture
of hydrogen gas that has $\sim 20$\% helium by number.

It is generally assumed that the disk is gravitationally stable to fragmentation. We monitor the stability 
of the disk in our code through the Toomre parameter \citep{toom64}
\begin{equation}
\label{equ:toomre}
Q_{\rm{T}} = \frac{c \Omega}{\pi G \Sigma}.
\end{equation}

\noindent
Values of $Q_{\rm{T}}$ less than unity lead to disk fragmentation, but for values 
$Q_{\rm{T}}\lesssim 2$ the disk is described 
as weakly unstable and may lead to the formation of clumps. However,
these clumps only form if the disk is atypically massive or cold \citep{raf05}, and unless there are
processes to keep the disk unstable, it is assumed that weak gravitational instabilities quickly lead to 
stabilization of the disk
via the excitation of spiral density waves. The process is thus self-limiting to the extent that these
waves carry away angular momentum that spread the disk, lowering $\Sigma$. If Toomre-unstable conditions arise, 
we can introduce a moderately large value of $\alpha_{\rm{t}}$ until stable conditions resume. In practice, however, this
does not happen in the disk models presented in this paper. 
 
We derive initial conditions for our disk models using the analytical expressions from \citet{LBP74},
as generalized by \citet{har98}, which are parametrized by some initial disk mass $M_{\rm{D}}$ and 
radius $R_0$. In determining the initial gas surface density
and radial velocity, the value and radial dependence of the turbulent viscosity is expressed as
a general power law of the form $\nu = \nu_0(R/R_0)^\beta$ \citep[e.g.,][]{har98}. These assumptions
lead to simple 1-D, vertically integrated and averaged expressions which can be readily derived from
\citet{har98} given here for $t = 0$:

\begin{equation}
\label{equ:insgas}
\Sigma(R,0) = \frac{M_{\rm{D}}}{\pi R^2_0}\left(\frac{2-\beta}{2}\right)\left(\frac{R}{R_0}
\right)^{-\beta}e^{-(R/R_0)^{2-\beta}};
\end{equation}

\begin{equation}
\label{equ:invgas}
V_{\rm{g}}(R,0) = -\frac{3\nu_0}{2R}\left(\frac{R}{R_0}\right)^{\beta}\left[1 - (4-2\beta)
\left(\frac{R}{R_0}\right)^{2-\beta}\right],
\end{equation}

\noindent
where $\nu_0 = \alpha_{\rm{t}} c^2_0/\Omega_0$, with all quantities evaluated at $R_0$.
Typical ranges for the
power law exponent that characterize plausible extremes of nebula radial variation are $0\leq \beta
\leq 3/2$ \citep[e.g.,][]{cuz03}. \citet{har98} favor $M_{\rm{D}} = 0.2$ M$_\odot$,
$R_0 = 10$ AU and $\beta = 1$ based on young star statistics and we use these as fiducial
values in this paper. The scale parameter $R_0$ can also be chosen to match solar system  specific angular momentum
\citep[$R_0=$4.5]{cuz03} or by other criteria \citep[$R_0=$53]{yc12}. Indeed the nominal value of $R_0$ has more often been set at larger values 20-60AU \citep{cc06,gar07,bra08,ha12,yc12} than chosen here but the rationale is not always clear. We note that these choices simply provide initial conditions
for the disk's surface density profile in terms of the initial disk mass $M_{\rm{D}}$, and that the
viscosity of the disk will in general not follow a single power law distribution - either initially, or as a 
function of time due to particle growth, changing opacity, and our self-consistent calculation of the disk temperature 
(see Sec. \ref{subsec:therm}). Thus, at subsequent timesteps the actual viscous evolution equations are 
integrated using the actual local properties. However, as noted in section \ref{sec:discuss}, these choices do influence disk evolution at early times of interest.

 
\subsection{Evolution of Dust and Vapor}
\label{subsec:diffadv}

A self consistent calculation of the disk temperature requires that we also know the distribution of
dust and vapor within the disk, and the initial distribution is established during our first calculation
of the disk temperature. Our model is capable of tracking any number of  species in both 
the vapor and solid phase. We define the concentrations of each species $i$ such that
\begin{equation}
\label{equ:alphas}
\alpha^{\rm{v,d}}_i = \frac{\Sigma^{\rm{v,d}}_i}{\Sigma},
\end{equation}

\noindent
is the fractional mass of constituent $i$, or the ratio of surface density of solids or vapor to the
gas surface density. In this paper, we include only five different species which
are listed in {\bf Table 2} along with their corresponding condensation temperatures $T_i$, densities $\rho_i$
and initial concentrations $x_i$ in the solid state. We denote by 
$f_{\rm{d}}$, $f_{\rm{v}}$ and $f_{\rm{m}}$ the total fractional amounts of the 
dust, vapor, and migrator (see Sec. \ref{subsubsec:mig}) components within the disk. For instance,
$f_{\rm{d}} = \sum_i \alpha^{\rm{d}}_i$. These quantities are updated locally at every time step 
$\Delta t$, and globally at every synchronization step $t_{\rm{sync}}$.

\subsubsection{Particle Stokes Number and Stopping Times}
\label{subsubsec:Standts}

Treating the growth and redistribution of solids in the nebula concomitant with its gas and thermal
evolution is a key aspect of this work, and is essential for a self-consistent model of the nebula. 
Initially, the dust-to-gas ratio in the
disk is assumed to be of order $\sim 10^{-2}$ which is consistent with values for the ISM. At this average abundance, the mass volume density of solid material $\rho_{\rm{d}}$ does not affect 
the motion  of the gas unless settling of dust particles leads to a layer in the midplane with density 
$\rho_{\rm{d}} \geq \rho$, where $\rho$ is the gas mass density \citep{nak86}. In fact, we can define an equivalent condition to Eq. (\ref{equ:toomre}) for the dust $Q_{\rm{d}} = \Omega^2/\pi G \rho_{\rm{d}}$,
such that if $Q_{\rm{d}} \lesssim Q_{\rm{T}}$, the condition for gravitational instability of the dust layer 
may be satisfied \citep{saf91,cuz93}. However, even for $Q_{\rm{d}} \gg Q_{\rm{T}}$, collectively the 
dust can play a significant role in the disk's thermal evolution and thus affect $\Sigma$ and $V_{\rm{g}}$ through
the temperature distribution, if a large fraction of the dust particle sizes remain small such that the 
opacity remains high (Sec. \ref{subsubsec:opac}). In particular, differences
in the growth and migration rate, as well as the possibility for evaporation and condensation of solid 
grains, can lead to significant radial variation in the distribution of solids.

For a large range of solids-to-gas ratios, the influence of the nebula gas on the motion of the particles 
over the full spectrum of sizes can be described by the dimensionless Stokes number

\begin{equation}
\label{equ:stokes}
{\rm{St}} \equiv \frac{t_{\rm{s}}}{t_{\rm{ed}}},
\end{equation}

\noindent
where $t_{\rm{s}}$ is the particle stopping time, and $t_{\rm{ed}}$ is some eddy turnover time, which we choose here to be the integral scale, or turnover time of the largest eddy  
in the turbulence. This is generally assumed (and has been shown) to be $\Omega^{-1}$ for global turbulence \citep{cuz01,joh07,car10}. The particle stopping time is defined as the time needed for the gas drag force to dissipate a particle's
momentum relative to the gas. The drag force on a particle of radius $r$ depends on the size of
the particle relative to the molecular mean free path $\lambda_{\rm{mfp}}$, and can be separated into
two distinct flow regimes:

\begin{equation}
\label{equ:tseps}
{t_{\rm{s}}} = \frac{\rho_{\rm{p}}}{\rho}\frac{r}{c}\,\,\,\,\,{\rm{if}}\,{r} \leq (9/4)\lambda_{\rm{mfp}};
\end{equation}

\begin{equation}
\label{equ:tssto}
{t_{\rm{s}}} = \frac{8}{3}\frac{\rho_{\rm{p}}}{\rho}\frac{r}{C_{\rm{d}}\Delta V_{\rm{pg}}}\,\,\,\,\,
\,{\rm{if}}\,{r} > (9/4)\lambda_{\rm{mfp}},
\end{equation}

\noindent
where $\rho_{\rm{p}}$ is the particle material density (section \ref{subsubsec:coag}).
Equation {\ref{equ:tseps}} describes the {\it Epstein flow} regime in which smaller grains are well coupled to the
gas flow and relative velocities between grains are small. In the larger particle {\it Stokes flow} regime 
(Eq. [\ref{equ:tssto}]), particles become increasingly less coupled to the gas flow and their stopping
times are affected by the drag coefficient $C_{\rm{d}}$ which depends on a particle Reynolds number
${\rm{Re}}_{\rm{p}}$ that is itself a function of the relative velocity $\Delta V_{\rm{pg}}$ 
between the particle and the gas \citep{wei77}.

In our code, we calculate the particle-to-gas and particle-to-particle relative velocities over the
evolving size distribution which can cover many decades in mass. As growth proceeds to larger sizes, some particles will remain in the Epstein flow regime, while larger ones will be subject to
Stokes flow. We use a bridging expression that provides a smooth transition between the two regimes
after \citet{pod88}. Our treatment of the stopping times is described in further detail in
Appendix A.4.

\subsubsection{Radial Diffusion-Advection}
\label{subsubsec:diffadv}

We determine the radial motion of the solid and vapor fractions of all compositional
species by solving the advection-diffusion equation

\begin{equation}
\label{equ:diffadv}
\frac{\partial{\Sigma^{\rm{v,d}}_i}}{\partial{t}} = \frac{1}{R}\frac{\partial}{\partial{R}}
\left\{R D_{\rm{v,d}}\Sigma \frac{\partial{\alpha^{\rm{v,d}}_i}}{\partial{R}} - RV_{\rm{v,d}}
\Sigma^{\rm{v,d}}_i \right\} + {\mathcal{S}}_i,
\end{equation}

\noindent
where $\Sigma^{\rm{v,d}}_i$ is the surface density for dust (d) or vapor (v) of species $i$, $V_{\rm{v,d}}$ is a net, inertial space advection velocity, 
and $D_{\rm{v,d}}$ is the diffusivity.
\clearpage
\begin{deluxetable}{l c c c}
\tablewidth{0pt}
\tablecaption{List of Species}
\tablehead{
\colhead{Species}	&	\colhead{$T_{\rm{sp}}$ (K)}	&
\colhead{$\rho_{\rm{sp}}$ (g cm$^{-3}$)}	&	\colhead{$x_{\rm{sp}}$ ($T<T_{\rm{sp}}$)}}
\startdata
Iron & 1810 & 7.8 & $1.26\times 10^{-4}$ \\
Silicates & 1450 & 3.4 & $3.41\times 10^{-3}$ \\
Troilite & 680 & 4.8 & $7.68\times 10^{-4}$ \\
Organics & 425 & 1.5 & $4.132\times 10^{-3}$ \\
Waterice & 160 & 0.9 & $5.55\times 10^{-3}$ \\
\enddata
\end{deluxetable}
\noindent
The separately treated term ${\mathcal{S}}_i$ represents sources
and sinks for ``dust'' and vapor for species $i$ which, in our treatment, includes the growth, 
radial drift and destruction of migrating material (Sec. \ref{subsubsec:mig}). The sign of the advection 
velocity in Eq. (\ref{equ:diffadv}) is 
such that $V_{\rm{v,d}} < 0$ indicates inward radial velocity, 
and $V_{\rm{v,d}} > 0$ outward radial velocity.

For the vapor phase, we assume $D_{\rm{v}} = D_{\rm{g}} = \nu/{\rm{Sc}_g}$ \citep{ha10}, where the gas
diffusivity $D_{\rm{g}}$ is the ratio of the viscosity to the Schmidt number Sc$_{\rm{g}}$ (which we take
to be unity in this work, see Appendix B) and the vapor advection velocity is just 
that of the gas; $V_{\rm{v}} = V_{\rm{g}}$ (equations \ref{equ:vgas}  and \ref{equ:invgas}).
The particle diffusivity $D_{\rm{d}}= D_{\rm{g}}/(1 + {\rm{St}}^2)$ \citep{yl07,car11} and $V_{\rm{d}}$ are determined at radius $R$ through a mass-weighted mean of all grain sizes in the dust population (indexed by $k$): 

\begin{equation}
\label{equ:ddiff}
D_{\rm{d}} = \sum_{k} \frac{\rho^{\rm{d}}_k}{\rho_{\rm{d}}}\left(\frac{D_{\rm{g}}}{1 + {\rm{St}}_k^2}\right);
\end{equation}
 
\begin{equation}
\label{equ:vrad}
V_{\rm{d}} = \sum_{k} \frac{\rho^{\rm{d}}_k}{\rho_{\rm{d}}} V^{\rm{rad}}_k,
\end{equation}

\noindent
where $\rho_{\rm{d}} = \sum_k \rho^{\rm{d}}_k$ is the total mass volume density of solids in the 
subdisk layer (see Sec. \ref{subsubsec:subdisk}). The radial velocity $V^{\rm{rad}}_k$ in Eq. 
(\ref{equ:vrad}) properly takes into account the radial drift of particles {\it relative to the gas} due to the local pressure 
gradient \citep[e.g.,][]{nak86,tl02} and is described in more detail in Sec. \ref{subsubsec:drift} 
(Eq. [\ref{equ:vdrift}]).

The sign of the mean radial drift velocity depends on
the particle size (large particles drift inwards and small ones are advected outwards). We utilize a power law distribution in
particle mass for the dust population with a typical exponent $q=11/6$, which is representative of a 
collisional population (see Sec. \ref{subsubsec:coag}), so that most of the mass will be in the largest sizes and the 
mean velocity will always be inward. However, the mass average approach by itself  as written in Eq. (\ref{equ:vrad}) would fail to take into account 
that the smallest grains would still move with the gas, and in the outer parts of the disk, this flow 
may be outward. In order to account for this in our code, we treat the dust as two separate populations 
and define a mean velocity for each, $V^+_{\rm{d}}$ which characterizes the mass fraction in small 
particles that move with the gas, and $V^-_{\rm{d}}$ which characterizes the larger grains that 
radially drift inward. 
Before calculating the mean drift velocities, we determine the grain size that separates
the two populations, and define the total fractional mass in each population which is used as a weighting
factor when solving the advection-diffusion equation (see Appendix A.2). 

\subsubsection{Vertical Diffusion and Subdisk Height}
\label{subsubsec:subdisk}


We treat the vertical diffusion and settling of dust grains in the ``+1D" part of our code using an analytical solution combining elements of \citet{dub95,cw06} and \citet{yl07} to calculate the particle distribution as a function of height $z$ in the disk.  The vertical scale
height $h_{\rm{d}}$ of particles with mass $m$ is defined as 
\begin{equation}
\label{equ:subdisk}
h_{\rm{d}}(m) = H(1 + {\rm{St}}/\alpha_{\rm{t}})^{-1/2}.
\end{equation}
This solution bears a resemblance to that of \citet[][her Eq. 22]{gar07} but includes a number of subtleties (see Appendix B).




 
To calculate a characteristic scale height $h_{\rm{d}}$ for the local particle subdisk as a whole, we specify 
a representative particle mass as either half the fragmentation barrier mass $m_*$ or the largest particle mass 
$m_{\rm{L}}/2$ (if no particles have yet reached $m_*$; see Sec. \ref{subsubsec:stick}),  
based on focused 2D calculations that follow growth as a function of height at a given radius $R$. In future refinements, a value of $h_{\rm{d}}$ could be tracked for all particle sizes, giving a complete size distribution as a function of altitude (as for instance in Sec. \ref{subsubsec:relvel}, Eq. [\ref{equ:rhod}]). The
relative velocities (Sec. \ref{subsubsec:relvel})  are also defined at the same representative $h_{\rm{d}}$. 
The layer of thickness $h_{\rm{d}}$ defines the volume accessible to particles growing in the midplane.


\subsection{Disk Thermal Evolution}
\label{subsec:therm}

\subsubsection{Temperature}
\label{subsubsec:temp}

We calculate the protoplanetary disk midplane and 
photosphere temperatures self-consistently, assuming that the nebula is heated by a combination of 
internal viscous dissipation $\dot{E}_\nu$, at a rate proportional to $\Sigma$ and $\nu$ 
and thus primarily near the midplane, and external illumination $\dot{E}_\star$ by the stellar luminosity 
$L_\star$ (which can vary with time; see below). The stellar luminosity heats the upper layers of the disk on both sides, 
and thus indirectly the material beneath \citep[e.g.,][]{rp91,wc99,cie10}. 
The thermal energy of the disk is radiated into space
from the photosphere at $T_{\rm{ph}}(R,t)$, the altitude where the optical depth at thermal wavelengths measured
vertically outwards is roughly unity. For cosmic abundance and a standard ISM-MRN grain size distribution,
the photosphere lies at a distance $H_{\rm{ph}} \sim \pm \,3-5$ H from the midplane. On each face of 
the nebula, we can express the energy balance as $\sigma_{\rm{SB}}
T^4_{\rm{ph}} = \dot{E}_\nu/2 + \dot{E}_\star$, where $\sigma_{\rm{SB}}$ is the Stefan-Boltzmann constant.
In reality, the disk has an optically thin hot exosphere \citep{cg97,dul02}
which indirectly warms the disk photosphere to $T_{\rm{ph}}$, but it can be shown that modeling direct
deposition of solar energy at, and thermal radiation by, the disk photosphere leads to the same
$T_{\rm{ph}}$. Note that Eq. (12a) of \citet{cg97} has an
extraneous factor of $2^{1/4}$ which is removed by integration over all angles of the optically thin 
emission from the superheated exosphere slab, as for instance in \citet{nn94}.

We are primarily interested in the midplane temperature $T$, because planetesimals and boulder-size
drifting particles which transport solids radially and feed EFs lie mostly near the midplane and not at
high altitudes. The midplane temperature is influenced both by $\dot{E}_\nu$ and $\dot{E}_\star$. The
external irradiation is both deposited and re-radiated at the disk photosphere, producing zero net
vertical flux through the rest of the disk in steady state and thus a vertically constant temperature
below the photosphere in the absence of other energy sources. The energy produced by viscous dissipation
must flow vertically away from the midplane (assuming it is primarily produced there, as we do) to the photosphere to be radiated away, leading to a 
vertical thermal gradient. The resulting temperature distribution depends on whether the disk is
optically thick or thin. \citet{nn94} suggest a ``bridging'' expression that covers both
of these regimes (their Eq. A15), which can be simplified to:

\begin{equation}
\label{equ:bridge}
\sigma_{\rm{SB}}T^4 = \left(\frac{3\tau}{8} + \frac{1}{2\tau}\right)\frac{\dot{E}_\nu}{2} + \dot{E}_\star,
\end{equation}

\noindent
where $\tau$ is the full optical depth of the nebula at thermal wavelengths. Between $\pm H_{\rm{ph}}$, 
$\tau = \kappa \Sigma/2$, where $\kappa$ is the average thermal opacity. Energy from infall shocks was also 
included by \citet{nn94}, but we do not treat it in this paper.

Comparing Eq. (\ref{equ:bridge}) to that for the disk surface shows that the midplane temperature responds
differently to internal and external energy sources. Regarding internal sources, most previous workers
have treated only the optically thick limit ($3\tau/8$ term above). For instance, \citet{cas94},
and \citet{wc99} generalized the classical radiative transfer solution for a layer of
optical depth $\tau \gg 1$ which they defined from the midplane outwards as $T^4 = (3/4)\tau T^4_{\rm{ph}}$. This is the classical
Eddington solution where all the energy is produced at the bottom of the layer, and is thus only half 
of the value in Eq. (\ref{equ:bridge}). Cassen (1993; and
others subsequently) merely stated that the leading factor of $3/8$ applies to the more realistic
situation where the energy production is vertically distributed and roughly proportional to the
mass density. The derivation of the $3/8$ factor for the regime of interest can be extracted from
\citet{ss73}, and assumes constant $\kappa$ which, for large optical depths, is
the Rosseland mean opacity (see Sec. \ref{subsubsec:opac}).

The $1/2\tau$ term in Eq. (\ref{equ:bridge}) treats the opposite limit where significant internally
generated energy must be radiated away, but the nebula opacity capable of doing this is low, for
instance, if most of the grains have evaporated. In this optically thin regime the emitted flux from
each face is given by $2\tau \sigma_{\rm{SB}}T^4$, where $\tau$ is the full optical depth, and the factor
of 2 comes from integrating intensity over all angles. In this regime, temperature is roughly independent
of altitude ($T \sim T_{\rm{ph}}$). 

We assume a local, vertically integrated viscous disspation rate $\dot{E}_\nu = (9/4)\Sigma
\nu \Omega^2$ which is closely related to the mass accretion rate \citep{lp85}. \citet{ss73} 
note that the local energy
production rate $\dot{E}_\nu$ differs by a factor of three from the local rate of release of
gravitational potential energy. The stellar flux on each disk face is given by

\begin{equation}
\dot{E}_\star = \frac{L_\star \phi}{4\pi R^2} = \phi \sigma_{\rm{SB}} T^4_\star \left(\frac{R_\star}{R}
\right)^2,
\end{equation}
 
\noindent
where $\phi$ is some grazing incidence angle depending on disk geometry, and the stellar luminosity
is $L_\star = 4\pi R^2_\star \sigma_{\rm{SB}} T^4_\star$, with $R_\star$ and $T_\star$ the stellar
radius and photospheric temperature. The incidence angle $\phi$ which determines the solar disk heating
differs considerably between so-called ``flat'' disks, having photospheric height $H_{\rm{ph}}$ a 
constant fraction of the distance $R$ from the star, and ``flared'' disks where $dH_{\rm{ph}}/dR$
increases with $R$, which leads to much larger thickness. Whether the disk is flared or flat
has implications for the gas mass densities we use to determine particle interactions, growth and
drift (Sec. \ref{subsubsec:coag}). For the more realistic flared disks, values of $\phi$ were
derived by \citet{kh87}, and \citet{rp91}, and reiterated by \citet{cg97}, who also derive the general radial variation

\begin{equation}
\label{equ:phi}
\begin{split}
\phi(R) \sim 0.4\frac{R_\star}{R} + R\frac{d}{dR}\left(\frac{H_{\rm{ph}}}{R}\right) \sim \\
0.005R^{-1}_{\rm{AU}} + 0.05R^{2/7}_{\rm{AU}},
\end{split}
\end{equation}

\noindent
where $R_{\rm{AU}}$ is radial distance in AU, and the first term on the RHS is the ``flat disk''
value \citep[e.g.,][]{as86}. The second term on the RHS can be understood as a manipulated
version of the incidence angle at the photosphere of a flared disk given more obviously by
$dH/dR - H/R$. Thus the illumination term is dominated by flared disk geometry in general. Iterative
treatment of $\phi(R)$, depending on $T_{\rm{ph}}$ and $L_\star$, is left for future work.

In one subtle difference, \citet{cg97} and \citet{rp91} appear to assume that each disk face sees only half of the stellar flux, while \citet{cie09,cie10}
assumes that the entire star is visible and the flux normal to itself at the disk
face is simply $L_\star/4\pi R^2 = \sigma_{\rm{SB}}T^4_\star (R_\star/R)^2$. For the
entire star to be visible from some point at $(R,H_{\rm{ph}})$, the opaque disk must vanish inside of
a radius $R_{\rm{min}}(R)$, where the photosphere height is $H_{\rm{ph,\,min}}$, such that

\begin{equation}
\frac{H_{\rm{ph,\,min}} + R_\star}{R_{\rm{min}}} \sim \frac{H_{\rm{ph}} + R_\star}{R},
\end{equation}

\noindent
so using $H_{\rm{ph}}/R \sim 0.2R^{2/7}_{\rm{AU}}$ \citep{cg97} at both $R$ and
$R_{\rm{min}}$, we find

\begin{equation}
R_{\rm{min}}(R) = R\left[1 + 0.2\frac{R}{R_\star}R^{2/7}_{\rm{AU}}\right]^{-1}.
\end{equation}

\noindent
That is, even from a point on the disk photosphere at 3 AU from the star, the entire stellar disk can be seen unless the opaque nebula disk extends further inwards than $3R_\star$, and the stellar disk becomes
more visible at larger distances. Thus we will assume the full stellar flux illuminates each
face of the disk (this is possible because the disk is flared and not planar). On the other hand, the value of $\phi$ adopted by \citet{cie09,cie10} is
several times smaller than the values given by Eq. (\ref{equ:phi}) above, reducing the
stellar flux accordingly. The equation determining the midplane temperature must be solved iteratively using a root 
finding algorithm: 

\begin{equation}
\label{equ:temp}
\sigma_{\rm{SB}}T^4  = 
\frac{9}{8}\nu \Sigma \Omega^2\left(\frac{3\tau}{8}+\frac{1}{2\tau}\right) +
\frac{L_\star \phi}{4\pi R^2},
\end{equation}

\noindent
where the Rosseland mean opacity $\kappa$ is a 
function of $\tau$ through the $\tau-$dependent ($T$-dependent) dominant wavelength, the evolving particle size 
distribution, and the solids fractions of all species $i$. 

We employ a time variable luminosity
$L_\star(t)$ using the model of \citet[][Table 3]{dm94} for a 1 M$_\odot$ star. We fit a polynomial
to these authors' tabulated values, which cover a time scale of $7\times 10^4 - 1\times 10^8$ years
after collapse. This gives an initial stellar luminosity of 
$L_\star \approx 12$ L$_\odot$ at $7\times 10^4$ years, which we use as the starting
time for our simulations, dropping to perhaps 3 L$_\odot$ at $10^6$ years.


\subsubsection{Rosseland Mean Opacity}
\label{subsubsec:opac}

The expressions from Sec. \ref{subsubsec:temp} relating the midplane temperature to the sources of
energy assume a wavelength-independent or grey opacity. The transfer of thermal radiation in regions of
high optical depth is maximized at wavelengths where opacity is {\it low}. The standard treatment is to
define a Rosseland mean opacity $\kappa$ from the basic wavelength-dependent opacity
$\kappa_\lambda$, weighting the inverse (the transparency) at wavelength $\lambda$ by the derivative of 
the Planck function $dB_\lambda(T)/dT$ which manifests the local flux gradient:

\begin{equation}
\label{equ:kappar}
\kappa^{-1} =  \frac{\int \kappa_{\lambda}^{-1}\frac{dB_{\lambda}}{dT}\,d\lambda}
{\int \frac{dB_\lambda}{dT},d\lambda} = \frac{\pi}{4 \sigma_{\rm{SB}}T^3}\int \kappa_\lambda^{-1}
\frac{dB_\lambda}{dT}\,d\lambda.
\end{equation}

\noindent 
The Planck opacity, which is preferred for low optical depth regions \citep[e.g.,][]{nn94}, can be calculated 
similarly through a straight average of the wavelength-dependent opacities $\kappa_\lambda$ as weighted 
by the Planck function $B_\lambda(T)$. \citet{pol94} show that the distinction between the
Rosseland and Planck opacities is not large for small solid particles, and thus as previously stated we 
include only the Rosseland mean opacity in our temperature calculations. 

We utilize a new opacity model in order to determine the $\kappa_\lambda$, 
which is fully described in Cuzzi et al. (2014; their Appendix A contains a derivation of Eq. [\ref{equ:kappar}]), which can be easily incorporated into evolutionary models at little
computational cost. We utilize realistic material refractive indices for a cosmic abundance suite that
likely characterizes nebula solids: water ice, silicates, refractory organics, iron sulfide and metallic
iron (summarized in Table 2). These indices and relative abundances are taken
from \citet{pol94}, but alternate tabulations can readily be used in our code \citep[e.g.,][]{dl84,hen99}.
We briefly summarize how the opacity model of \citet{cuz14a} is applied
in our model in Appendix A.3.

\subsection{Solid Body Growth}
\label{subsec:growth}

Global modeling of the aggregation and radial evolution of solids in the protoplanetary nebula is
required over time scales of millions of years in order to understand many key aspects of primitive
bodies. A key feature of our model is the capability of modeling particle growth over many decades of particle
size, from sub-micron-sized dust to meter-sized and larger boulders which can radially drift large distances
as they grow. 

Growth by sticking in the protoplanetary disk starts with sub-micron grains which are
dynamically coupled to the nebula gas, and proceeds incrementally through larger sizes that collide at larger relative velocities (Sec. \ref{subsubsec:relvel}). 
The growth of aggregates continues at a rate determined by local nebula conditions  
until some ``bouncing barrier" is reached where collisional energy cannot be dissipated, but remains inadequate to destroy the colliding particles (see Sec. \ref{sec:intro}). The particle size where this occurs depends on assumed particle strength and the local value of turbulent $\alpha_{\rm{t}}$. 

The bouncing barrier is not impermeable, but merely slows growth by restricting collision partners. Growth beyond the bouncing barrier may continue by accretion of sufficiently smaller particles, through a transition regime of ``sub-migrators" where 
particles are highly susceptible to mutual collisional destruction (fragmentation) as they drift radially, to 
``migrators'' which have grown large enough to have a much lower probability of destruction \citep{ec09}. 
The latter particles can drift large distances and grow further, perhaps even into planetesimals.

The treatment of growth up to the bouncing and/or fragmentation barriers has until recently presented the most 
significant challenge, because a full-scale solution to the problem of dust coagulation for a size spectrum at 
every spatial location in the disk, and as a function of time, has been computationally prohibitive, and 
as a result previous models have been limited in different ways (see Sec. \ref{sec:intro}).
Growth through the transitional sub-migrator and migrator regimes between dust and planetesimals, and
associated radial drift over long times, is even less thoroughly studied. 
In our model, we use the moments method of \citet{ec08} to greatly accelerate coagulation modeling up to the bouncing
barrier, as described below in Sec. \ref{subsubsec:coag}. Growth beyond the bouncing barrier through the transitional regimes
is treated explicitly in the traditional way, as described in Sec. \ref{subsubsec:mig}.

In this paper we do not treat collective concentration or sweepup effects such as streaming instabilities 
\citep[][see however, Sec. \ref{subsec:mass}]{gp00,yg05}, turbulent concentration \citep{cuz08,cuz10}, or ``pebble accretion" \citep{ok10,lj12}.
All these subsequent processes depend strongly on initial local conditions (primarily, particle size and abundance) which are determined by sticking for some given $\alpha_{\rm{t}}$ in the presence of drift, and it is those conditions which are the outcome of the models described here.

\subsubsection{Coagulative Grain Growth}
\label{subsubsec:coag}

The standard approach to modeling dust coagulation involves solving some form of the
collisional coagulation equation \citep{smo16}

\begin{equation}
\label{equ:scce}
\begin{split}
\frac{df(m,t)}{dt} =
\frac{1}{2}\int_0^m K(m-m^\prime,m^\prime)f(m-m^\prime,t) \times \\
f(m^\prime,t)\,dm^\prime -
\int_0^\infty K(m,m^\prime)f(m,t)f(m^\prime,t)\,dm^\prime
\end{split},
\end{equation}

\noindent
where $f(m,t)$ is the particle number density per unit mass for mass $m$, and the collisional kernel
$K(m,m^\prime) $ contains all of the relevant information about the interacting masses $m$ and $m^\prime$,
such as their mutual relative velocities, collisional cross-sections, and sticking efficiencies, and can
include other properties such as particle porosities \citep[e.g., see][]{ec08}. The difficulty
with the application of Eq. (\ref{equ:scce}) that has made global models
of nebula evolution intractable is that the calculation is computationally expensive.
Coagulative grain growth can span many decades of particle mass which may require $\sim 10^2-10^3$ bins 
or more depending on desired accuracy, and at least the first integral of Eq. (\ref{equ:scce}) must be solved for {\it each} mass bin, thus solving Eq. (\ref{equ:scce}) at every spatial location and 
at every time step can become intractable. It has been shown that taking shortcuts with the number of mass
bins risks producing artificial growth.

\citet{ec08,ec09} developed a
scheme for modeling coagulative growth that overcomes these difficulties and that lends itself quite nicely
to problems in which one is interested in globally simulating the evolution of solids over a wide range
of sizes where the detailed size distribution {\it at small sizes} both has a generally predictable form, and is of secondary interest. This approach uses a finite 
number of moments ${\mathcal{M}}_l$ of the particle mass distribution, defined by

\begin{equation}
\label{equ:mom}
{\mathcal{M}}_l = \int_0^\infty m^lf(m,t)\,dm,
\end{equation}

\noindent
to track general properties of the particle population over time under the assumption
that the general form of the particle mass distribution up to some fragmentation mass $m_*$ is known.  
Note that, by this definition of $f(m)$, the mass volume density in a bin of width $dm$ is $m f(m)\,dm$. In the
moments method, the coagulation equation is reduced to a set of ordinary differential equations

\begin{equation}
\label{equ:dmom}
\begin{split}
\frac{d{\mathcal{M}}_l}{dt} = \int_0^\infty \int_0^\infty \left[\frac{1}{2}(m+m^\prime)^l - m^l\right]\times \\
K(m,m^\prime)f(m,t)f(m^\prime,t)\,dm\,dm^\prime
\end{split},
\end{equation}

\noindent
which leads to a closed set of equations as long as the total number of powers of $m$ in the expression
on the RHS is $\leq l$, which is true even for the realistic collisional kernel we use.
In this work, we assume that the dust mass distribution is a power law 
$f(m) \propto m^{-q}$ with (potentially variable) exponent $q$. The assumption of a power law distribution 
for the dust population is motivated by a number of detailed models \citep[e.g.,][]{wei97,wei00,dd05,bra08,bir10,bir11,bir12a} that show distributions with nearly constant mass per decade up to some upper limit which grows with time until a 
frustration limit is reached \citep[see][]{ec08}.

Under the assumption of a powerlaw with fixed exponent $q$, we can derive an equation for the time 
rate of change of the particle mass $m_{\rm{L}}$ that characterizes the largest mass in the distribution,  
until $m_{\rm{L}}$ gets as large as the fragmentation mass (see next section) \citep[Eq. (28) of][]{ec08}:
 
\begin{equation}
\label{equ:mL}
\begin{split}
\frac{dm_{\rm{L}}}{dt} = (3-q)(2-q)\rho \Gamma_2 \biggl[(3-q)\times\left(m_{\rm{L}}^{2-q}-m^{2-q}_{\rm{min}}\right)
\times \\ 
m_{\rm{L}}^{2-q} - (2-q)\left(m^{3-q}_{\rm{L}} - m^{3-q}_{\rm{min}}\right)
m^{1-q}_{\rm{L}}\biggr]^{-1}.
\end{split}
\end{equation}

\noindent
For $q < 2$, $m_{\rm{L}}$ is represented by the upper end of what we will refer to as the ``dust'' mass distribution. 
For $q=2$, 
there is equal mass per decade, whereas for $q>2$, most of the mass is in the smaller particles and the mass 
$m_{\rm{L}}$ will depend on the smallest size in the distribution. In Eq. (\ref{equ:mL}), $\Gamma_2$ is 
an integral over the collisional kernel derived from Eqns. (\ref{equ:mom}-\ref{equ:dmom}) for $l=2$
\citep[see][]{ec08}:

\begin{equation}
\label{equ:gamma2}
\Gamma_2 = \int_{m_{\rm{min}}}^{m_{\rm{L}}(t)}\int_{m_{\rm{min}}}^{m_{\rm{L}}(t)}
K(m,m^{\prime})m^{1-q}m^{\prime 1-q}\,dm\,dm^\prime.
\end{equation}

\noindent
The definition of the kernel $K(m,m^\prime)$ is discussed in Sec. \ref{subsubsec:stick}. We solve
Eqns. (\ref{equ:mL}) and (\ref{equ:gamma2}) using a 4th order Runge-Kutta method.

%
%


Because we assume a fixed power law representation of the dust particle mass distribution, its mass histogram 
is defined by the moments at any time, the index $q$, and for any resolution, by its lower and upper bounds.
We define a particle radius distribution logarithmically spaced from lower bound $r_{\rm{min}}$ to upper 
bound $r_{\rm{L}}$:

\begin{equation}
r_k = r_{\rm{min}}(r_{\rm{L}}/r_{\rm{min}})^{\frac{k-1}{n_{\rm{p}}-1}},
\end{equation}

\noindent
where $n_{\rm{p}}$ defines the number of points per decade of radius. \citet{dra13,dra14} have emphasized 
the importance of maintaining good mass resolution in brute-force solutions of the coagulation equation, finding that
$\sim 40$ bins per mass decade (or $\sim 120$ per radius decade) are generally required. However, such
a large number of bins is not required for the moments method, given our assumption of a power law distribution in the dust component.
We typically use $20-100$ bins per radius decade to calculate the relative velocities, which is more than
sufficient. The particle masses can then 
be determined using the average dust particle material density $\rho_{\rm{p}}$, which we determine from the
fractional masses of solid species:

\begin{equation}
\label{equ:density}
\rho_{\rm{p}} = \frac{\sum_i \alpha^{\rm{d}}_i}{\sum_i \alpha^{\rm{d}}_i \rho_i},
\end{equation}

\noindent
where $\rho_i$ is the material density of species $i$ (see Table 2). 
In this paper, we always assume that
the initial distribution has $r_{\rm{min}} = 0.1$ $\mu$m, and $r_{\rm{L}}(t=0) = 1$ $\mu$m at all $R$.
Though the density of dust particles
evolves as their composition changes, and thus $m_{\rm{min}}$ is not a true constant, the 
minimum {\it radius} $r_{\rm{min}}$ is a
constant in our evolutions. Much like modeling a variable $q$, we could in the future model a different or variable $r_{\rm{min}}$
by employing the appropriate number of moments \cite[see][]{ec08}. The logarithmic binning in $r_k$ (and
thus $m_k$) is used as the basis for calculations of relative velocities (Sec. \ref{subsubsec:relvel}), for 
characterizing the reservoir of ``dust'' material that migrators may sweep up (Sec. \ref{subsubsec:mig}), 
and to compute the opacity (Sec. \ref{subsubsec:opac}). 

We also note that, it is fairly straightforward to implement particle porosity in the growth parts of our 
code \citep[see][]{ec08}. Particle porosity may be very important \citep[see, e.g.][]{orm08,zso10,oku12} 
in allowing for particles to effectively grow to larger sizes while maintaining low St. Thus, their radial drift times would be longer perhaps allowing them to circumvent the radial drift barrier (see Sec. 
\ref{subsubsec:drift} and \ref{subsec:psize}). Furthermore, the particle porosity can itself evolve with 
size which would also affect the opacity in addition to stopping times. Our code can handle this modification, 
but we leave this further layer of complexity for a later paper.

\subsubsection{Particle Sticking, Bouncing and Fragmentation}
\label{subsubsec:stick}

The collisional kernel $K(m,m^\prime)$ can be factored into three components:

\begin{equation}
\label{equ:kernel}
K(m,m^\prime) = \sigma(m,m^\prime)\Delta V_{\rm{pp}}(m,m^\prime)S(m,m^\prime),
\end{equation}

\noindent
where $\sigma(m,m^\prime) = K_0(m^{1/3}+m^{\prime 1/3})^2$ is the collisional cross section between masses
$m$ and $m^\prime$, $K_0 = \pi(3/4\pi\rho_{\rm{p}})^{2/3}$, and $\Delta V_{\rm{pp}}(m,m^\prime)$ is their 
relative velocity. The kernel properties depend on the size distribution of particles, the individual particle
densities, the total mass fraction of solids, and ambient nebula conditions. 

We define sticking coefficients  $S(m,m^\prime)$, which depend on the particle masses and relative velocities, to capture both the ``bouncing barrier'' and the ``fragmentation barrier'' as follows. The first barrier encountered by growing particles is the bouncing barrier, when $S_{\rm{b}}(m,m^\prime) \rightarrow 0$ according to:

\begin{equation}
\label{equ:bounce}
S_{\rm{b}}(m,m^\prime) = 1 - \frac{m}{m + m^\prime}\frac{\Delta^2 V_{\rm{pp}}(m,m^\prime)}{V^2_{\rm{b}}} \geq 0.
\end{equation}

\noindent
We adopt a bouncing prescription using the second row of Fig. 11 of \citet{gut10} in which the threshold velocity for bouncing
collisions between similar-sized compact silicate aggregates can be approximated as $V_{\rm{b}} = (C_0/m^\prime)^{1/2}$, 
where the constant $C_0 = 10^{-7}$ g cm$^2$ s$^{-2}$. Particle pairs that have $S_{\rm{b}}=0$ are considered to have sufficient energy to avoid sticking, but 
insufficient energy to fragment. 

In a similar way we adopt a 
condition for the fragmentation of a target with mass $m^\prime$ by a projectile of mass $m$:

\begin{equation}
\label{equ:stick}
S_*(m,m^\prime) = 1 - \frac{m}{m+m^\prime}\frac{\Delta^2 V_{\rm{pp}}(m,m^\prime)}{Q_*} \geq 0.
\end{equation}
Equations (\ref{equ:bounce}) and (\ref{equ:stick}) account for bouncing 
and fragmentation criteria implicitly \citep[e.g.,][]{win12a,win13}  by smoothly decreasing $S_{{\rm{b}},*}(m,m^\prime)$ from 1 at zero 
collision velocity to 0 for a particle pair colliding with a mass-dependent collision velocity. 

\noindent
The particle fragmentation strength is captured through the parameter $Q_*$, which has the dimensions of velocity squared as in Eq. (\ref{equ:bounce}). We follow \citet{sl09} and \citet{bei11} for weak silicate particles of comparable mass, colliding at low relative velocity. We include a compositional variation in $Q_*$, motivated by recent results that
suggest that icy particles are ``stickier'' or stronger, and might grow larger, faster, and with higher porosities than silicate
particles \citep[see also Sec. \ref{sec:intro}]{wad09,wad13,oku12}. We determine the local value of $Q_*$ by a 
mass weighted average over the species making up the composition of a grain 

\begin{equation}
\label{equ:Qstar}
Q_*(R) = \sum_i Q_i\alpha^{\rm{d}}_i(R)/f_{\rm{d}}(R),
\end{equation}

\noindent
Because we consider icy particles to be stickier, we also scale the bouncing threshold velocity by a factor
of 10 (or 100 in energy) as we do for the fragmentation. Although the kernel can also include particle 
porosities \citep[see][]{ec08}, we do not include them explicitly for this paper. In practice this leads to 
$Q_*\simeq 4\times 10^5$ outside the ice line, and $Q_*=10^4$ inside. 

The fragmentation barrier mass $m_*$ is determined by the condition that the energy per unit mass in a collision 
between a target particle of mass $m^\prime = m_*$ and some other particle $m \leq m_*$ exceeds the strength
of the particle $Q_*$. In turbulent conditions, $m_*$ is the mass of a particle that is destroyed upon colliding with a comparable mass particle. However, the radius of the particle that satisfies the condition in Eq. (\ref{equ:stick})
may be smaller than $m_*$ under low turbulence conditions where headwind-drag-driven velocities
dominate (See next section). 

Once the fragmentation barrier is reached for a target particle, our code provides a reservoir of material
(with an associated {\it creation rate}) from which growth may proceed by sweep up of smaller grains, 
but which remains subject to destruction by particles of comparable or smaller sizes.  These particles represent the lower
end of the submigrator population.
Particles that are fragmented release their mass into the background ``dust'' population with 
its current local size distribution. We treat the fragmentation of particles statistically (see
Sec. \ref{subsubsec:dprob}) as well as mass transfer between them (Sec. \ref{subsubsec:mig}).
Thus we believe our model captures the essential physics of recent experimental
outcomes and models regarding collisional sticking, bouncing and fragmentation
\citep{gut09,gut10,zso10,zso11,wei12,win12a}.

 
%
%
%
%

\subsubsection{Relative Particle Velocities}
\label{subsubsec:relvel}

We include a variety of sources for the particle relative velocities: Brownian motion, pressure gradient, vertical settling and turbulence. We briefly summarize them here, while giving a more detailed description
of how we calculate them in Appendix A.4.

The thermal motion of particles, Brownian motion, is dependent upon the masses of the particles $m$ and
$m^\prime$, and the ambient nebula temperature

\begin{equation}
\label{equ:brown}
\Delta V_{\rm{Bro}}(m,m^\prime) = \sqrt{\frac{8k_{\rm{B}}T}{\pi}\frac{(m+m^\prime)}{m m^\prime}},
\end{equation}

\noindent
and is only effective for the smaller particles near the lower bound of our mass distribution.

The pressure-induced, or systematic dust velocities result from the gas in the nebula orbiting at slightly
less than the local Kepler velocity. In a rotating frame, a parcel of gas experiences an outward directed 
pressure gradient force that counters the inward force of solar gravity. The dust particles do not feel this radial pressure gradient force directly, but experience an azimuthal drag force from the more slowly rotating gas, leading to size-dependent radial and azimuthal velocities. In cases where the local solids fraction is high and affects the gas velocity, particle and gas relative velocities must be determined iteratively. To ensure correct results in all regimes, we routinely solve for these 
components of the velocity from a set of equations generalized from \citet{nak86} by \citet{ec08} for a particle size distribution \citep[though, also see][]{tan05}:

\begin{equation}
\label{equ:Ui}
\frac{\partial{U}_k}{\partial{t}} = -A_k\rho(U_k - u) + 2\Omega V_k;
\end{equation}

\begin{equation}
\label{equ:Vi}
\frac{\partial{V}_k}{\partial{t}} = -A_k\rho(V_k - v) - \frac{1}{2}\Omega U_k,
\end{equation}

\begin{equation}
\label{equ:ug}
\frac{\partial{u}}{\partial{t}} = -\sum_l A_l\rho^{\rm{d}}_l (u - U_l) + 2\Omega v - 
\frac{1}{\rho}\frac{\partial{p}}{\partial{R}},
\end{equation}

\begin{equation}
\label{equ:vg}
\frac{\partial{v}}{\partial{t}} = - \sum_l A_l\rho^{\rm{d}}_l (v - V_l) - \frac{1}{2}
\Omega u,
\end{equation}

\noindent
where $(U_k,V_k)$ are the radial and azimuthal velocity components for particles of mass $m_k$, $(u,v)$ are those for
the gas, $A_k = (\rho t^{\rm{s}}_k)^{-1}$, and

\begin{equation}
\label{equ:rhod}
\rho^{\rm{d}}_k = \sqrt{\frac{2}{\pi}}\frac{\Sigma^{\rm{d}}_k}{2h^{\rm{d}}_k}e^{-\frac{1}{2}
(z/h^{\rm{d}}_k)^2}
\end{equation}

\noindent
is the mass volume density of mass bin $k$ where $h^{\rm{d}}_k$ is the scale height of
$m_k$ (see Sec. \ref{subsubsec:subdisk}). 

Equations (\ref{equ:Ui}-\ref{equ:vg}) represent a set of $2n+2$ equations in $2n+2$ unknowns
where $n$ is the number of particle bins in the distribution given that there are
$n_{\rm{p}}$ particle bins per decade radius (see Sec. \ref{subsubsec:coag} and Appendix B). 
We solve this system of equations using a matrix method as
defined in Appendix A.4. The pressure gradient in Eq. (\ref{equ:ug}), where $p$ is the gas pressure, can 
be expressed in terms of the more familiar $\eta$ parameter as $-(1/\rho)\partial{p}/\partial{R} =
2\Omega \eta V_{\rm{K}}$, where

\begin{equation}
\label{equ:eta}
\eta(R,t) = -\frac{1}{2}\left(\frac{c}{V_{\rm{K}}}\right)^2 \frac{R^{5/2}}{\Sigma T^{1/2}}\frac{\partial}
{\partial{R}}\left[R^{3/2}\Sigma T^{1/2}\right],
\end{equation}

\noindent
and $V_{\rm{K}}$ is the local Kepler velocity \citep{nak86,cuz93}. Pressure gradients 
can be quite steep near the outer
edge of the disk, and can lead to rapid inward migration of even very small particles.
In most cases, when the local particle density is small compared to the gas density, the particle drift velocity $U_k$ can be well approximated by (Weidenschilling 1977, Cuzzi and Weidenschilling 2006): 

\begin{equation}\label{equ:Udrift}
U_k = \frac{2 {\rm St}_k \eta V_{\rm{K}}}{1 + {\rm St}_k^2 }.
\end{equation}

For the turbulence-induced velocities, we use the closed form prescriptions for a particle size
distribution of \citet{oc07}. 
%
%
The relative velocity with
respect to the gas is then $V^2_{\rm{pg}} = v^2_{\rm{t}} - V^2_{\rm{t}}$ \citep{ch03,oc07},
where the turbulent gas velocity is given by $v_{\rm{t}} = \alpha_{\rm{t}}^{1/2}c$ 
\citep[see, e.g.,][]{cuz01}, and $V_{\rm{t}}$ is the average inertial space particle velocity due to turbulence. The particle-to-particle turbulent relative velocities $\Delta V_{12}$
\citep[Eq. (16)][]{oc07} are less straightforward because of the different coupling that exists between particles and eddies of 
different sizes. 
Recent numerical simulations have obtained results differing from these parametrizations by a factor of order unity depending on St \citep{hub12,pp13}. It remains unclear how much of this difference arises from the relatively low Reynolds number of the numerical simulations, relative to the inertial range turbulent kinetic energy spectrum of the actual nebula which is assumed by Ormel and Cuzzi (2007). 
For more discussion, see \citet{ch03}.

From the various contributions, we calculate the particle-to-gas relative velocities for a particle of
mass $m$ which we use to calculate the particle stopping times;

\begin{equation}
\label{equ:deltavpg}
\Delta V_{\rm{pg}} = \left[(U - u)^2 + (V - v)^2  + W^2 + V^2_{\rm{pg}}\right]^{1/2},
\end{equation}

\noindent
where $W = -\Omega^2 z t_{\rm{s}}$ is the particle vertical settling velocity, in which the vertical coordinate
$z$ is generally taken to be the subdisk scale height $h_{\rm{D}}$ (see Sec. \ref{subsubsec:subdisk}). 
The particle-to-particle relative velocities are then given by

\begin{equation}
\label{equ:deltavpp}
\Delta V_{\rm{pp}}(m,m^\prime) = \left[\left(\Delta V_{\rm{Bro}}\right)^2 + \left(\Delta V_{\rm{pre}}
\right)^2 + \left(\Delta V_{12}\right)^2\right]^{1/2},
\end{equation}

\noindent
where $\left(\Delta V_{\rm{pre}}\right)^2 = (U-U^\prime)^2 + (V-V^\prime)^2 + (W-W^\prime)^2$. If we 
are strictly in the Epstein flow regime, then the stopping times do not depend on $\Delta V_{\rm{pg}}$, 
so $\Delta V_{\rm{pp}}$ can be calculated in a straightforward manner from the velocity components.
However, if there are particles in the Stokes flow regime, then the stopping times depend on 
$\Delta V_{\rm{pg}}$ and iterations are needed to converge to a proper solution (see Appendix A.4).
 
We note that in Eq. (\ref{equ:deltavpg}), we have summed the mean relative velocity between the gas and 
particle in quadrature with that of the fluctuating turbulent relative velocity which is not strictly correct.
A similar argument can be made for the inclusion of $\Delta V_{\rm{pre}}$ in
Eq. (\ref{equ:deltavpp}) which we take to be the mean collision velocity. 
It has been argued by \citet{gar13} that this construct is not accurate when the dominant velocities are systematic.
However, in the models we present here, the turbulent velocities 
dominate the systematic ones so that we do not expect that any differences will be significant. We leave a more 
proper treatment for future work.

\subsubsection{Migrators and Growth Beyond the Fragmentation Barrier by Mass Transfer}
\label{subsubsec:mig}

Once the fragmentation barrier $m_*$ has been reached, we employ a more sophisticated, semi-analytical and
statistical algorithm for the further growth of particles. In this regime, 
we track mass and radial drift explicitly for a 
population ranging from ``submigrators'' which are subject to mutual collisional destruction from particles 
of similar size or smaller, to ``migrators'', which have grown large enough to 
have a much lower probability of destruction and may grow by mass transfer between them and particles of smaller
size \citep{win12a,gar13}. Disrupted submigrators and migrators are distributed back into the 
dust population with the same $q$. The migrator distribution is defined on a continuous, time-dependent
mass grid where the width of each mass bin can vary due to different rates of growth for different
sized particles. Thus the evolving mass histogram for $m > m_*$ does not follow a power law in general.
 
Beyond the fragmentation barrier,  particles can still grow through incremental means, by the
sweepup of smaller material. The simplest expression for incremental growth assumes perfect sticking 
between a particle and (smaller) feedstock particles so that, schematically, 

\begin{equation}
\label{equ:dmdt}
\frac{dm}{dt} = \sigma \Delta V_{\rm{d}} \rho_{\rm{d}},
\end{equation}

\noindent
where $\sigma$ is the collision cross section and $\Delta V_{\rm{d}}$ is some characteristic relative velocity between $m$ and the feedstock population \citep[see, e.g.,][]{cuz93,bra08}. We generalize this to a size distribution, in which sticking is not assumed to be perfect, to define the growth rate of a migrator $m_k > m_*$:

\begin{equation}
\label{equ:dmkdt}
\frac{dm_k}{dt} = \sum^{n_*}_{l=1} \sigma_{k,l} S_{k,l}\Delta V^{\rm{pp}}_{k,l}\rho^{\rm{d}}_l \,\,+
\sum^{k}_{l=n_*+1} \sigma_{k,l} S_{k,l}\Delta V^{\rm{pp}}_{k,l}\rho^{\rm{m}}_l,
\end{equation}

\noindent
where $S_{k,l}$ are the sticking coefficients, the index $n_*$ refers to the fragmentation barrier mass bin, 
and $\Delta V^{\rm{pp}}_{k,l}$ is the relative velocity between $m_k$ and $m_l$. The first term 
on the RHS is due to the dust population, and the second term the migrator population. 
The volume density of migrators $\rho^{\rm{m}}_k$ is defined in terms of the total mass of solids in a 
migrator mass bin, $M^{\rm{m}}_k$, and the total volume of the particle sublayer, of vertical thickness 
$2 h_{\rm{d}}$ near the midplane, from where migrators can accrete other material 

\begin{equation}
\label{equ:rhosm}
\rho^{\rm{m}}_k = \frac{M^{\rm{m}}_k}{2{\mathscr{A}} h_{\rm{d}}},
\end{equation}

\noindent
where ${\mathscr{A}}$ is the surface area of the radial bin. Equation (\ref{equ:dmkdt}) 
allows for  a migrator $m_k$ 
to accrete other particles of $m_l \leq m_k$, but the sticking will be zero for a large range of size 
pairs when the collision specific energy exceeds $Q_*$ (or $C_o$).

On the other hand, there are outcomes of high-velocity collisions which lead to growth of the target particle, which have been observed and studied
experimentally \citep[e.g.,][]{wur05,kot10}. Specifically, the impact velocity may be insufficient to fragment the larger target 
particle, but sufficient to fragment the smaller particle and allow deposition of mass
 on the larger particle - if the efficiency of accretion $\epsilon_{\rm{ac}}$ exceeds that of 
erosion $\epsilon_{\rm{er}}$. We use the model of \citet{win12a,win12b},  in which the 
threshold for fragmentation is both mass and velocity dependent, to account for this process:

\begin{equation}
\label{equ:mufrag}
\mu(m,V_{\rm{cm}}) = C_{\rm{w}} m^{-0.068} V^{-0.43}_{\rm{cm}},
\end{equation}

\noindent
where $V_{\rm{cm}}$ is the relative velocity of the target and impactor in the center of mass frame, the
coefficient $C_{\rm{w}}=3.27$ g$^{-0.068}$ and masses and velocities are in cgs units. In order to
account for the stickiness of icy particles (see Sec. \ref{subsubsec:stick}), we scale $C_{\rm{w}}$
by a factor $(Q_*/10^4)^{0.21}$. The fragmentation threshold is reached when $\mu = 1$. We apply Eq.
(\ref{equ:mufrag}) to both $m_l$ and $m_k$ where the center of mass velocities are given by

\begin{equation}
\label{equ:vcm}
V^{\rm{cm}}_k = \frac{\Delta V_{\rm{pp}}}{1 + m_k/m_l};\,\,\,\,
V^{\rm{cm}}_l = \frac{\Delta V_{\rm{pp}}}{1 + m_l/m_k}.
\end{equation}

Fragmentation with mass transfer only occurs when $\mu_k \geq 1$ and $\mu_l < 1$ \citep[see][]{win12a}. If
this condition is satisfied, we then calculate the efficiency of accretion from \citep{bei11,win12a}

\begin{equation}
\label{equ:eac}
\epsilon_{\rm{ac}} = -6.8\times 10^{-3} + 2.8\times 10^{-4} \left(\frac{m_l}{4.1\,{\rm{g}}}\right)^{0.16}
\Delta V_{\rm{pp}},
\end{equation} 

\noindent
and the efficiency of erosion \citep{win12a}

\begin{equation}
\label{equ:eer}
\epsilon_{\rm{er}} = 9.3\times 10^{-6} \left(\frac{m_l}{m_0}\right)^{0.15} \Delta V_{\rm{pp}} - 0.4,
\end{equation}

\noindent
where $m_0 = 3.5\times 10^{-12}$ g is a monomer mass. The erosion efficiency is an interpolation between
the experimental results of several workers \citep{par07,tw09,sb11}. A successful transfer of mass to the
larger particle $m_k$ occurs then if $\Delta m_k /m_l = \epsilon_{\rm{ac}} - \epsilon_{\rm{er}} > 0$, which
is assigned to the value of $S_{k,l}$ in Eq. (\ref{equ:dmkdt}).
Inspection of Eqns. (\ref{equ:eac}) and Eq. (\ref{equ:eer}) demonstrates that for a given impactor mass $m_l$,
higher relative velocities are required to initiate erosion versus accretion. As an example, for a $m_k = 1$ g
target particle being impacted by a projectile with $m_l = 0.05$ g for a relative velocity of $5$ m s$^{-1}$,
the target particle accretes roughly 6\% of the impactor with no net erosion. On the other hand, for the same
pair impacting at 15 m s$^{-1}$, the target particle accretes over 20\% of the impactor, but also experiences
$\sim 7$\% erosion. In our treatment, the remaining projectile mass, which by definition is fragmented, as well 
as the eroded mass are assumed to be returned to the dust population.

Finally, once migrators are large enough that $\Delta^2 V_{\rm{pp}} \lesssim 2Q_*$, they can accrete
migrators as large as themselves in pairwise mergers if they collide.  A check is made to ensure that no more mass is ``accreted" in a timestep than actually exists locally (for more detail, see Appendix A.5).


\vspace{0.2in}
\subsubsection{Radial Drift and Evaporation Fronts (EFs)}
\label{subsubsec:drift}

Due to variable coupling with the nebula gas, particles of different size will
drift radially  at different rates with respect to the gas. The radial drift velocity for a particle of mass $m_k$ has two contributions \citep[e.g., see][]{tl02,bir10}:

\begin{equation}
\label{equ:vdrift}
V_k^{\rm{rad}} = \frac{V_{\rm{g}}}{1 + {\rm{St}}^2} + \Delta U_k.
\end{equation}

\noindent
The first term is directly
imposed by the radial motion of the gas that moves with advective velocity $V_{\rm{g}}$ 
(section \ref{subsec:gasevol}). The second term
($\Delta U_k = U_k-u < 0$) is the radial drift velocity of the particle with respect to the gas
(section \ref{subsubsec:relvel}). 
This drift increases with particle size until 
$t_{\rm{s}} \sim \Omega^{-1}$ or ${\rm St} = t_{\rm{s}} \Omega \sim 1$, but then decreases for larger sizes (see Eq. [\ref{equ:Udrift}]). In the terrestrial planet region, meter size particles drift most rapidly, but further out in the disk where gas
densities are low and the pressure gradient can be very strong, much smaller particles drift inward the most rapidly \citep{bra08,ha10}.

In our code we track the radial drift across radial bin boundaries for individual migrator mass bins.
We do this by calculating the
inward drift time of the migrator across (logarithmically spaced) radial bins of width $\Delta R$:

\begin{equation}
\label{equ:trad}
t_k^{\rm{rad}} = - \frac{R}{V_k^{\rm{rad}}} \left(1 - e^{-|\Delta R|}\right),
\end{equation}

\noindent
where the radial drift velocity for larger particles is generally a negative quantity (consistent
with our sign convention, see Sec. \ref{subsec:gasevol} or \ref{subsubsec:diffadv}). This allows us to
calculate for every migrator mass bin $k$ how much mass is drifting out of the local radial bin in a time 
$\Delta t(R)$:

\begin{equation}
\label{equ:mdrift}
M^{\rm{drift}}_k = M^{\rm{m}}_k \frac{2 \Delta t(R)}{t^{\rm{rad}}_k},
\end{equation}

\noindent
where the factor of 2 assumes that the drift time
from the bin's midpoint is representative. Equation (\ref{equ:mdrift}) is the maximum amount of mass that can drift
out of a radial bin in time $\Delta t$, and does not take into account the fragmentation of some particles (see Sec. 
\ref{subsubsec:dprob}). 

It is possible that some
migrators with $m \sim m_*$ have positive outward drift if, for example, $\alpha_{\rm{t}}$ is sufficiently
large (which leads to more vigorous collisions and smaller $m_*$) that the gas advection term (first term on the RHS of Eq. [\ref{equ:vdrift}])
can overwhelm the radial component of the headwind-driven drift velocity. Under such circumstances, the redistribution of
these migrators is done using the radial diffusion-advection equation (Sec. \ref{subsubsec:diffadv}). 

Self-consistently modeling a globally evolving nebula requires that we consider the presence of Evaporation Fronts (EFs) 
which are regions in the disk where phase changes between solids and vapor can occur. 
The  evaporation of radially drifting 
material, and subsequent recondensation of outwardly diffusing vapor, can have three main effects. First, it can increase the
abundance of vapor inside an EF, with implications for chemistry and mineralogy. Second, it can increase 
the fractional abundance of solid material available {\it just outside} the EF, perhaps by a factor of
$\sim 10$ or more \citep[e.g.,][]{cz04,cc06,gar07}, with implications for accretion. Third, it can
significantly change the composition of solid material outside the EF from ``cosmic abundance'' values. 
We will illustrate all these effects in this paper; a more detailed study will await future publications. 

\begin{figure}[t]
 \resizebox{\linewidth}{!}{%
 \includegraphics{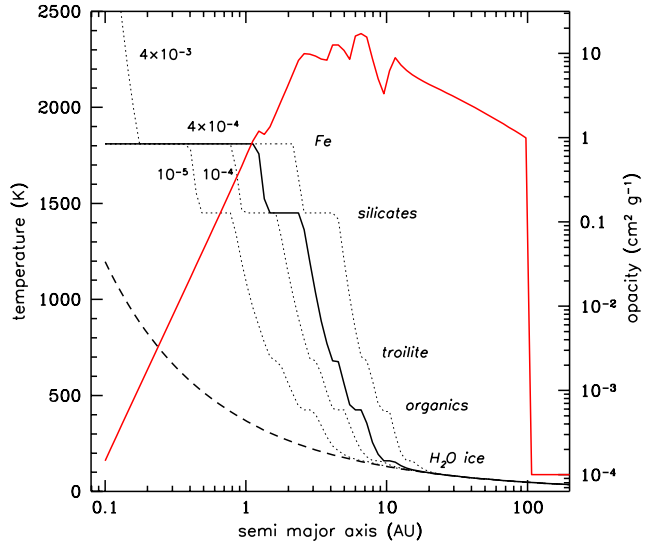}}
\caption{Initial profiles of the Rosseland mean opacity (red curve), and temperatures 
at the midplane (solid black curve) and photosphere (dashed curve) for our fiducial model with 
$\alpha_{\rm{t}}=4\times 10^{-4}$. Additional midplane temperature curves are shown (dotted curves) for 
different values of $\alpha_{\rm{t}}$. The nearly flat regions in the temperature profiles correspond to 
the location of EFs as labeled.}
\label{fig:initcond}
\end{figure}

Rather than treating an EF as a sharp boundary, 
we allow for evaporation (or condensation) to occur over a small range of radii covering a  midplane 
temperature range $2\Delta T_{\rm{EF}} \sim 0.1-1$ K
relative to the nominal species evaporation temperature $T_i$ (see Table 2). The fractional solid 
abundance $\alpha^{\rm{v,d}}_i$ of a species $i$ (with density 
$\rho_i$) at some radial location $R$ and at some local temperature $T$, is transformed into vapor linearly as the calculated midplane temperature changes over the temperature range 
$T_i-\Delta T_{\rm{EF}} \leq T \leq T_i+
\Delta T_{\rm{EF}}$ (Sec. \ref{subsubsec:temp}). 
This gradual radial transition in solids abundance, which can span a significant radial range, 
mimics the anticipated effect (as described below) of ``buffered'' temperature changes across an EF as
material is evaporated or condensed, rather than allowing unrealistic (and numerically
problematic) abrupt radial changes in opacity and temperature just inside an EF. 
This simple numerical treatment captures the essence of the actual condensation process in which material first evaporates at the midplane, and then at increasingly higher altitudes 
with decreasing distance from the star as the disk gets warmer \citep{dav05,min11}.  The effect is seen in the 
constant midplane temperature regions in our simulations (e.g., see Fig. \ref{fig:initcond}). 

In our code, 
dust grains are effectively treated as aggregates of chemically distinct monomers (although 
see Sec. \ref{subsubsec:coag}) whose fraction of species $i$ can be quickly removed or emplaced. Larger particles 
that are followed explicitly (Sec. \ref{subsubsec:mig}) are assumed to lose that fraction of their material 
that is of evaporating or condensing species $i$ as they migrate through the EF for species $i$, and their masses and mean
densities are adjusted accordingly.  Their evaporated material is then added to
the local vapor inventory. Better models of the largest particles whose interiors are somewhat insulated
from ambient nebular conditions, and could potentially transport very volatile species from
very cold to warmer regions \citep[e.g., see][]{est09}, will require physical evaporation rates and internal structure models.
We do not treat the kinetics of evaporation and condensation here \citep[see discussions in][]{cuz03,cc06}. We plan to include these effects in future work.

\vspace{0.3in}
\subsubsection{Probability of Destruction and ``Lucky Particles''}
\label{subsubsec:dprob}

In our treatment of growth, the particle mass such that the sticking coefficient first approaches zero for equal mass particles represents
the ``bouncing'' barrier. That is, such a collision is energetic enough to prevent
any sort of sticking, but not energetic enough to fragment the particle \citep[e.g.,][]{gut10,zso10}.
Our nominal fragmentation mass $m_*$ (section \ref{subsubsec:stick}) is also a convenient reference value, at which particles of equal mass fragment each other under local nebula conditions.  In practice, we
employ a statistical scheme for the probability of destruction of a migrator. The scheme utilizes a Gaussian
PDF of relative velocities for a given impactor mass, with rms velocity equal to the mean \citep{car10,hub12,pp13}.

The collision rate of target migrators of mass $m^\prime$ with projectile particles of masses $m \leq m^\prime$ is given
by ({\it cf.} Eq. [\ref{equ:kernel}])

\begin{equation}
\label{equ:rcoll}
{\mathscr{R}}_{\rm{coll}}(m^\prime,m) = \pi(r^\prime + r)^2 b(m)\Delta V_{\rm{pp}}(m^\prime,m),
\end{equation}

\noindent
where $b(m) = \rho_{\rm{d,m}}/m$ is the number density of particles of mass $m$, 
and the subscripts refer to either the volume density of dust or migrators. The probability that
a migrator of mass $m^\prime$ will suffer fragmentation as it grows during time $\Delta t$ can then be obtained from

\begin{equation}
\label{equ:pfrag}
\begin{split}
{\mathscr{P}}(m^\prime) = \int_{t}^{t+\Delta t} dt^\prime \int_{m_0}^{m^\prime} \pi (r^\prime + r)^2
\Delta V_{\rm{pp}}
(m^\prime,m)\\ \times \zeta(m^\prime,m) f(m,t^\prime)\,dm
\end{split},
\end{equation}

\noindent
where $\zeta(m^\prime,m)$ is an integral over a Gaussian distribution of relative velocities covering a
range equal to or greater than the critical impact velocity $V_{\rm{c}} = 
\sqrt{2Q_*(m+m^\prime)/m}$:

\begin{equation}
\label{equ:zeta}
\begin{split}
\zeta(m^\prime,m) = \frac{1}{\sqrt{2\pi}\Delta V_{\rm{pp}}(m^\prime,m)} \int_{V_{\rm{c}}(m)}^{\infty}
e^{-\frac{(V^\prime - \Delta V_{\rm{pp}})^2}{2\Delta V^2_{\rm{pp}}}}\,dV^\prime \\
 = 
\frac{1}{2}\left[1 - {\rm{erf}}\left(\frac{V_{\rm{c}}-\Delta V_{\rm{pp}}}{\sqrt{2}\Delta V_{\rm{pp}}}\right)
\right]
\end{split}. 
\end{equation}

\noindent
We discretize Eq. (\ref{equ:pfrag}) for use in our calculations as described in Appendix A.6. Although other
workers have utilized a Gaussian scheme as we do here, others have argued that the distribution should be
Maxwellian \citep[e.g.,][]{gal11,win12b}. \citet{gar13} have developed a more
detailed model in which she shows that the mean and rms square velocities are not the same. Furthermore, our
approach to calculating $\zeta$ is more akin to that of \citet{win12b}, whereas \citet{gar13}
argues that the relative velocity (in our notation $V^\prime$) should be included in the integral over the
PDF in order to recognize that collisions of particles with larger collision velocities have a higher collisional 
frequency than those with lower collision velocities. We intend to examine the more detailed model of Garaud et 
al. in future work.

Using our formalism, we calculate the fraction of migrators within every migrator mass bin that is
destroyed during every growth step. 
The amount of mass returned to the dust
distribution for any mass $m^\prime$ is then $M_{\rm{dest}} = {\mathscr{P}}(m^\prime)M_{\rm{m}}(m^\prime)$
where  $M_{\rm{m}}(m^\prime)$ is the total mass in migrators with mass $m^\prime$.
This mass is subtracted from the total mass contained within the $m^\prime$ mass bin prior to determining
how much mass drifts out of a radial bin (and thus a factor $1-{\mathscr{P}}$ is included in Eq.
[\ref{equ:mdrift}]).

We find quite generally that growth stalls at masses slightly-to-moderately larger than the fragmentation 
barrier mass $m_*$ and radius $r_*$. We will represent this particle radius where growth stalls to be 
$r_{\rm{M}}$ which also is the particle containing the bulk of the total mass of migrators in a radial
bin. This result of stalled growth is fairly robust unless the
nebula turbulence is very small, in which case incremental growth can proceed to large sizes
because the particle layer becomes very dense near the midplane, and collision velocities are low. 

As was seen by \citet{win12a,win12b} and \citet{gar13}, we also find that a small fraction of 
migrators are able to grow to larger sizes
because they have grown large enough to have a low probability of destruction. We refer to these
migrators as ``lucky particles'', and characterize the {\it largest} particle in each radial bin by  radius 
$r_{\rm{L}}$. 
However, previous studies (except for Drazkowska et al 2013, who imposed a ``pressure bump" to prevent loss by drift) have not allowed for the significant radial drift of such particles, which greatly reduces their abundance. While rare, such particles can transport material over large radial distances. Treating the growth of these ``lucky" migrating particles simultaneously with their radial drift and nebula evolution is a significant advantage of our code over previous global models.

\section{Results}
\label{sec:results}

\subsection{Baseline model}
\label{subsec:initcon}

We choose as our fiducial model a nebula with a stellar mass of $M_\star = 1$ M$_\odot$, an initial
disk mass of 0.2 M$_\odot$ and a scaling factor $R_0 = 10$ AU (see Sec. \ref{subsec:gasevol}). The
radial grid in our evolutions spans a range from $0.5-1000$ AU using 96 logarithmically spaced bins,
chosen to make optimal use of the Haswell CPUs (12-core, 24 processors) at the NAS Pleiades
cluster used in our simulations. In our code, radial bin width is only a concern for radial drift of 
migrator particles, because other properties that communicate across the grid such as gas and dust evolution 
are fully implicit\footnote{A minimum requirement we impose is that the timestep used in our simulations
is much smaller than the drift time of the fastest migrating particle across a radial bin. For all
of our simulations presented here, this is easily satisfied.}. All simulations in this paper span up
to $2\times 10^5$ years, and use variable stellar luminosity $L$ and particle strength $Q_*$ (except in sections \ref{subsec:conQstar} and \ref{subsec:L1} where these parameters are kept constant).

We employ a constant turbulent viscosity parameter $\alpha_{\rm{t}} = 4\times 10^{-4}$ as our baseline,
though we explore a model that uses a constant $\alpha_{\rm{t}}=4\times 10^{-3}$. As noted in Section \ref{sec:intro}, recent studies of purely hydrodynamical instabilities suggest that $\alpha_{\rm{t}} > 10^{-4}$ (Nelson et al 2013, Marcus et al 2013, 2015; Stoll and Kley 2014). We do not model disks with lower values
of $\alpha_{\rm{t}}$. Such models have been explored in detail \citep[e.g.,][]{wei97,wei00,wei04,wei11}; vanishingly small $\alpha_{\rm{t}}$ leads to
rapid {\it in situ} growth by sticking and/or collective instabilities rather than the delayed growth, and extensive redistribution of solids, which is the primary focus for
this paper and arguably in better agreement with observations (Sec. \ref{sec:intro}). 

In the following sections, we consider  models with fragmentation (F) only, then bouncing and 
fragmentation (BF) and finally add mass transfer (MTBF) as described in Sections 
\ref{subsubsec:stick} and \ref{subsubsec:mig}. In forthcoming papers, we will explore radially and vertically variable 
$\alpha_{\rm{t}}$ profiles \citep[e.g.,][]{bs11,kd15}, which introduce another layer of complexity.

In this work, we choose to cut off the {\it initial} distribution of solids at 100 AU (where the initial gas surface
density $\Sigma$ quickly drops off to $\ll 1$ g cm$^{-2}$). Other workers have
chosen similar cutoffs \citep[e.g.,][]{bra08,bir10}; while arbitrary, this initial condition simply provides a convenient reference point from which to follow the solids evolution. The motivation is that the strong pressure gradients at the outer edge 
of the nebula will lead to rapid inward migration of even the smallest grains unless
one assumes sufficiently high values of $\alpha_{\rm{t}}$ which leads to outward transport of solids as the gas 
disk spreads in spite of the presence of a strong pressure gradient (e.g., see Sec. \ref{subsec:HA}).

In {\bf Figure \ref{fig:initcond}} we show the initial midplane temperature and opacity profile for our fiducial
model (solid curves). The corresponding photosphere temperature is shown by the dashed curve. The initial
disk temperature is sufficiently hot at the midplane for our high $\alpha_{\rm{t}}$ model (Sec. \ref{subsec:HA}) that 
all of the EFs for the species used in this paper (see Table 2)
are present. Although midplane temperatures in EF regions appear flat, there is a small gradient in temperature (Sec. \ref{subsubsec:drift}) over which evaporation of volatiles is assumed to take place. It
is over these transition radial ranges that the opacity decreases linearly. For comparison, we plot the
temperature profiles for different values of $\alpha_{\rm{t}}$ (dotted lines) to the show the sensitivity
to the choice of this parameter. Even for a value of $\alpha_{\rm{t}} = 10^{-5}$, the disk is relatively
hot in the inner regions because the initial stellar luminosity is $\sim 12$ L$_\odot$ \citep{dm94}. In this work we adopt a constant value of $10^{-4}$ cm$^2$ g$^{-1}$ for the gaseous opacity. In the hot, innermost 
regions this may be an underestimate because gaseous opacity is a function of temperature 
\citep[e.g.,][]{fer05}. However, for our models, dust is usually present at all locations in the disk so 
that the Rosseland mean opacity is dominated by this component. In future work, we will incorporate a gaseous 
opacity table into our models.

\begin{figure}
 \includegraphics[trim = 120 0 -120 0,clip,width=6.0in,height=6.0in,angle=0.0]{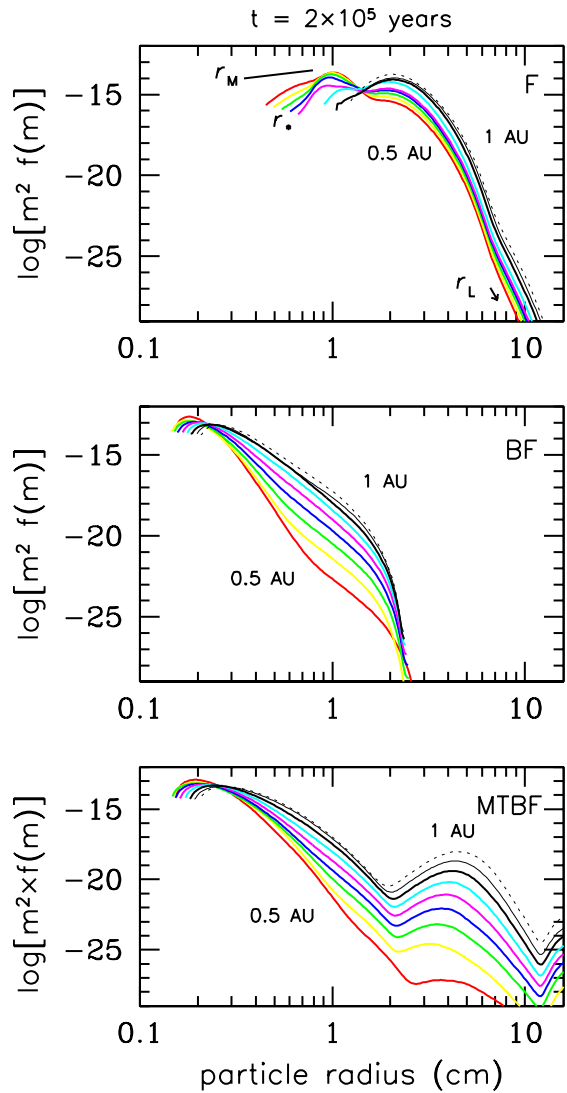}
\caption{This figure shows several of the {\it actual} histograms of particle size distributions, as mass 
volume density per bin of width $dm$, for three different cases and several distances from the star. The lower boundary of these plots aligns with the upper boundary of the ``dust" power law distribution at $r_*$, that extends downward to 0.1$\mu$m radius. The method of determining these distributions is described in section \ref{subsec:sizedist}. As is specifically indicated in the upper panel, the peak in this distribution represents the radius $r_{\rm M}$ which characterizes most of the net mass. The {\it largest single particle} in each distribution, at the far right end, provides our value of $r_{\rm L}$. It is clear that the mass contained in these ``lucky" particles is extremely small. In the lower panel one sees secondary peaks in 
$m^2f(m)$ (g cm$^{-3}$), which are likely due to mass transfer as suggested by \citet{win12a} and \citet{gar13}. Though we might regard these as ``breakthrough" particles, they are actually not {\it that} much larger than $r_{\rm M}$, and {\it still} contain much less mass than particles of size $r_{\rm M}$. This figure should be kept in mind for interpreting the remaining figures where only $r_{\rm M}$ and $r_{\rm L}$ are shown.}
\label{fig:sizedist}
\end{figure}

\subsubsection{Typical particle size distributions}\label{subsec:sizedist}
Before discussing the different cases in detail, we show some typical particle size distributions in {\bf Figure \ref{fig:sizedist}}. Three of our cases (discussed below) are shown in the three panels. The curves in each panel are actual calculated distributions of particle mass volume density as a function of particle radius for particles with $r > r_*$, calculated with 100 bins per decade radius; recall that particles with $r<r_*$ are modeled by a ``dust" distribution which is a powerlaw extending downward and to the left of the values shown. Each family of curves is for several different subsolar distances between $0.5-1$ AU. Values of the fragmentation radius $r_*$, the mass-containing radius $r_{\rm{M}}$ (shown for the red curve only), and the largest ``lucky" particle radius $r_{\rm{L}}$ are indicated in the top panel. The downward arrow indicates that the largest
particle is actually below the range plotted which covers seventeen orders of magnitude in mass volume 
density. In future plots we will only show these three characteristic values, but the distributions of Fig. \ref{fig:sizedist} underlie them all. In the lower panel, secondary peaks are seen which are most likely due to mass transfer as will be discussed
later.

\subsection{Fiducial model with fragmentation (``F")}
\label{subsec:F}

In our model, the fragmentation size is defined by that particle mass $m^\prime=m_*$ that gives $S_*=0$ in
Eq. (\ref{equ:stick}). The value of $m_*$ will depend on ambient nebular conditions, as well as the particle strength $Q_*$, which is  composition dependent (Eq. [\ref{equ:Qstar}]) for the bulk of 
the simulations we present here. 

\begin{figure}
 \includegraphics[trim = 120 0 -120 0,clip,width=6.0in,height=6.0in,angle=0.0]{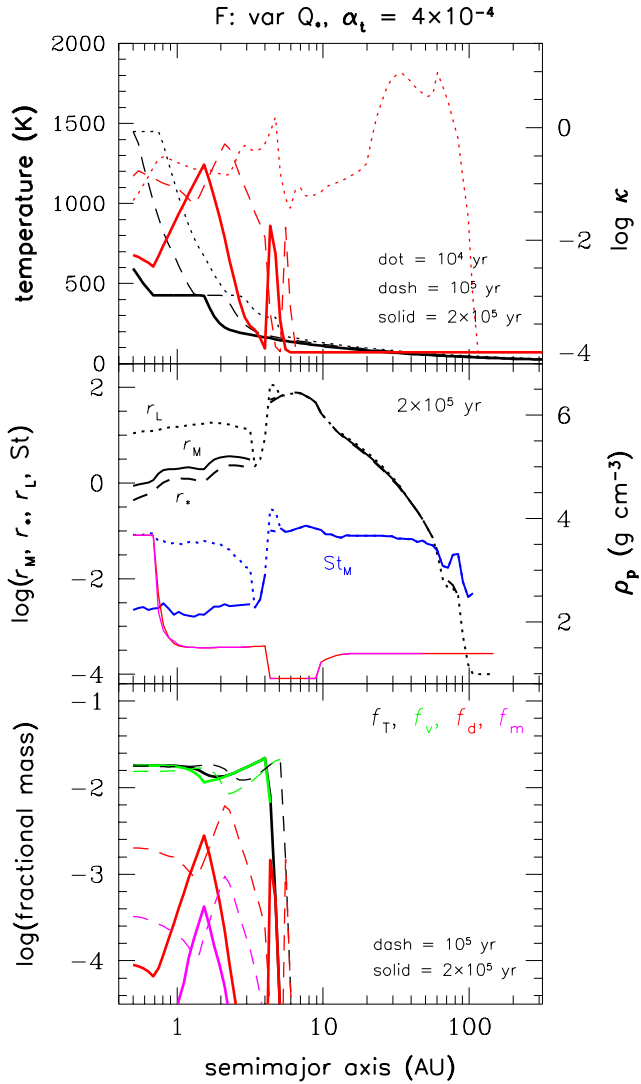}
\caption{Fiducial model with fragmentation only (F). \underline{upper panel:} 
Temperature (black) and opacity (red) plotted at three different times. \underline{middle panel:} 
Fragmentation radius $r_*$ (dashed), particle radius that carries most of the mass $r_{\rm{M}}$ (solid black) and largest particle radius $r_{\rm{L}}$ (solid black curve). The Stokes numbers 
for $r_{\rm{M}}$ and $r_{\rm{L}}$ 
are shown in blue. Also plotted are the mean particle densities for dust (red) and migrators (magenta).
\underline{lower panel:} Fractional masses shown at two different times. The F model is characterized by 
rapid growth and drift, leading to depletion of the outer disk and only modest enhancements in the inner nebula.
However, the largest particle sizes can be $> 1$ m. See text for details.}
\label{fig:3panF}
\end{figure}

\begin{figure}
 \includegraphics[trim = 120 0 -120 0,clip,width=6.0in,height=6.0in,angle=0.0]{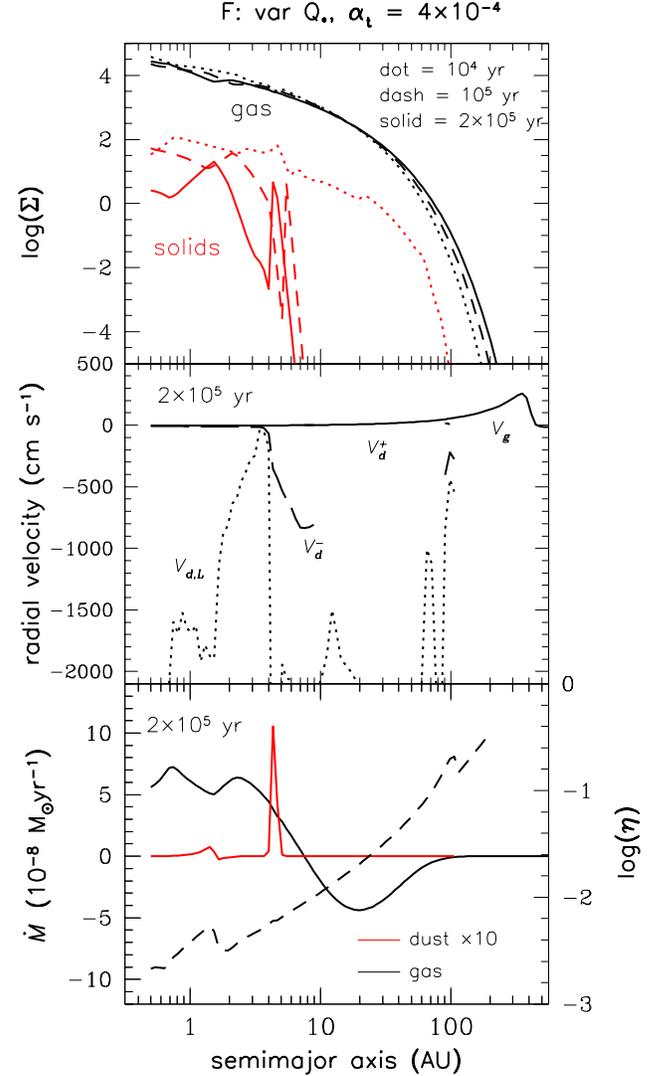}
\caption{Fiducial model with fragmentation only (F). \underline{upper panel:} Gas (black)
and dust (red) surface densities at three different times. \underline{middle panel:} Radial velocities of the
gas (solid), largest particle (dotted), and mean inward (long dashed) and outward (short dashed) components of
the dust population at $2\times 10^5$ years. \underline{lower panel:} Mass accretion rates for the gas (solid 
black) and the dust component (red curve). The vertical scale for the dust component is enhanced by a factor of ten for clarity. The 
dashed curve denotes the pressure gradient. For the fiducial choice of $\alpha_{\rm{t}}$, the evolution of the
gas is modest, but substantial changes in the solids arise due to rapid growth and inward migration. See text for
details.}
\label{fig:3pangasF}
\end{figure}

{\bf Figures \ref{fig:3panF}  and  \ref{fig:3pangasF}} show the results of a fragmentation-only model "F" in our nominal, moderately turbulent disk. The temperature (black curves) and Rosseland mean opacity
(red curves) are plotted in the top panel of Fig. \ref{fig:3panF} at evolution times of $10^4$, $10^5$ and 
$2\times 10^5$ years. Early on, the disk remains relatively hot in the inner regions as seen by the persistence
of the silicate EF at 1 AU even after $10^5$ years, but the temperature quickly drops to much lower values by
$2\times 10^5$ years. This cooling is a direct consequence of the lower opacity associated with rapid particle growth when only limited by fragmentation. Evaporation fronts can clearly be seen to 
evolve over time, with the organics and water ice EFs at 425 K and 160 K, respectively, being the most 
prominent at later times (solid curve).

Fractional masses of different components are shown in the lower panel of Fig. \ref{fig:3panF}. Here, the
red curves refer to the dust, magenta to the migrator and green to the vapor fractions, respectively. Peaks appear 
in the fractional masses of the dust and migrator populations outside of EFs, because migrating particles
passing through them release their volatiles into the vapor component, and some of this vapor subsequently 
diffuses outward across the EF to recondense onto grain surfaces. This effect has been seen in previous simulations 
\citep{cz04,cc06,gar07}. Relatively strong peaks in the fractional masses of solids at $10^5$ years can be 
associated with the organics and water ice EFs,
whereas at $2\times 10^5$ years these peaks have migrated inwards considerably as the disk cools. A less
prominent silicate peak is also seen at $10^5$ years, that has all but vanished at later times, but this is 
because the silicate EF is no longer on our computational grid, and the inner boundary cannot relay information
about what is occurring inside itself. The organics 
enhancement\footnote{For this introductory work we assume that ``organics" are the moderately refractory, carbon-rich materials seen in comet Halley and primitive carbonaceous chondrites, which volatilize at roughly 400K \citep{gw03,pol94}. In what is probably an oversimplification, we further assume that these materials behave like normal volatiles, able to reversibly evaporate and recondense. In reality, their ``evaporation" might be closer to an irreversible decomposition into simpler molecules such as CO and/or CH$_4$. This refinement is deferred to future work, but there will be implications for opacity, solids mass enhancement and nebula chemistry.} is then largest at around 1.5 AU, while the water ice peak lies near 5 AU. The solid phase mass enhancements map well
with the the opacity shown in the top panel. That is, the highest opacities in the inner disk at all times are 
associated with the peaks in solids.
A steady transport of material from the outer regions to the inner regions maintains a significant population of
smaller particles, which keeps the opacity high even as particles grow, and maintains a high temperature. This can be seen by noting that at $10^4$ years, there is still 
a lot of material in the outer disk out to 100 AU (large opacities and fractional masses there).  By $10^5$ years, for these specific models, the bulk of the material in the outer nebula has effectively been transported 
into the inner nebula, though some material still remains in the outer disk (at fractional masses of $\lesssim 10^{-6}-10^{-5}$). We see overall enhancements in the inner regions of a factor of $\sim 2-3$, though it is mostly 
present in the vapor component (green curves). 
The inner nebula situation 
changes significantly by $2\times 10^5$ years where much of the material at $R\lesssim 1$ AU has drifted inwards 
leaving much lower fractional masses of solid material, and thus much lower opacity leading to cooler overall
temperatures.

The relatively rapid transport of solids to the inner regions occurs because a fragmentation-only model
allows larger particle sizes to be reached relatively quickly in the outer nebula, and in relatively 
large numbers. In the middle panel of Fig. 
\ref{fig:3panF} we plot particle sizes of interest (black curves), particle Stokes numbers for $r_{\rm{M}}$ and
$r_{\rm{L}}$ (blue curves) and the mean particle density of the dust (red) and migrators (magenta) 
at $2\times 10^5$ years.
The first thing to notice is that the fragmentation barrier size $r_*$ has been reached almost everywhere 
in the disk. This is indicated by the presence of a heavy dashed curve at all distances (actually the situation was reached at around $5\times 10^4$ years). Migrator particles also form
early on, which accelerates inward solids transport. Once the fragmentation barrier has been
reached, further incremental growth is hampered by the gradual destruction of particles $r > r_*$ with time,
and the size of particle that carries most of the mass in the migrator size distribution  ($r_{\rm{M}}$) does not
stray far from the fragmentation size.  The characteristic ``hump'' outside of $3-4$ AU corresponds to the
transition between weaker and less sticky silicate particles inside the water ice EF, with the stronger and stickier icy particles
(with larger $Q_*$) outside the EF. Indeed for this case, outside the ice EF, most of the mass has reached 
$r_{\rm{M}} \sim$ meter size.  Although incremental growth by sweepup of particles too small to fragment large targets leads to ``lucky" particles approaching $r_{\rm{L}}\sim$ meter size even for some distance inside the ice EF, these particles are very few in number and contain a negligibly small portion of the mass (see Fig. \ref{fig:sizedist}). This distribution of sizes essentially
stays the same between $5\times 10^4$ and $2\times 10^5$ years, although there is some variation in transition boundaries  
due to the radial motion of the EFs.
 
The particle Stokes numbers (blue curves in the middle panel of Fig. 
\ref{fig:3panF}) help to demonstrate why there is systematic mass transfer from the outer nebula to the inner nebula. The solid Stokes number curve refers to $r_{\rm{M}}$ and the dotted Stokes curve refers to the largest particle
in the size distribution which include the ``lucky" 
particles $r_{\rm{L}} > r_{\rm{M}}$. They diverge inside the ice line, where only rare lucky particles have such large values of ${\rm{St}}$ and $r_{\rm{L}}$. Even while, near the ice EF, particles of about 1 m radius can be found, the high gas densities mean that ${\rm{St}} < 1$ everywhere except at the outermost
edge of the solids disk ($\sim 200$ AU) where the gas density is very small.   Recall from Eq. (\ref{equ:Udrift}) that for ${\rm St} < 1$, the radial drift velocity is proportional to ${\rm St }$. Inside the ice EF, ${\rm St_M }$ drops by almost two orders of magnitude because of the smaller $Q_*$, causing the bulk of particles to drift inward more slowly than outside the ice EF. This explains why there is a pileup of solid 
material inside the ice EF. The fact that $r_{\rm{L}}$ gradually decreases from just outside the water 
ice EF to the inner nebula is due to the fact that even the lucky particles are not growing fast enough to overcome 
the radial drift barrier. This is discussed in more detail in Sec. \ref{sec:discuss} (see Fig. \ref{fig:stokes}).
The maximum drift rate for particles is at St = 1 (Eq. [\ref{equ:Udrift}]), but particles with 
${\rm{St}} \lesssim 1$ are still drifting faster as they grow larger, so these particles are suffering a gradual but steady loss inwards. Because the radial drift barrier cannot be overcome means that eventually most all material
will be lost at later times.
 
Interestingly, the maximum Stokes number St$_{\rm{L}}$ is roughly {\it constant} outside the water ice EF, at around a value of $\sim 0.1$. A simple relationship explains this behavior in the fragmentation-dominated regime 
\citep{wei88,cw06,bir09,bir11}.
One merely sets the collision speed $\Delta V_{\rm{pp}}$ equal to the fragmentation speed $u_f \sim \sqrt{2Q_*}$. 
Since $\Delta V_{\rm{pp}}^2 \sim 2{\rm St} \alpha_{\rm t} c^2$ for identical particles 
\citep[][note Birnstiel et al. drop the numerical factor]{oc07},
the largest particle in fragmentation equilibrium has Stokes number 
\begin{equation}
\label{equ:stfrag}
{\rm St}_* = \frac{u_f^2}{2\alpha_{\rm{t}} c^2} = \frac{Q_*}{\alpha_{\rm{t}} c^2}, 
\end{equation}

\noindent
and $c$ is nearly constant across the outer disk at least. Even in the inner disk where $c$ varies somewhat more significantly, ${\rm St}$ is nearly constant, but at a smaller value commensurate with the lower $Q_*$ for silicates. 

In the Epstein flow regime, which applies to most of the particles shown in Fig. \ref{fig:3panF}, the Stokes number can be
written as \citep[e.g.,][]{bir11}

\begin{equation}
\label{equ:StE}
{\rm{St}} = \frac{2 r\rho_{\rm{p}}}{\Sigma}.
\end{equation}

\noindent
For a roughly constant Stokes
number as a function of semimajor axis, $r \propto \Sigma$ implying that the decrease in $r_*$ and $r_M$ 
in the outer regions simply mirrors the decrease in gas surface density. 
However, the behavior in the {\it inner} nebula is not reliably predicted from these formulae because the particle sizes are not strictly in the Epstein regime, but often in the transitional or full Stokes flow regime (which is generally ignored in other models). Though we cannot easily find simple expressions for expressing St in terms of the transitional formula we use, we 
can still give estimates for St when $r\gtrsim \lambda_{\rm{mfp}}$ (see Appendix A.4):

\begin{equation}
\label{equ:StS}
\begin{split}
{\rm{St}} = \frac{16}{3}\frac{\rho_{\rm{p}}}{\Sigma}\frac{r c}{C_{\rm{d}}\Delta V_{\rm{pg}}} \simeq 
\,\,\,\,\,\,\,\,\,\,\,\,\,\,\,\,\,\,\,\,\,\,\,\,\,\,\\
\begin{cases}
\frac{2}{9}\frac{\rho_{\rm{p}}r^2 V_{\rm{K}}}{R \mu_{\rm{m}}} &\,{\rm{Re}_p} < 1;\\
\frac{1}{4}\frac{\rho_{\rm{p}}^{5/6}r^{4/3}}{(\alpha_{\rm{t}}^{1/2}\Sigma)^{1/3}}\left(\frac{V_{\rm{K}}}{R\mu_{\rm{m}}}
\right)^{1/2} &\,{\rm{Re}_p} < 800,\\
\end{cases}
\end{split}
\end{equation}

\noindent
where in the case ${\rm{Re}_p} > 1$ the particle-to-gas relative velocities $\Delta V_{\rm{pg}}$ are assumed to be due to turbulence. In this
work, the largest particles we grow have near unity particle Reynolds numbers. In general, for relatively smaller particles,
the assumption of turbulent driven relative velocities is probably adequate unless $\alpha_{\rm{t}}$ is very small. However, as particles grow larger or in the outer nebula, their relative velocities may be driven by headwind effects and closer to $\sim \eta V_{\rm{K}}$. Later in Sec. \ref{sec:discuss} we will compare these estimates to our 
simulations directly.
 
Also in the middle panel of Fig. \ref{fig:3panF}, the particle internal densities are plotted for both the migrator and dust particles as a function of
semimajor axis. The ``cosmic  abundance" particle density of $\sim 1.5$ g cm$^{-3}$ is maintained in the
outermost nebula, but significant changes in mean particle density have occurred in regions where EFs are or were found. The smoothness of the particle density profiles reflects the process of sublimation from drifting particles at EFs, 
followed by outward diffusion, recondensation, and advection. The most striking example of this process is
evident in the region between $5-9$ AU where the density of local solid material has essentially dropped to the density
of water ice.  
The radial extent of this water-ice enhanced region is partly a result of outward diffusion of vapor and
advection of icy grains, but (in this case) also of the inward evolution over time of the water ice EF to its current location at $\sim 4$ AU.
The magnitude of this enhancement may be slightly overestimated {\it if} some amount of the drifting water ice were not evaporated inside the EF, but somehow buried and trapped in migrators to be carried further inwards. Finally, we note that 
slight differences in the density of migrators versus the ``dust" population outside of EFs are due to the fact that we assume 
that condensation of vapor follows the surface area of particles (dominated by the smaller particles).

For the same case ``F", {\bf Figure \ref{fig:3pangasF}} shows the evolution of gas (black curves) and solids (red curves) surface density at 
three different times (top panel), and (at $2\times 10^5$ years) the radial velocities of solids and gas (middle panel) and the gas and
solids accretion rates (lower panel). The gas radial velocities (middle panel, solid curve) are
relatively low due to the choice of $\alpha_{\rm{t}}$, and change from inward ($V_{\rm{g}} < 0$) to
outward flow ($V_{\rm{g}} > 0$) at around 7 AU (see below). Also in the middle panel we plot the radial velocities of the largest particle 
$V_{\rm{d,L}}$ (dotted curve, corresponding to the dotted curve in the middle panel of Fig. \ref{fig:3panF}) 
which shows how quickly in most places these particles are drifting inwards. Variations in $V_{\rm{d,L}}$ are due to
differences in the size (and mass) of particles as well as transitions through EFs (Sec. \ref{subsubsec:drift}). 
We also plot the radial
velocities of the inward ($V^-_{\rm{d}}$, long dash) and outward ($V^+_{\rm{d}}$, short dash) drifting components 
of the dust population (Sec. \ref{subsubsec:diffadv}). The (small) outward flow of the smallest
grains (due to advection and diffusion) explains why there are solids further out than the initial $100$ AU solids 
cutoff. The inward drift
is strong in the outer disk and explains how the outer disk has been cleared of most of its material.
We note that $V^-_{\rm{d}}$ decreases sharply in the inner disk and eventually becomes similar to $V_{\rm{g}}$.
This decrease can be associated with the drop in St inside the water ice EF, and leads to 
the subsequent pileup of solid material in the inner nebula.

Reversal of the radial mass flow driven by viscosity occurs
around 7 AU, with outward flows beyond this value, but in spite of this, by $2 \times 10^5$ years the bulk of the outer nebula solids have collapsed to within this radius. This
can be seen in the surface density plot (top panel) where although the gas surface density has not
evolved considerably due to the low $\alpha_{\rm{t}}$, the changes in the solids over time are pronounced (it is even more apparent in the lower panel of Fig. \ref{fig:3panF}). At $10^4$ years there is still a considerable amount of solid material out to 100 AU, but 
by $2\times 10^5$ years, peaks in $\Sigma_{\rm{d}}$ outside the organics and water ice EFs are well defined, supplied by inwardly drifting material.  Inside 7 AU, the advection of material is inward, and although some vapor may diffuse outward to recondense, this 
appears to remain well enough inside the radius of mass flow reversal so that there is no longer an 
effective outward transport mechanism and material will remain confined to the inner nebula. 

Other effects are more subtle. In the lower panel of Fig. \ref{fig:3pangasF}, sharp decreases in radial drift and an ensuing pileup of solids  at the 
water ice EF are associated with a spike in the total 
accretion rate of solids  $\dot{M}_{\rm{d}}$ (red curve). Just inside the water ice peak one finds that there is a bump in the gas accretion rate
that corresponds to a local minimum in the pressure gradient (dashed curve). The radial drift of the lucky particles drops at 1.5 AU; this is associated with a local decrease in $r_{\rm{L}}$, related to a peak in the total migrator {\it mass} and the locally increased erosion that results.

\subsection{Fragmentation and bouncing (``BF")}
\label{subsec:BF}

\begin{figure}
 \includegraphics[trim = 120 0 -120 0,clip,width=6.0in,height=6.0in,angle=0.0]{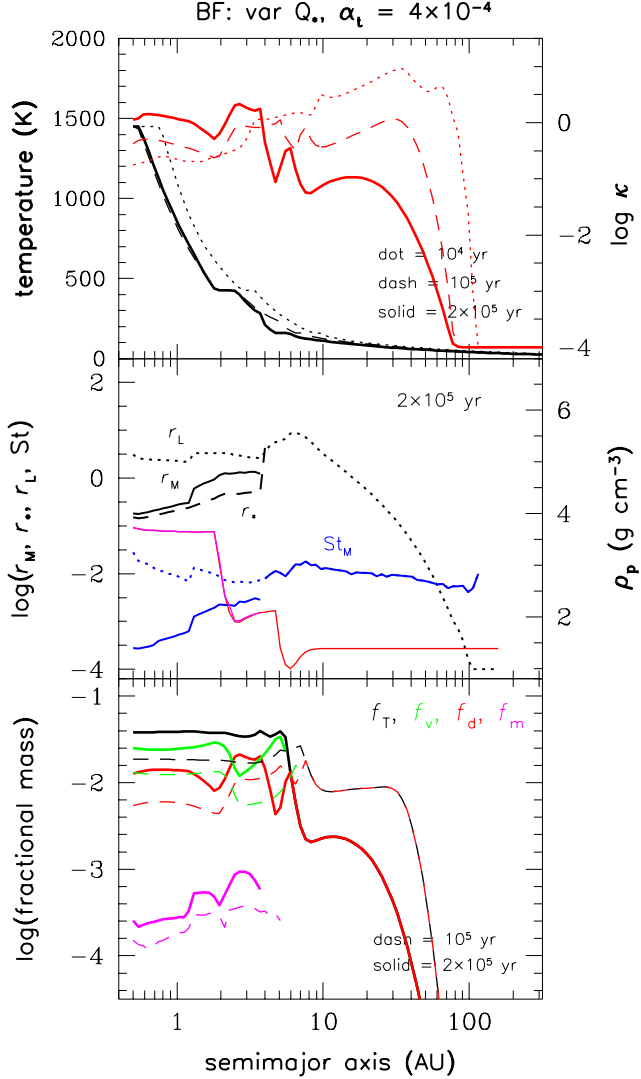}
\caption{Fiducial model with bouncing and fragmentation (BF). Panel descriptions same
as Fig. \ref{fig:3panF}. When the bouncing barrier is included in our simulations, particles grow more slowly
and remain relatively small over the period of the simulation ($r \geq r_*$ inside $\sim 5$ AU only). Lucky 
particles are also smaller compared to the F case. As a result the temperature remains hot in the inner nebula, 
and opacities are generally larger there as well. Particle growth in the outer disk is still sufficient to lead 
to significant inward transport, though much less than the F case. Smaller particles also lead to relatively
large enhancements in dust and vapor in the inner nebula. See text for details.}
\label{fig:3panBF}
\end{figure}

When the so-called ``bouncing barrier" is included in our fiducial model, the situation changes considerably. Because the particles are growing more slowly and remain smaller (see below) the opacity is larger and the temperature
in the inner nebula remains high even after $2\times 10^5$ years (see {\bf Figure \ref{fig:3panBF}}, top panel), 
remaining more or less unchanged from its value at $10^5$ years.
Also, the water ice EF is found further out, at around 7 AU. The opacities are still decreasing with time in 
the outer nebula and increasing in the inner nebula, but remain generally larger than in the pure fragmentation case F.  The fractional masses (lower panel, Fig. \ref{fig:3panBF}) again show enhancement peaks in the solids near the organic and water ice EFs corresponding to local peaks in the opacity, but not so much the silicate EF. Between $10^5$ and $2\times 10^5$ years, the EFs evolve only slightly in this case, so unlike the fragmentation-only case F, the enhancements {\it in the solids} are higher outside EFs - in fact, exceeding the enhancements of the vapor for
the organics EF. The {\it total}  
enhancement of condensibles, in both solid and vapor form - in the inner regions is as high as $\sim 5-6$ 
(black curve) by $2\times 10^5$ years.
The overall fractional mass of migrators is similar ($\sim 10^{-3}$) to the fragmentation case, but they are now 
present everywhere in the innermost regions. A sharp cutoff in the migrator population around 5 AU is because
particle sizes have not yet reached the fragmentation barrier outside this location. Interestingly, though no peak
in the dust outside the silicates EF is quite discernable, the migrators do show a noticeable enhancement.

Particle sizes plotted in the middle panel of Fig. \ref{fig:3panBF} demonstrate that in most of the disk, the largest particle has not reached the fragmentation size
($r_{\rm{L}} < r_*$) - so the heavy dashed curve for $r_*$ does not appear. The bouncing barrier has slowed the growth rate considerably because sticking becomes zero for equal mass particles at much smaller sizes than in the fragmentation-only case. 
In general, the bouncing barrier (BB) restricts eligible collision partners for growth to a narrower range of smaller projectile sizes - causing {\it all} particles to grow more slowly. Still, as long as a population of small particles remains, the BB is not impermeable and we expect that the fragmentation barrier will be reached at later times.

After $2\times 10^5$ years, only
particles inside the water ice EF have reached $r_*$ (heavy dashed curve appears) which is understandable since the larger $Q_*$ outside
this EF means it will take much longer to reach the larger fragmentation limit $r_*$  ({\it cf.} Fig. \ref{fig:3panQ107}, Sec. \ref{subsec:conQstar}). It is apparent that $r_*$, $r_{\rm{M}}$, and $r_{\rm{L}}$ in the inner nebula are smaller in this case than in case F. This seems counterintuitive since the strength remains the same; however, regarding $r_{\rm{M}}$  and $r_{\rm{L}}$, the reason is that
the number density of dust particles is much higher and the probability of destruction for particles of size $r_M$ and 
$r_{\rm{L}}$ is proportional to the number densities of
the smaller particles they are colliding with (see Eq. [\ref{equ:rcoll}]). St$_{\rm{M}}$ is lower than St$_*$ which
we argue is due to a slow growth or drift related effect which we discuss in Sec. \ref{subsec:psize}.
 
The smaller and inwardly decreasing value of $r_*$ is explained in a different way. The Stokes numbers (blue curves) are
smaller by an order of magnitude than in case F. A smaller ${\rm{St}}_*$ is explained by the higher opacity, 
temperature, and stronger radial temperature gradient than the lower and constant temperature in case F at $2 \times 10^5$ years (see Eq. [\ref{equ:stfrag}], also see discussion, Sec. \ref{sec:discuss}), and this leads to a smaller $r_*$. The bulk of particles in the inner nebula have
${\rm{St_M}} \lesssim 10^{-3}$ causing them to drift at a much slower rate, and to be retained 
for longer periods before being lost with an ensuing increase in mass enhancement. Outside the water ice EF, we see a similar trend in St$_{\rm{M}}$ as in case F, but at smaller values because the fragmentation size has not been reached.
 
\begin{figure}
 \includegraphics[trim = 120 0 -120 0,clip,width=6.0in,height=6.0in,angle=0.0]{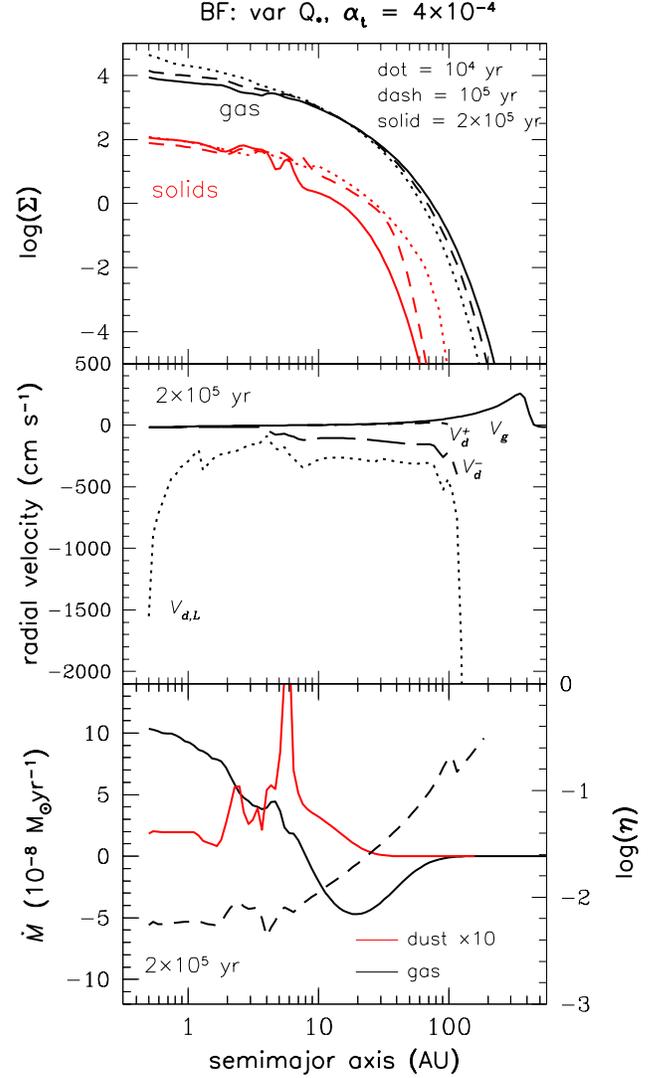}
\caption{Fiducial model with bouncing and fragmentation (BF). Panel descriptions same
as Fig. \ref{fig:3pangasF}. Because growth times are longer, and particles generally smaller, the dust surface
density has evolved much less than the F case after $2\times 10^5$ years. The radial velocities of the
particles sizes of interest all reflect this trend. The dust mass accretion rate is larger because more mass is
in the dust, and shows substantial variation due to interaction with the various EFs. See text for details.}
\label{fig:3pangasBF}
\end{figure}

Finally, the internal density of dust and migrator particles (red and magenta curves, middle panel)
retains interesting variability.
In general, the internal density profiles are characterized by 
decreases just outside the various EFs; this is due to enhancement in the evaporating species as it diffuses back across the EF and condenses on to dust and migrator particles. The radial range over which the density decreases is due not only
to vapor condensation but to the diffusion and advection of smaller dust grains. The interplay is quite complex though,
as we will see below from gas and dust accretion rates. In particular, the density
variation outside the water ice EF for case BF does not span the same radial extent as in case F,
because the EF has been relatively stationary over the simulation time in case BF. However, its edges are less
sharp because of the more significant outward advection and diffusion of smaller icy grains.

The radial velocities of the various particle sizes plotted in {\bf Figure \ref{fig:3pangasBF}} (middle panel) are
 lower than for case F, reflecting slower radial migration of smaller particles. The location where $V^-_{\rm{d}}$
becomes similar to the gas radial velocity is slightly further outward due to the smaller particle size, and
in fact occurs just outside the organics EF. The inward drift of lucky particles reaches a minimum speed at this point, 
and increases inwards up to the silicate EF which is at the inner edge of the computational grid.
The more complex behavior seen in the dust and gas accretion rates compared to case F again emphasizes that there is inward advection of material in the inner regions, although at speeds less than the gas velocity, but also outward 
diffusion at EFs further affecting the accretion rate.
The gas accretion rate shows variation through the organics and water ice EFs, because particles containing 
most of the mass are small and traveling with the gas. The mass accretion rates of both the 
gas and dust become more similar at these points implying that the dust is apparently influencing the gas evolution even though the solids-to-gas-ratio is not near unity.
The reason is probably because opacity peaks lead to temperature gradient fluctuations, that lead to radial viscosity variations through the sound speed $c$, and thus corresponding variations in the gas radial flow.

\begin{figure}
 \includegraphics[trim = 120 0 -120 0,clip,width=6.0in,height=6.0in,angle=0.0]{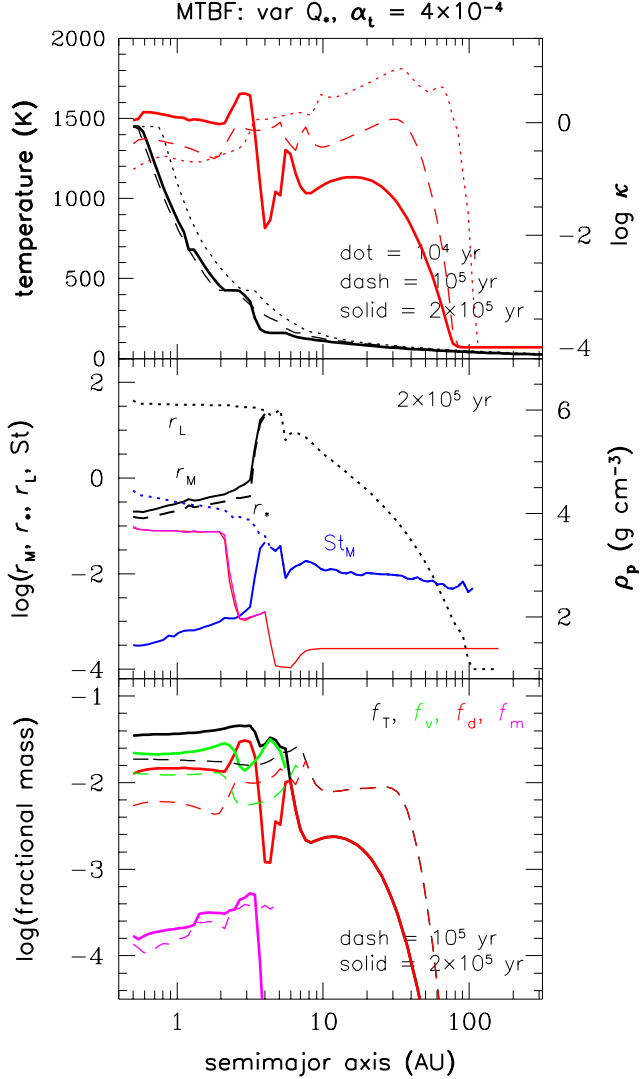}
\caption{Fiducial model with mass transfer, bouncing and fragmentation (MTBF). Panel
descriptions same as Fig. \ref{fig:3panF}. In general, the MTBF case is quite similar to BF with the exception
that growth in the migrators is slightly faster, and particle sizes are a bit larger. Overall, the enhancements
one finds in both the dust and vapor in the inner nebula are similar to the BF case, except that more mass 
appears in the dust and less in the vapor. The temperatures in the silicate rich region are slightly hotter as well. 
See text for details.}
\label{fig:3panMTBF}
\end{figure}

\begin{figure}
 \includegraphics[trim = 120 0 -120 0,clip,width=6.0in,height=6.0in,angle=0.0]{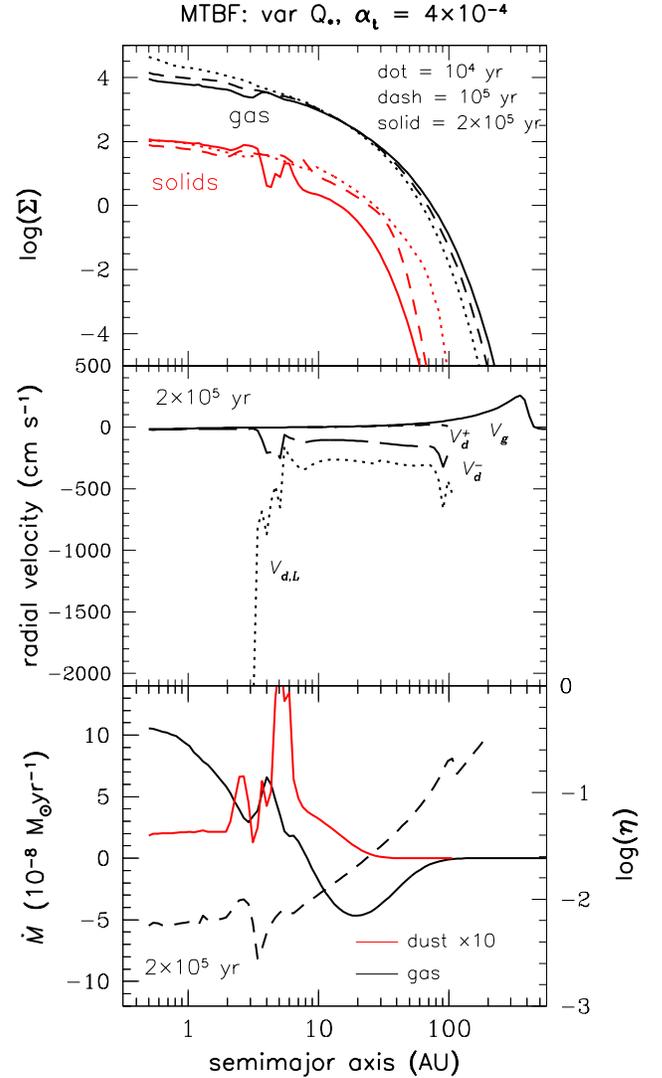}
\caption{Fiducial model with mass transfer, bouncing and fragmentation (MTBF). Panel
descriptions same as Fig. \ref{fig:3pangasF}. The overall trends are similar to the BF case, especially in the
outer nebula. Differences in the inner nebula can be attributed to the evolution of the particle size distribution
when mass transfer is included. See text for details.}
\label{fig:3pangasMTBF}
\end{figure}

\vspace{0.2in}
\subsection{Fragmentation, bouncing and mass transfer (``MTBF")}
\label{subsec:MTBF}

The process of mass transfer (in high velocity collisions) encompasses an abbreviated range of conditions (Sec. \ref{subsubsec:mig}), but can have a significant effect on certain aspects of particle growth over
long periods \citep{win12a,win12b,win13,gar13}. In {\bf Figures \ref{fig:3panMTBF} and \ref{fig:3pangasMTBF}} 
we show the results of our fiducial model that also includes mass transfer. The  evolution of temperature
(top panel, Fig. \ref{fig:3panMTBF}) somewhat differs between this case and the previous case BF, in that $T$ in 
the innermost nebula is slightly higher by $2\times 10^5$ years, even though the water ice EF has moved inwards. 
The opacities also are similar in magnitude, but their radial distribution is different. The water ice EF has a 
larger radial extent here. Compared to the BF case, there is a much larger drop in the opacity inside the ice EF 
and a much larger opacity enhancement outside it, suggesting (like in case F) that there is a larger
difference between particle sizes inside {\it vs.} outside the water ice EF. The large contrast across the water 
ice EF is also visible in the fractional mass plot (lower panel).  The outer disk remains similar to case BF, but 
differences in the inner regions can be associated with the evolution of the particle size distributions. Overall, the total enhancement in the inner nebula is similar to the BF case, a factor of $\sim 5-6$, though with slightly more
in the solids component and less in the vapor.

The contrast between particle sizes inside and outside the water ice EF is shown in the middle panel, where 
compared to case BF, growth at the larger sizes is faster and as a result the fragmentation size $r_*$ has been 
reached further out in semimajor axis by $2\times 10^5$ years. The effect of mass transfer appears to lead to
two things. A slightly larger value of $r_{\rm{M}}$ relative to $r_*$ most easily seen inside $R < 1.2$ AU, and
consistently larger lucky particles. These effects are much more evident when comparing the middle and lower panels of
Fig. \ref{fig:sizedist} which shows particle size distributions over this radial range. The lower panel, which
refers to the MTBF case, is characterized by secondary peaks which are indicative of particles that are benefitting
from particle collisions in which the growing particle accretes more mass than is eroded 
(see Sec. \ref{subsubsec:mig}).
A slight dip in $r_{\rm{L}}$ appears outside the the water ice EF which is associated with the large 
spike in fractional mass there. 
The particle mean density profiles are not much different from the BF case, except in the water ice enhanced region, with the differences tied to the different radial extents of the organic and water EFs in the two cases. In the ice-enhanced region just outside the ice EF, both $r_{\rm{M}}$ and $r_{\rm{L}}$ are larger when mass transfer is 
playing a role. 
 
The velocity trends in {\bf Figure \ref{fig:3pangasMTBF}} are qualitatively similar to the BF case. The most noticeable differences
appear in the radial velocities (middle panel). The presence of much larger lucky particles inside the water ice EF that are migrating 
inwards with relatively larger radial velocities (Eq. [\ref{equ:Udrift}]), accounts for the much steeper inwards increase in $V_{\rm{d,L}}$
compared to the BF case. Also, whereas $V^-_{\rm{d}}$ drops off gradually to its lowest value in the BF case between $\sim 3-4$ AU, a 
relatively sharp bump appears in the MTBF case because the fragmentation barrier has been reached and most of the mass here is in larger
particles which influence the gas. This coincides with a stronger peak in $\dot{M}$ than what was seen at the same location in the BF case.
The corresponding pressure gradient also a takes a much sharper drop here (lower panel).

\begin{figure}
 \includegraphics[trim = 120 0 -120 0,clip,width=6.0in,height=6.0in,angle=0.0]{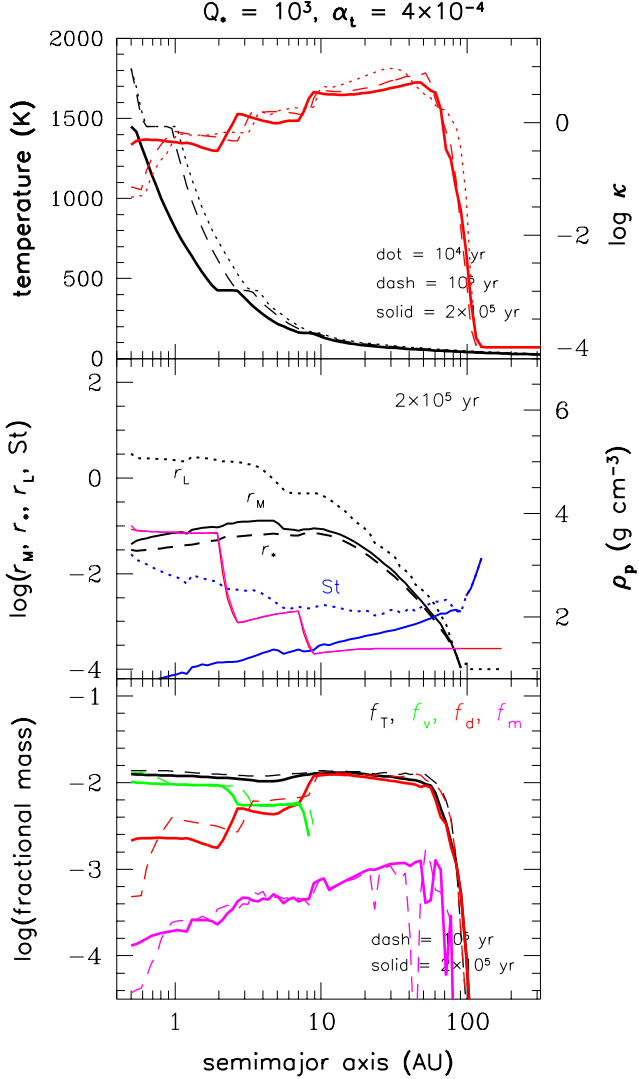}
\caption{MTBF model assuming a constant fragmentation energy $Q_* = 10^3$ (Q103). Panel
descriptions same as Fig. \ref{fig:3panF}.}
\label{fig:3panQ103}
\end{figure}

\begin{figure}
 \includegraphics[trim = 120 0 -120 0,clip,width=6.0in,height=6.0in,angle=0.0]{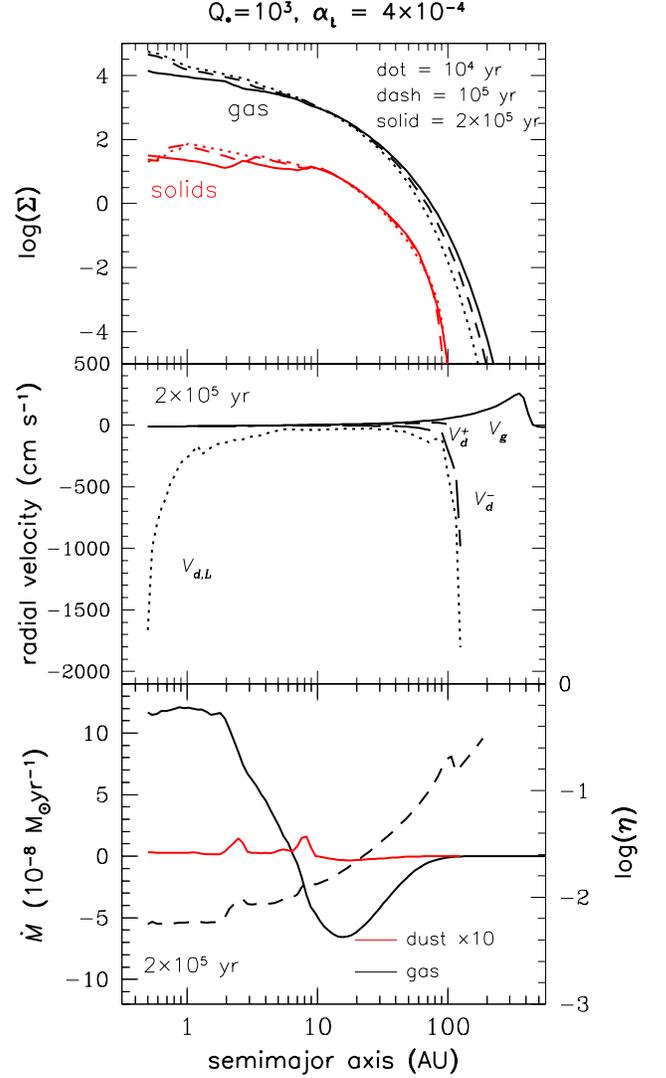}
\caption{MTBF model assuming a constant fragmentation energy $Q_* = 10^3$ (Q103). Panel
descriptions same as Fig. \ref{fig:3pangasF}.}
\label{fig:3pangasQ103}
\end{figure}

\subsection{Models with constant $Q_*$}
\label{subsec:conQstar}

We have employed a variable $Q_*$ (and bouncing barrier) in all of the models presented thus far, motivated by experimental results that suggest that icy aggregates are stickier and indeed stronger, at the velocities in question, than silicate aggregates (see Sec.
\ref{subsubsec:stick}). In this section, we present MTBF simulations that assume  constant values for material strength, choosing two extreme values for $Q_* = 10^3$ and $Q_* = 10^7$ which
would correspond to fragmentation velocities of $\sim 30$ cm s$^{-1}$ and $\sim 30$ m s$^{-1}$,
respectively. Of these two, results of \citet[][for silicates]{bei11} suggest that the larger value is especially unrealistic. 

\begin{figure}
 \includegraphics[trim = 120 0 -120 0,clip,width=6.0in,height=6.0in,angle=0.0]{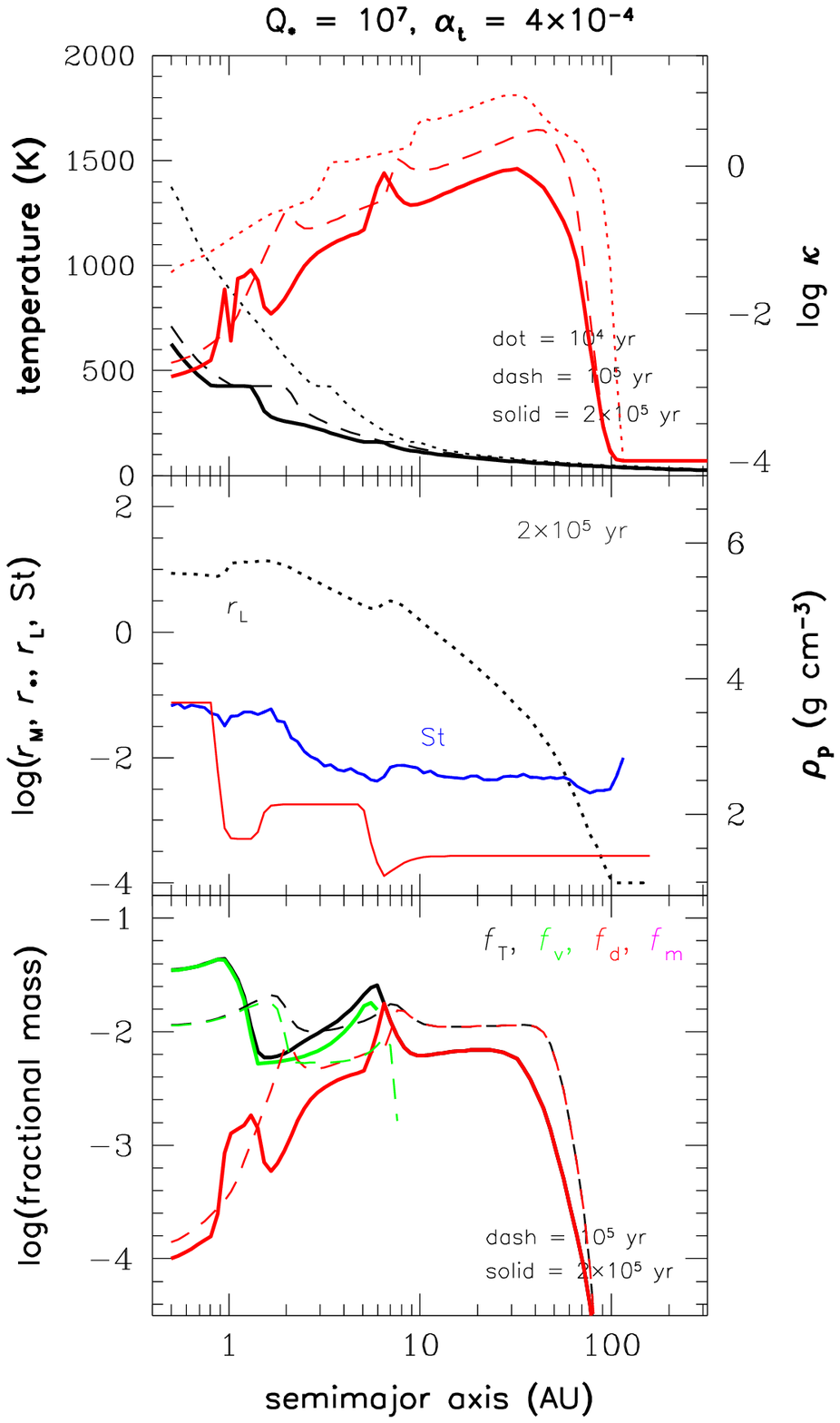}
\caption{MTBF model assuming a constant fragmentation energy $Q_* = 10^7$ (Q107). Panel
descriptions same as Fig. \ref{fig:3panF}.}
\label{fig:3panQ107}
\end{figure}

\begin{figure}
 \includegraphics[trim = 120 0 -120 0,clip,width=6.0in,height=6.0in,angle=0.0]{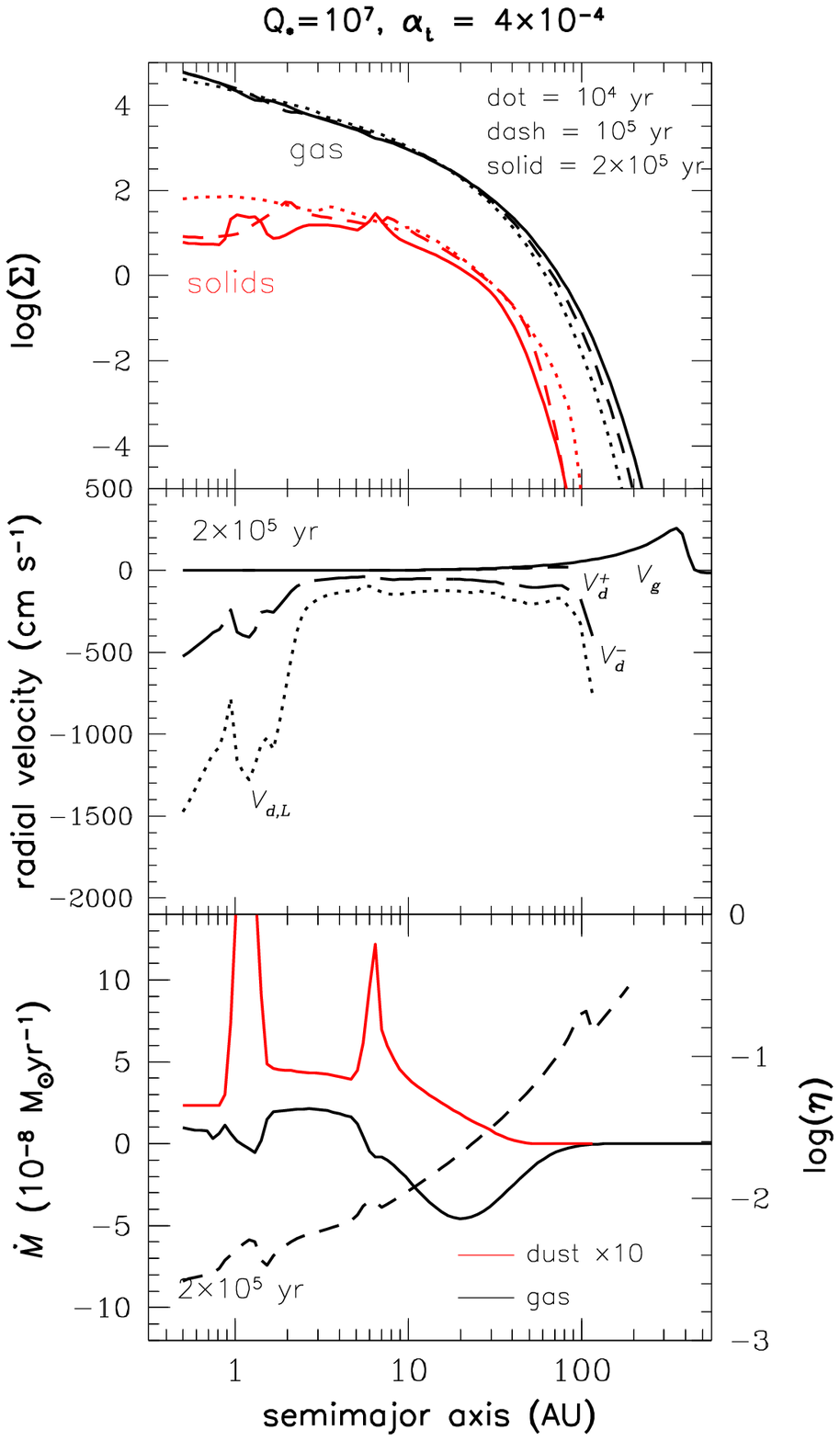}
\caption{MTBF model assuming a constant fragmentation energy $Q_* = 10^7$ (Q107). Panel
descriptions same as Fig. \ref{fig:3panF}.}
\label{fig:3pangasQ107}
\end{figure}

In {\bf Figures \ref{fig:3panQ103} and \ref{fig:3pangasQ103}}, we show our simulation results for $Q_* = 10^3$.
Not surprisingly, this small value of $Q_*$ means that particle sizes (middle panel, Fig. 
\ref{fig:3panQ103}) are much smaller throughout the
entire disk. As an example, $r_* \simeq 0.18$ cm at 1 AU in the MTBF case (where $Q_* = 10^4$ for silicate
grains) whereas $r_* \simeq 0.02$ cm for this case, nearly an order of magnitude smaller. The maximum
fragmentation size occurs around 10 AU. Because particles remain smaller, Stokes numbers
are smaller which means there is  less radial drift (and loss) of material over $2\times 10^5$ years.  The Stokes numbers for $r_{\rm{M}}$  decrease monotonically from
larger values in the outer disk to ${\rm{St_M}} < 10^{-4}$ in the innermost regions, and we do not see the
roughly constant values as we did in the previous cases. Interestingly, however, we do
see the St$_{\rm{L}}$ roughly constant, or $r \propto \Sigma$, relationship hold for the lucky particles in the outer nebula. 

The temperature and opacity (upper panel) and the fractional masses (lower panel) further 
emphasize the sluggish disk evolution this case results in. The disk has only cooled significantly in the last $10^5$ years, but still
remains relatively hot partly because of a gradual
increase in inner nebula opacity due to a small enhancement of solids. Unlike other cases we have seen previously, the outer 
edge of the water ice EF has moved very little, with a slight increase in radial extent. Because the
fragmentation barrier is at such small sizes, migrators appear throughout the disk at a fraction
of $\sim 10$\% of the dust component. Moreover, the fractional mass in the outer disk has remained
similar to its initial value because the gas and dust are still well coupled. Despite
the sharp pressure gradient at the outer edge (dotted curve, Fig. \ref{fig:3pangasQ103}, lower panel)
a significant amount of solid material has advected outward, beyond 100 AU. The radial velocities in 
Fig. \ref{fig:3pangasQ103} (middle panel) are very close to the gas velocity everywhere except approaching the
disk edges. In fact the inward drifting dust velocity component $V^-_{\rm{d}}$ diverges from nearly zero
only at the outer edge of the nebula solids disk. The surface densities (upper panel) also show generally
little evolution, and the dust accretion rate (lower panel) is muted. A low $Q_*$ model thus appears to be
a way to minimize the amount of solids loss over long timescales, as previously suggested by \citet{bir09}.

\begin{figure}
 \includegraphics[trim = 120 0 -120 0,clip,width=6.0in,height=6.0in,angle=0.0]{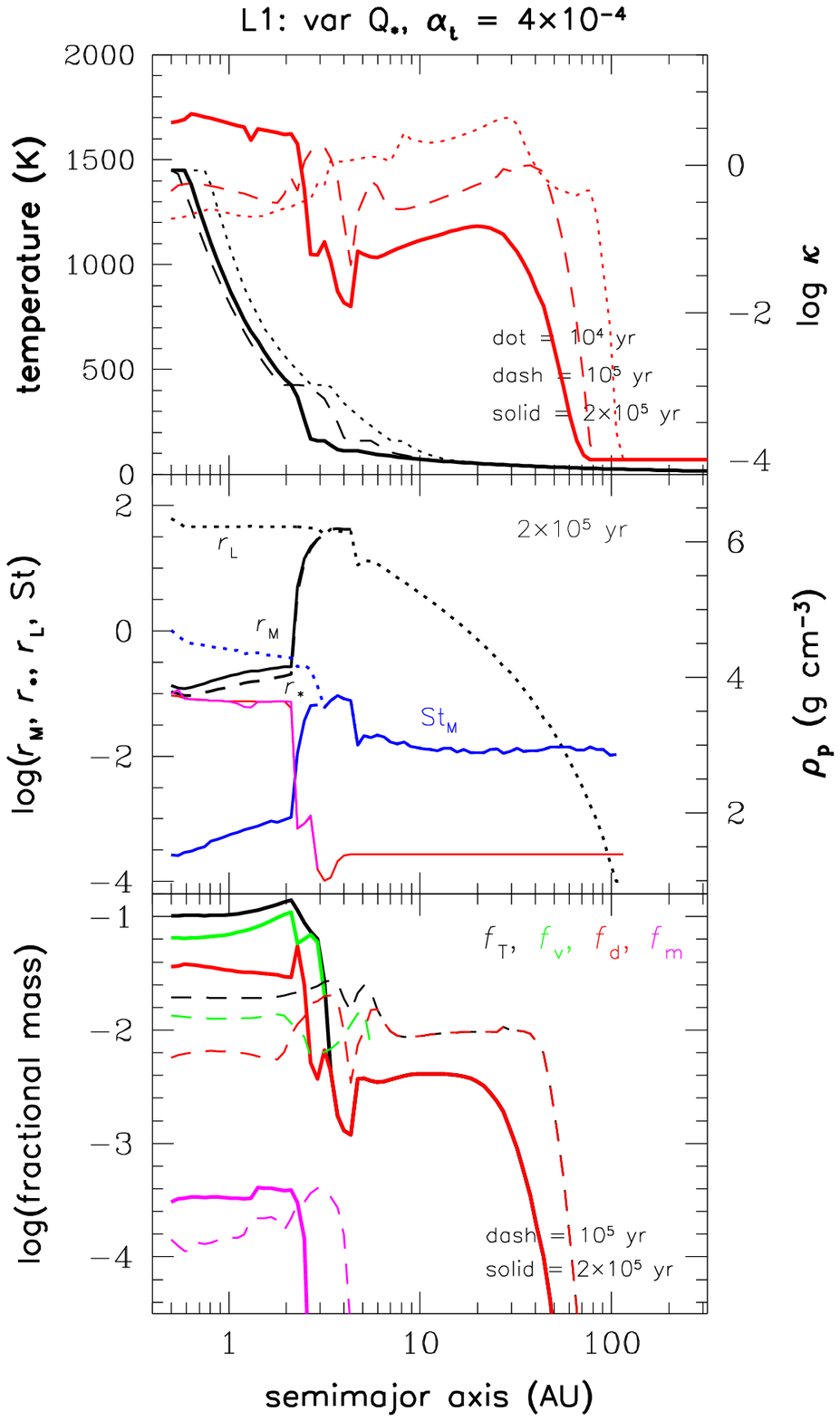}
\caption{MTBF model assuming a star with 1 solar luminosity (L1). Panel
descriptions same as Fig. \ref{fig:3panF}.}
\label{fig:3panL1}
\end{figure}

\begin{figure}
 \includegraphics[trim = 120 0 -120 0,clip,width=6.0in,height=6.0in,angle=0.0]{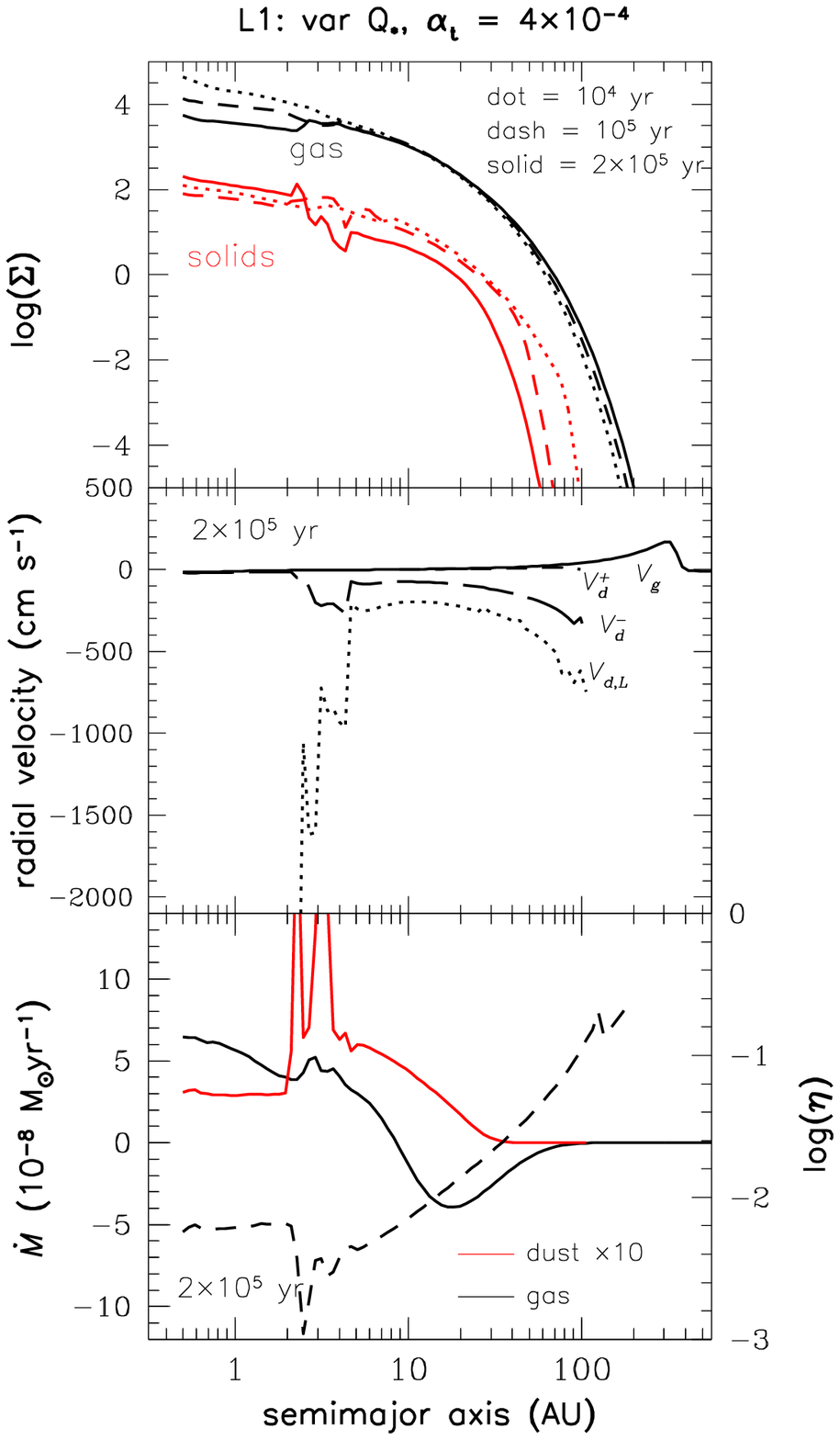}
\caption{MTBF model assuming a star with 1 solar luminosity (L1). Panel
descriptions same as Fig. \ref{fig:3panF}.}
\label{fig:3pangasL1}
\end{figure}

{\bf Figures \ref{fig:3panQ107} and \ref{fig:3pangasQ107}} for $Q_* = 10^7$ show an entirely different behavior.
In this case, the fragmentation size is so large that even after $2\times 10^5$ years, $r < r_*$ everywhere (heavy dashed curve does not appear),
even though $r_{\rm{L}} \sim r_{\rm M}$ is as large as $\sim 10$ cm in the inner nebula (Fig. \ref{fig:3panQ107}, 
middle panel, dotted curve). The Stokes number neatly follows the nearly constant theoretical ``equilibrium" value in the outer disk, giving $r \propto \Sigma$ (see Sec. \ref{subsec:F}) but diverges from this relationship in the inner regions possibly because these relatively large particles may be in the transitional Stokes regime. The disk 
temperature (top panel) decreases quickly because
the particles grow relatively quickly, and much of the material in the inner disk
is beginning to be lost, leading to the substantial decrease in opacity. Despite the systematic
loss of material, we still see two strong peaks in opacity due to the enhancements outside the organics and water 
ice EFs. These are also mirrored in the fractional masses (lower panel). We note an apparent peculiar difference
between this case and other cases we have shown so far: the outer nebula solids disk has not ``collapsed''
as far as other models, despite the fact that the higher $Q_*$ here allows growth to much larger particles, which
should drift more quickly. The explanation is that the bouncing barrier we assume here (as in the case of $Q_*=10^3$)
is also a constant set at the silicate value, but is {\it not} scaled with $Q_*$ (see Sec. \ref{subsubsec:stick}), 
so that the disparity between the bouncing and fragmentation barriers is especially considerable in this case. 
Not only does this mean that growth is slower since larger particles can only grow from particles smaller than
the bouncing size, but the number density of those bouncing barrier-or-smaller-sized particles is always decreasing
here because they are not being replenished by fragmentation.

Finally, looking at the radial velocities in Fig. \ref{fig:3pangasQ107} (middle panel), we see that the radial 
drift rates are larger in the inner nebula for both the largest particles ($V_{\rm{d,L}}$) and the dust ($V^-_{\rm{d}}$), explaining the significant loss of material from the inner nebula, the opposite of what was found in other cases. 
In this case, the gas surface density (top panel) has evolved the least of all the cases we have shown thus far, with values very
close to the initial profile even after $2\times 10^5$ years. In fact, in the innermost region, the gas density profile has become even steeper. This is because the rapid growth and loss of inner solar system solids has led to the lowest opacity, producing the fastest cooling and lowest viscosity. Remember that results like this depend on the $\alpha_{\rm{t}}$-model 
prescription for viscosity increasing with temperature.  The steady inward migration
of material is also evident from the dust accretion rate that remains positive throughout the disk. The gas
accretion rate is much less strong in this case in the inner regions because of 
the rapid cooling and steeper surface density. 

\subsection{A 1 L$_\odot$ model}
\label{subsec:L1}

We have explored an MTBF model in which we fix the luminosity at 1 L$_\odot$,  as 
presented in {\bf Figures \ref{fig:3panL1} and \ref{fig:3pangasL1}}. Previous models have all used the time
dependent luminosity model of \citet[Table 3,][]{dm94} where the initial luminosity is of order 
$\sim 12$ L$_\odot$ and decreases to $\sim 3$ L$_\odot$ after 1 Myr.
Thus here we can compare the effects of luminosity in the MTBF model. The primary effect one would expect
is on the disk temperature (top panel of Fig. \ref{fig:3panL1}). We find only 
subtle differences in the overall temperature profiles, which are primarily associated with the
radial extent of the water ice and organics EFs, and the overall steepness of the temperature profile inside this
location. In particular, the gradient is so steep that the organics EF is traversed within one 
radial bin. The inner nebula shows only small changes because its energetics is dominated through these stages by local viscous dissipation rather than stellar irradiation. 
Differences in the Rosseland mean opacity profiles suggest that growth is proceeding
a bit more rapidly in the outermost portions of the disk, at least at the earliest stages ($10^4$ years).
This is also evident in the fractional mass plots (lower panel). The $10^5$ year fractional mass curves look quite similar for this case and the fiducial MTBF case, but have slightly larger radial extent because of the location of the EFs. However, by $2\times 10^5$ years the vapor and
dust content in the inner regions are considerably higher than in the MTBF case, with an overall enhancement of an
order of magnitude over the initial profile.
The Stokes numbers corresponding to $r_{\rm M}$ and $r_{\rm L}$ 
(middle panel) are even flatter across the outer disk (and higher than in the MTBF case, which may explain 
why the outer disk evolved faster in the early stages here). In the inner nebula, there are also differences in the 
particle sizes relative to the MTBF case, which may derive from the increased number density of particles
making mass transfer only slightly more effective for particles growing from $r_*$ to near the size of $r_{\rm{M}}$.

In {\bf Figure \ref{fig:3pangasL1}}, the gas accretion rate profile is flatter, and its magnitude lower than the MTBF
case. This is once again due to the steeper temperature gradient. In particular, one finds that the pressure
gradient plummets much more sharply through the organics EF with this location corresponding to the sharp decrease 
in the solids accretion rate at $\sim 2$ AU just inside the strong organics peak.
The peak in $\dot{M}_{\rm{d}}$ near the water ice EF is more intense, meaning that more solids are being
delivered to the inner disk at this time compared to the MTBF model.
Although a higher St$_{\rm{M}}$ in the outermost part of the nebula implies faster radial drift rates, these are
hardly noticeable in the radial velocity plot (middle panel).
In general, the radial velocities are similar to the MTBF case. 
Overall, the 1 L$_\odot$ model makes little difference to disk temperature evolution at these early stages, though the 
outer disk is colder, and this may have led to the subtle changes that we see in this model compared with the variable
(higher) luminosity MTBF case. At later times, the higher (more realistic) luminosities may play a more important role 
in a relative sense.

\subsection{MTBF model with high $\alpha_{\rm{t}}$ (``HA-MTBF")}
\label{subsec:HA}

\begin{figure}
 \includegraphics[trim = 120 0 -120 0,clip,width=6.0in,height=6.0in,angle=0.0]{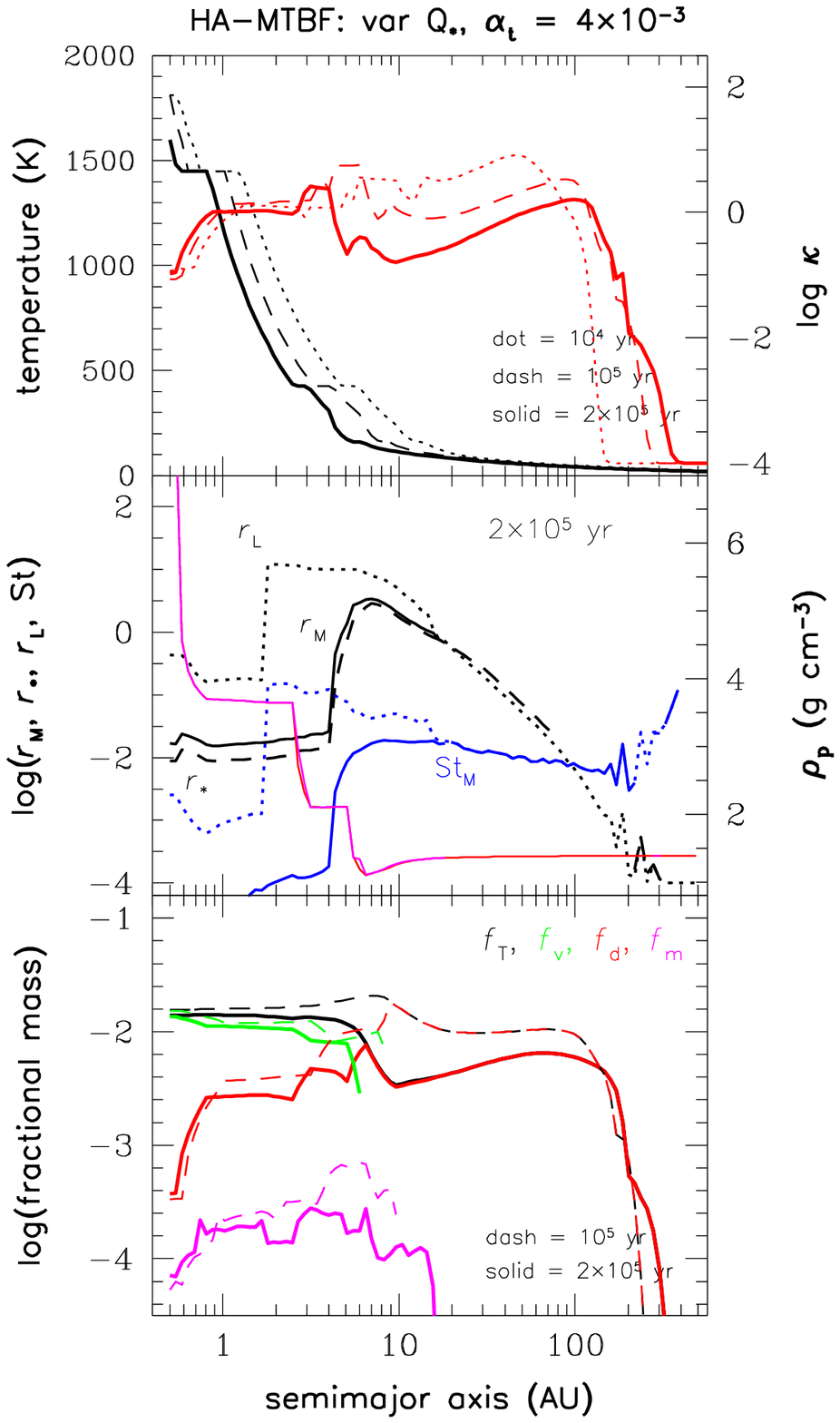}
\caption{MTBF model assuming $\alpha_{\rm{t}} = 4\times 10^{-3}$ (HA-MTBF). Panel
descriptions same as Fig. \ref{fig:3pangasF}.}
\label{fig:3panHA}
\end{figure}

\begin{figure}
 \includegraphics[trim = 120 0 -120 0,clip,width=6.0in,height=6.0in,angle=0.0]{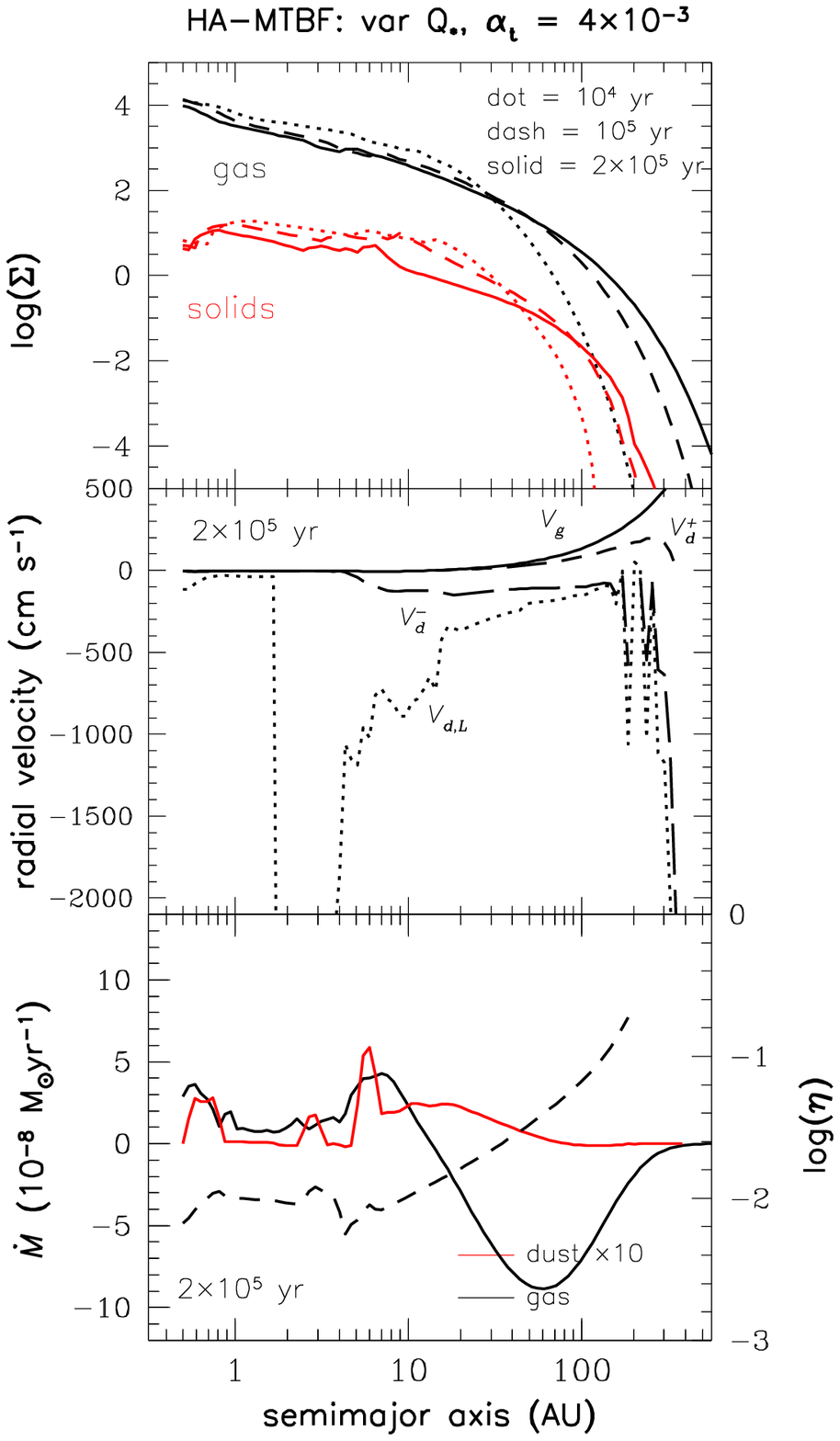}
\caption{MTBF model assuming $\alpha_{\rm{t}} = 4\times 10^{-3}$ (HA-MTBF). Panel
descriptions same as Fig. \ref{fig:3pangasF}.}
\label{fig:3pangasHA}
\end{figure}

As a final case, we compare a more viscous disk to our fiducial MTBF model; here the turbulence
parameter $\alpha_{\rm{t}} = 4\times 10^{-3}$ {\bf ({\bf Figures \ref{fig:3panHA} and \ref{fig:3pangasHA}})}. 
We expect that regions which are dominated by viscous heating should be hotter in this model, and indeed {\bf Fig. \ref{fig:3panHA}} (top panel) shows high temperatures extending to larger distances from the star at the nominal time of $2 \times 10^5$ years. It is interesting that the higher $\alpha_{\rm{t}}$ disk starts hotter than the nominal MTBF disk 
and cools to the current values, whereas in the fiducial MTBF model (and L1 as well) the temperature cools and then
actually increases to the value at $2 \times 10^5$ years.
The greatest differences we see between the two values of $\alpha_{\rm{t}}$, apart from the full silicates
EF being present, have to do with the radial extent of both the organics and water ice EFs. Both fronts are further out,
but are also smaller in radial extent than the MTBF model.
The systematically decreasing $T$ is naturally due to the decrease in Rosseland mean opacity as particles grow. Yet,
the MTBF case shows an increasing $T$ despite having larger particles because the enhancement leads to a higher
opacity. In the HA-MTBF model, there is very little enhancement as is seen in the fractional mass (lower panel).
Another clear effect of the more viscous disk is that the nebula gas evolves more strongly outwards, carrying a significant 
amount of solids with it {\bf (Fig. \ref{fig:3pangasHA}}, upper panel). 
The strong outflow of solid material is also evident in Fig. \ref{fig:3panHA} (lower panel) where after
$2\times 10^5$ years, the edge of the solids disk extends to about $\sim 300$ AU. 
Although the fraction of migrators is comparable between this and the MTBF case, the dust fraction in the innermost disk
is less by an order of magnitude. Most of the solids are in the vapor phase.

The particle sizes (middle panel) also differ in an understandable way from the fiducial MTBF model. The relative velocities
between particles are larger for the larger $\alpha_{\rm{t}}$, which means that the bouncing and fragmentation barriers are both reached for smaller particles. Comparing the middle panel of Fig. \ref{fig:3panHA} with that of Fig. \ref{fig:3panMTBF}, we see that the fragmentation size $r_*$ is now over an order of magnitude smaller. 
Interestingly, even while the {\it mass-containing} particles are smaller here, the {\it lucky} particles are not too 
much different in size from those in the fiducial MTBF case - so particles here can get even luckier relative to 
the nominal fragmentation barrier! This again may be due to the
radial drift barrier being the limiting factor in how large these particles can grow (see discussion on Stokes numbers, Sec. \ref{sec:discuss}). Increasing $\alpha_{\rm{t}}$ by an order of magnitude has led to a smaller ${\rm St_M}$ in the 
inner regions - also by an order of magnitude, consistent with Eq. (\ref{equ:stfrag}). However, ${\rm St_M}$ values
are hardly changed outside the water ice EF. This suggests that the inner nebula is obeying fragmentation-barrier limitations, but something else - probably radial drift - is playing a role in the outer nebula
(see Sec. \ref{subsec:psize}).

In Figure \ref{fig:3pangasHA}, top panel, we can see that the larger $\alpha_{\rm t}$ for this case 
has evolved the gas density much more, with $\Sigma$ already at values typical of or smaller than a minimum mass 
nebula \citep[e.g.,][]{ha81} outside $\sim 4-5$ AU, while inside the evolution has not proceeded as quickly and 
is similar to the MTBF value at $2\times 10^5$ years because of the hotter, steep temperature gradient disk.
In fact, the accretion rate of the gas (lower panel) shows that $\dot{M}$ lingers close to zero between $\sim 1-4$ AU
indicating very small gas advection velocities\footnote{This is mainly due to the steepness of the temperature gradient.  
From inspection of Eq. (\ref{equ:vgas}), we can see how $V_{\rm{g}}$ can become positive. For example, if the temperature 
$T \propto R^{-s}$, it is easy to show that a reversal in sign will occur if, ignoring a term in the surface density 
$\propto \partial{\,{\rm{ln}}\,\Sigma}/\partial{\,{\rm{ln}}\,R} < 0$ for simplicity, $2-s \lesssim 0$, where $s$ is the 
power law exponent.}. The mass flux reversal occurs much further out in the disk than the 
fiducial MTBF model, as one might expect with a much more viscous disk. The variations in $\dot{M}_{\rm{d}}$ are
due to the various EF transitions but are more muted than the fiducial MTBF case again because of a low level of mass
and opacity enhancement. The `bumps' in solids as seen in the fractional mass profile may still be
building up, however. Finally, the radial velocities (middle panel) demonstrate that outward transport is more
significant than the lower viscosity models. The gas radial velocity $V_{\rm{g}}$ increases more steeply in the outer 
nebula, and this affects the sizes (and amount) of particles that have positive (outward) velocities $V^+_{\rm{d}}$. The
inward drifting solids component $V^-_{\rm{d}}$ is stronger than in the fiducial MTBF case, suggesting that 
at later times, vast regions of the outer nebula (between the water ice EF and $\sim 100$ AU) the disk
may become depleted in solids. 

\begin{figure*}
 \resizebox{0.943\linewidth}{!}{%
 \includegraphics[angle=-90.0]{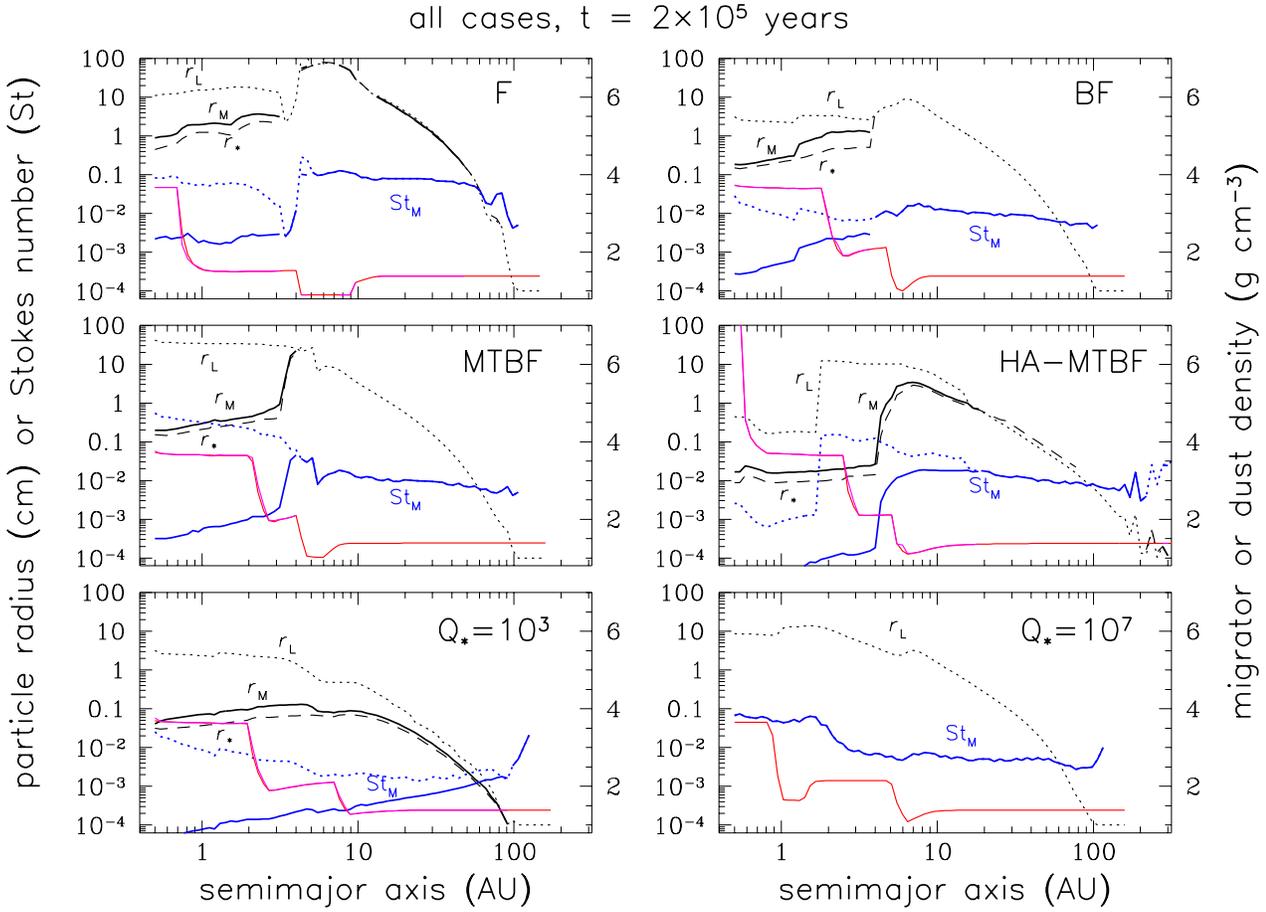}}
\caption{Summary of a the particle size distributions after $2\times 10^5$ years for a
suite of models we have presented in this paper.}
\label{fig:6panel}
\end{figure*}

\section{Discussion}
\label{sec:discuss}

In this section we 
review the effects of different sticking/bouncing/fragmentation schemes on particle size distributions, and note some implications regarding mm-cm wavelength observations of extrasolar disks (Sec. \ref{subsec:psize}). Also in 
Sec. \ref{subsec:psize}, we digress into how the tradeoff between growth and drift in {\it evolving} nebulae may explain a small but interesting difference between our results and previous work. Next, we look more closely at general implications of growth and drift for radial redistribution of mass (Sec. \ref{subsec:mass}), and note implications for meteorites, planetesimal formation, and perhaps even the radial distribution of planets.  In Sec. \ref{subsec:EFs} we look more closely at the effects of Evaporation Fronts, specifically on producing large objects enriched in, {\it e.g.}, water ice. We next discuss implications of our self-consistent opacities, in particular for mechanisms of turbulence generation and nebula gas evolution (Sec. \ref{subsec:opac}). We conclude with some caveats and a short discussion of future work.

\vspace{0.2in}
\subsection{Particle size distributions; the drift and bouncing barriers}
\label{subsec:psize}

In {\bf Figure \ref{fig:6panel}} we collect the particle size distribution results after $2\times 10^5$ years for the suite of models described in Sec. \ref{sec:results}. Though there is considerable variation in particle sizes across the models, there are
specific trends that are characteristic of all of these results. 

In all figures, particle radii are shown in black and the corresponding Stokes numbers in blue. Particle internal densities are shown by the red/magenta curves. Recall that $r_*$ (dashed lines) is the fragmentation radius, and 
$r_{\rm{M}}$ (solid lines) is the radius dominating the total particle mass. When $r_*$ and $r_{\rm{M}}$ are plotted, the dotted line ($r_{\rm{L}}$) refers to the largest ``lucky" particle in a radial bin. Since particles of radius $r_{\rm{L}}$ contain a negligible amount of mass (section \ref{subsec:sizedist}), we emphasize the mass-dominant particle $r_{\rm{M}}$. Recall that, if the heavy black lines for $r_*$ and $r_{\rm{M}}$ do not appear, it indicates that the largest particle $r_{\rm{L}}$ (which in this case {\it does} contain most of the mass) has not yet reached the fragmentation size $r_*$. 

The critical role of composition-dependent strength $Q_*$ for limiting particle size is clear. In all the nominal cases where we allow icy particles to be stronger than ice-free particles, a matching step-function is seen in $r_{\rm{M}}$ and St$_{\rm{M}}$. The constant $Q_*$ cases, naturally, do not show this step. This simple result is explained, to first order, by the significant change in the fragmentation barrier (equation \ref{equ:stfrag}) when $Q_*$ changes so significantly (see below for more discussion). In section \ref{subsec:mass} we discuss how this size difference  leads to strong mass enhancement/depletion differences between the inner and outer nebula. Another general result is that while ``lucky" particles $r_{\rm{L}} \gg r_{\rm{M}}$ can be found throughout the inner nebula (although carrying negligible mass), the outer nebula is devoid of such particles. That is, in the outer nebula the size distribution cuts off more sharply at its largest size. The same result is manifested in the Stokes number plots; this is discussed below. The most realistic cases (MTBF and HA-MTBF) show a peak in St$_{\rm M}$ and $r_{\rm M}$ just outside the ice EF, probably because of the locally enhanced solids mass. However, this does not hold true in the inner nebula where there is little if any correlation with the inner EFs and the distributions in $r_{\rm{M}}$ and St$_{\rm{M}}$, with
perhaps the exception of case F. The near-constant values of St$_{\rm M}$ in most of the plots in Fig. \ref{fig:6panel} suggest fragmentation-dominated regimes, as can be understood from equation \ref{equ:stfrag} in section \ref{subsec:F} (recalling that $T$ is only weakly varying in, at least, the outer nebula).

For our most realistic models (BF and MTBF), the inner nebula where the asteroids and terrestrial planets form (where the particle strength is lower because ice is absent) is capable of supporting growth only to $0.1-$several mm radius, as has been shown previously \citep{zso10}.  This size is typical of ``chondrules" which dominate unmelted meteorites. Our particles are not themselves ``chondrules", which {\it most} workers believe have been melted as freely-floating nebula objects, but represent aggregates of the right size to serve as their precursors. Just outside the ice EF, radii range from several dm to almost a meter. This might seem surprising in view of the iconic ``meter-size barrier", which is generally placed at St $\sim 1$ in a Minimum Mass Solar Nebula \citep{cw06}, but notice from the Stokes number curves that St$_{\rm{M}}$ remains in the range $0.01 - 0.1$ because at this early stage the nebula gas density remains fairly high. This means that radial drift velocities and collisional velocities are still {\it increasing} as these particles grow. The sizes of particles in the outermost nebula depend most strongly on the strength of turbulence and the assumptions of the sticking model. Around 30 AU for case F, particles can reach 1 cm radius, but for other, more plausible cases, sizes at 30 AU are closer to 1 mm because of the inhibiting role of bouncing. Thus, even at this early stage, ``bouncing" has played an important role in limiting the sizes of particles detectable by mm-submm observations (for more discussion see ``Digression" below). 

Our results thus generally agree with the results of previous workers \citep[e.g.,][]{bra08,bir09,bir10,zso10} that the fragmentation and bouncing barriers (along with radial drift; see below) limit particle growth. When only fragmentation is included (F), growth is quite rapid and stalls at particle sizes that are the largest amongst all of these cases. Yet, because particle growth reaches larger sizes, the disk mass drains most rapidly into the inner nebula due to radial drift (see Fig. \ref{fig:3panF}, lower panel), with important consequences, as discussed below.  The bouncing barrier (BF case), even if mass transfer is added (MTBF case), effectively reduces the rate of growth because a target particle is limited in the size of particles it can grow from. Thus in Fig. \ref{fig:6panel}, we see that the fragmentation barrier has not been reached in the outer nebula by $2 \times 10^5$ years for the BF and MTBF cases. Like previous workers \citep[e.g.,][]{win12a,win12b,gar13,dra13,dra14} we find that incorporating high-speed mass transfer (see Fig. \ref{fig:sizedist}) can allow the growth of a small number of ``lucky particles" to sizes $r_{\rm{L}} > r_*$; however, the mass contained in this population is extremely small at best, and  they are absent in regions dominated by drift (see below). 

\begin{figure}
 \includegraphics[trim = 120 0 -120 0,clip,width=6.0in,height=6.0in,angle=0.0]{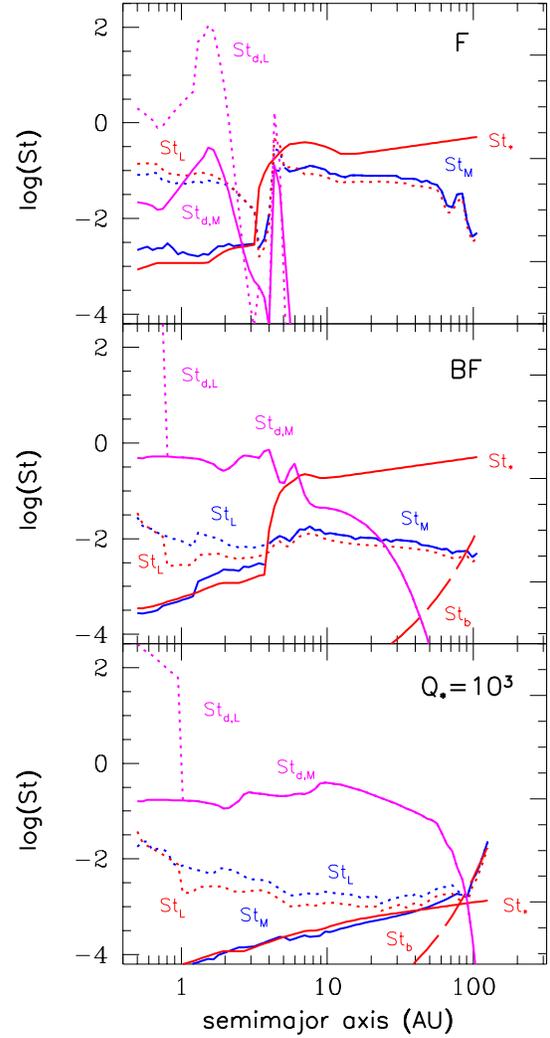}
\caption{Comparison of our estimates for the fragmentation, bouncing, and radial drift
Stokes numbers to the computed Stokes numbers for models F, BF and Q103 after $2\times 10^5$ years. Radial
drift dominates the outer disks in models F and BF. St$_* >$ St$_{\rm{M}}$ because systematic velocities
are comparable to the turbulent velocities for these sizes. Only in model Q103 is $t_{\rm{d}} > t_{\rm{gr}}$
in the outer disk. See text for details.}
\label{fig:stokes}
\end{figure}

{\bf Figure \ref{fig:stokes}} compares several important metrics for a subset of these models to the observed, mass-dominant value St$_{\rm M}$ (solid blue). As noted above, the near-constancy of the mass-dominant particle Stokes number St$_{\rm M}$ through regions where $Q_*$ is constant is largely explained by Eq. (\ref{equ:stfrag}, red solid curve for ${\rm St_*}$). However, while all the models follow the general trend of ${\rm St_*}$, and quantitative agreement is excellent for the moderately realistic F, BF, and MTBF cases in the inner nebula (and at all radii under conditions of weak particles for $Q_*=10^3$), quantitative agreement is poor in the outer nebula once bouncing and Mass Transfer are introduced (in the sense that St$_{\rm M} < {\rm St_*}$). This disagreement arises because Eq. (\ref{equ:stfrag}) {\it assumes} that the collision velocity is dominated by turbulent relative velocities, and it turns out that, at least for the models presented here, given the small values of St$_{\rm{M}}$, the large values of $\eta$ in the outer nebula drive systematic drift-and-headwind related velocities that exceed the turbulent velocities, so Eq. (\ref{equ:stfrag}) is an overestimate of the fragmentation barrier size. When bouncing is included, the discrepancy (vertical offset) increases since approach to the fragmentation barrier is now slower (particles larger than $r_{\rm{b}}$ can now only grow from the small fraction of particles $r < r_{\rm{b}}$), and St$_{\rm{M}}$ is ``quenched" at smaller, non-equilibrium values by the evolving radial drift barrier, as described below. In the low-strength $Q_*=10^3$ case, the fragmentation barrier is reached quickly everywhere with St$_{\rm{M}} \sim$ St$_*$. In the {\it inner} solar system, for cases F, BF, and MTBF, turbulent relative velocities dominate so the nominal St$_*$ expression appears valid in all three cases. However, looking deeper, the situation actually presents a subtle puzzle, and to understand this we now estimate ``limiting" Stokes numbers for bouncing and drift.

{\it Digression: role of the ``Drift Barrier" in setting $r_{\rm{M}}$ and $r_{\rm{L}}$:} We can derive a limiting condition for bouncing by setting $S_{\rm{b}}(m=m'=m_{\rm{b}})=0$ in equation (\ref{equ:bounce}), leading to $V_{\rm{b}}^2 = C_0/m_{\rm{b}}$. As in deriving the fragmentation limit, we set $V_{\rm{b}}$ equal to the typical collision velocity in turbulence for some radius $r_{\rm{b}}$: $\Delta V_{\rm{pp}} = c \sqrt{{\rm St_{\rm{b}}}\alpha_{\rm{t}}}$, where the particle size can be related, approximately, to the corresponding St$_{\rm{b}}$ \citep[using Eq. (\ref{equ:StE}); also see, e.g.,][]{bir11}. We thus get the bouncing barrier Stokes number St$_{\rm b}$:

\begin{equation}
\label{equ:Stb}
{\rm{St}}_{\rm{b}} \simeq \left(\frac{6 C_0 \rho_{\rm{p}}^2}{\pi \alpha_{\rm{t}} c^2 \Sigma^3}\right)^{1/4}, 
\end{equation}
\noindent
for which, in typical conditions, the strongest dependence is on the gas surface density. 

While the importance of a ``radial drift barrier" has been known since the early work of Weidenschilling (1988 {\it et seq.}), it has only recently been quantified and its influence is still being assessed. 
For instance, \citet{bra08} simply assumed that ${\rm{St}}=1$ represented the radial drift barrier; this was echoed by \citet{bir09, bir10}. \citet{bir12a} used a more realistic approach based on finding the particle size for which growth and drift timescales were equal. We follow the approach of \citet{bir12a} and generalize it in several ways. 

We derive our expression for the
radial drift limited Stokes number St$_{\rm{d}}$ by comparing the radial drift time $t_{\rm d } = R/U_r \sim R/ \eta V_{\rm K}{\rm St }$ to the particle growth time $t_{\rm{gr}} = m/\dot{m}$. As in Eq. (\ref{equ:dmdt}), 
$\dot{m} \sim \pi r^2 \rho_{\rm{d}} \Delta V_{\rm{pg}} S $, where $S < 1$ is some average sticking coefficient which we will take as $\leq 0.5$ for specifics. Approximating $\Delta V_{\rm{pg}} \sim c \sqrt{\alpha_{\rm{t}} {\rm St}}$ as the turbulent relative velocity, whether for same-size or different-size particles \citep{oc07}, and the near-midplane density to be $\rho_{\rm{d}} = \Sigma_{\rm{d}}/2h_{\rm{d}} = f\Sigma\sqrt{{\rm{St}}/\alpha_{\rm{t}}}/2H$ ($f=f_{\rm{d}}$ here), we get $t_{\rm{gr}} = 2/3f\Omega$, essentially the same as \citet[ their Eq. 13]{bir12a}. Then, we simply define the particle radius $r_{\rm d}$  as the ``drift barrier",
such that $t_{\rm{d}} \lesssim t_{\rm{gr}}$:
\begin{equation}
\label{equ:rdfrift}
r_{\rm{d}} \gtrsim \frac{3\sqrt{2}}{8}\frac{f S \Sigma}{\rho_{\rm{p}}\eta}.
\end{equation}
\noindent
Note that the ``drift barrier" is not simply determined by a drift velocity, but also depends on the {\it growth rate} through the amount of material available locally and the typical sticking coefficient. We relate $r_{\rm{d}}$ to a corresponding drift-limited Stokes number St$_{\rm d}$, in either the Epstein or Stokes regime, using Eqns. (\ref{equ:StE}) or (\ref{equ:StS}): 
\begin{equation}
\label{equ:Std}
{\rm{St}_d}\simeq 
\begin{cases}
\frac{3\sqrt{2}}{4}\frac{f S}{\eta} &\, r_{\rm{d}} < (9/4) \lambda_{\rm{mfp}};\\
\frac{1}{16}\frac{V_{\rm{K}}}{\rho_{\rm{p}}R\mu_{\rm{m}}}\left(\frac{f S \Sigma}{\eta}\right)^2 &\,
{\rm{Re}_p} < 1;\\
\frac{1}{10}\left(\frac{V_{\rm{K}}}{\rho_{\rm{p}}R\mu_{\rm{m}}}\right)^{1/2}\left(\frac{f S}{\eta}\right)^{4/3}
\frac{\Sigma}{\alpha_{\rm{t}}^{1/6}} &\, {\rm{Re}_p} < 800,\\
\end{cases}
\end{equation}
 
\noindent
If particles have St $>$ St$_{\rm{d}}$, they are drifting faster than they can grow. The first expression, valid for the Epstein regime, corresponds to equation (17) of \citet{bir12a}; however, our expression includes a factor of $S<1$ and  retains the full value of $\eta$, which is not always well represented by $c^2/V_{\rm{K}}^2$ as in equation (17) of \citet{bir12a}. In both cases, a higher local mass density makes it easier for growth to dominate drift (increases St$_{\rm{d}}$), and a stronger radial pressure gradient makes drift more dominant (decreases St$_{\rm{d}}$).

We can now compare the calculated Stokes numbers (in blue), for the mass-dominant (St$_{\rm M}$; solid) and largest (St$_{\rm L}$; dotted) particles with our estimated limiting ``barrier" values for fragmentation (St$_*$; red solid), bouncing (St$_{\rm{b}}$, dashed red) and 
radial drift (St$_{\rm{d}}$, solid and dotted magenta as related to $r_{\rm{M}}$ and $r_{\rm{L}}$ respectively). As an indication of where the Stokes-Epstein regime transition lies, we also plot 
our approximation Eq. (\ref{equ:StS}) for the Stokes regime
\citep[dotted red, see][and Sec. \ref{subsec:F}]{cw06}.
Our actual models (blue curves) use a smoother bridging expression, averaging over all particle sizes; the discrepancy between the red and blue dotted curves indicates that the effects of Stokes drag actually extend over a broad range of nebula radii. The bouncing barrier is active almost everywhere (St$_{\rm{b}} <$ St$_{\rm{M}}$); as mentioned earlier, St$_{\rm{b}}$ does not provide an impermeable barrier, but merely slows growth for larger particles, which in principle could still reach the fragmentation barrier $r_*$ given a long enough time (in the absence of drift). 


The subtle puzzle in Fig. \ref{fig:stokes} is that while \citet{bir12a,bir12b} found, in general, that the radial distribution of the {\it size} of the mass-dominant particles in the outer nebula corresponded very well to the radial variation of St$_{\rm{d}}$ when drift dominated growth (actually they present the corresponding $r_{\rm{d}}$ instead of St$_{\rm{d}}$), our results show St$_{\rm{M}}$ to have the shape of St$_*$ rather than that of St$_{\rm{d}}$, even when drift dominates (${\rm{St}} > {\rm{St_d}}$).
 

\begin{figure}
 \includegraphics[trim = 120 0 -120 0,clip,width=6.0in,height=6.0in,angle=0.0]{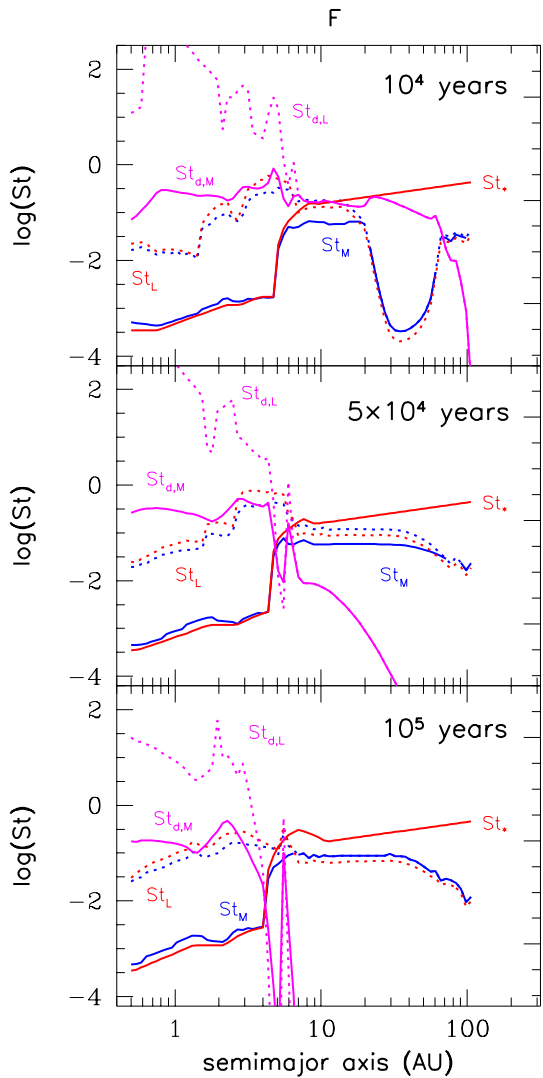}
\caption{Comparison of the fragmentation, bouncing, and radial drift
Stokes numbers to the computed Stokes numbers as a function of time, in case F. Here the transition from
the growth dominated to the radial drift dominated regime is clearly shown. In a relative short time, the
outer nebula solids are effectively transported to the inner nebula. See text for details.}
\label{fig:std_var}
\end{figure}

We believe the explanation lies in a non-equilibrium behavior to which the nebula gas evolutionary models presented here lend themselves, but perhaps those adopted by \citet{bir12a,bir12b} do not. {\bf Figure \ref{fig:std_var}} shows the contents of Fig. \ref{fig:stokes} for case F, at three different times. The ``drift barrier" St$_{\rm{d}}$ is larger than St$_{\rm{M}}$ throughout the nebula at early times, so growth dominates everywhere. The fragmentation barrier is reached at larger distances as time goes on because at early times $f$ is nearly constant, and growth is more rapid at smaller distances where $\Omega$ is smaller (see $t_{\rm{gr}}$ above Eq. \ref{equ:rdfrift}). By $2-3 \times 10^4$ yrs, St$_*$ is reached nearly everywhere, while  St$_{\rm{d}} >$ St$_{\rm{M}}$. Meanwhile, growth to large values of St$_{\rm{M}}$ leads to rapid inward mass loss by radial drift, which decreases $f$, which in turn decreases St$_{\rm{d}}$, so by $5 \times 10^4$ yrs, St$_{\rm{d}}$ has fallen well below St$_{\rm{M}}$ in the outer disk and by $10^5$ years even for $R>5-7$ AU, and growth has ceased. There is, however, nothing to {\it decrease} particle sizes because St$_{\rm{M}} <$ St$_*$ still. Essentially the fragmentation-limited {\it size} of earlier times becomes frozen in, a fossil of an earlier era, while mass outside the ice EF continues to be lost to the inner nebula. A similar crossover behavior is seen for St$_{\rm{d}}$ and St$_{\rm{M}}$ in the BF and MTBF cases, at different times. 

This behavior has interesting implications for mm-submm observations of disks. If it is true that the particle size distribution faithfully mirrors the instantaneous, underlying surface mass density, as in the models of \citet{bir12a,bir12b}, then measurements of size through spectral index determinations should be consistent with inferences about total mass. Yet, \citet{bir10b} actually come up with inconsistencies when comparing models with observations; they found that ad hoc reduction in the surface mass density is required relative to the model. While  case F (Fig. \ref{fig:std_var}) is an extreme and indeed implausible case (the dominance of drift means that almost no mass is left over to form ice giants), the general decoupling of particle size distribution from simultaneous surface mass density provides a cautionary note.  Other models (such as the MTBF case) also show an apparently fragmentation-limited size distribution represented by St$_*$ nearly constant, but also show significant mass depletion relative to cosmic abundance. We believe that \citet{bir12a,bir12b} found a different behavior than do we because their underlying gas nebula, while formally evolving, is extremely large, so evolves in a much more leisurely way than does ours. It suggests that choice of initial disk size is something worthy of careful thought. As noted in section \ref{subsec:gasevol}, different studies have adopted very different characteristic initial nebula lengthscales, and these differences may have unforeseen but interesting implications. For instance, it might be that if one {\it does} measure the solids surface mass density and particle size distribution of a disk simultaneously, one may be able to infer something about the evolutionary history of the region, and this is an exciting possibility. 
 
Finally, the changing role of St$_{\rm{d}}$ {\it vis-a-vis} St$_{\rm{M}}$ helps us understand the variable abundance of ``lucky" particles (the variable behavior of St$_{\rm{L}}$ {\it vis-a-vis} St$_{\rm{M}}$) as seen across all the models of Figs. \ref{fig:stokes} and \ref{fig:std_var}, in a simple way. In all cases, lucky particles vanish when St$_{\rm{d}} <$ St$_{\rm{M}}$. This is because growing ``lucky" particles relies on long periods of inefficiently sweeping up much smaller particles at a slow rate. Particles simply do not reside long enough in a region where drift dominates to exceed St$_{\rm{M}}$, so their growth must have occurred entirely in environments encountered earlier, which are no longer present. As they drift inwards, they do grow, although too slowly to reach the fragmentation barrier. {\it Only} in a growth-dominated regime where drift is not a factor, can ``lucky" particles St$_{\rm{L}} >$ St$_{\rm{M}}$ appear. Inspection of Fig. \ref{fig:std_var} indicates that this rule holds across all three cases (as well as others not shown here). Lucky particles are common (although a negligible fraction of the total mass) in the inner nebula in all cases, because St$_{\rm{d}} >$ St$_{\rm{M}}$ (Fig. \ref{fig:std_var}). We note, however, that the frequency and abundance of 
lucky particles may be significantly increased if we implement porosity (Sec. \ref{subsubsec:coag}). 
\citet{oku12} find that if the particle porosity can get extrememly high (volume density of $\sim 10^{-5}$)
then drift behavior can change qualitatively. We explore this in future work.
\subsection{Radial redistribution of mass, and some implications}
\label{subsec:mass}
We find that enrichment of the inner nebula in solids and vapor is a general, but not universal, result which seems dependent on the different $Q_*$ associated with the inner outer nebula. As we have seen in our
simulations, the higher fragmentation energy of icy particles allows them to grow large enough to drift radially in significant
abundance over relatively short time scales, even with the bouncing barrier imposed. The lower fragmentation 
energy of silicate particles in the inner disk means that particles originating in, or drifting into, that region, after losing their ice, can only reach or maintain smaller sizes which do  not drift quickly. Thus the high flux of material migrating into the inner regions from outside exceeds the loss flux towards the sun, leading to significant enhancements in residual solids. \citet{cz04} identified the same effect with respect to vapor, which can become enhanced inside EFs because its inward removal rate by gas advection is smaller than its supply rate, in the form of solids drifting from outer regions.  

It is illuminating to compare the mass enrichment/depletion results of the F, MTBF, and HA-MTBF cases (focusing on the red solid and dashed curves in the bottom panels of Figs. \ref{fig:3panF}, \ref{fig:3panMTBF}, and \ref{fig:3panHA}; the MTBF and BF cases are fairly similar). The most obvious difference is in the outer nebula, where adding the bouncing barrier keeps particles, and associated radial drift loss, smaller. In case F, solids drain almost entirely from the outer nebula in $1-2\times 10^5$ years, clearly problematic for outer planet formation. The HA-MTBF case represents another end of the spectrum, where the higher $\alpha_{\rm t}$ keeps particles and their drift-loss smaller than the baseline F or MTBF cases, retaining outer nebula material longer and in higher abundance. 

In the innermost regions, both the F and HA-MTBF cases show the least buildup of material but for different reasons. In the
F case, particles become sufficiently large that they can be removed by drift on short timescales, even while growing. This effect becomes very strong between $10^5$ and $2\times 10^5$ years, after the outer nebula supply has ended, and even the inner nebula starts to drain away. Contrast this with the HA-MTBF case in which particle growth to larger sizes is limited by
strong turbulence, Stokes numbers are low, and thus so is particle drift. As we pointed out before, the temperature
gradient is also very steep so that the gas advection velocity is also small. Thus the solids distribution changes much
more slowly. In the outer regions, strong outflow transports mass outwards countering the inflow from drift
of larger particles. These effects conspire to keep enhancement low, though a peak is beginning to grow at the water ice
EF suggesting that at later times enhancements may increase. 
For both these cases, the solids enhancements never gets large enough to lead to breakthrough growth to larger sizes 
(as in \citet{bra08}). On the other hand, the slower particle growth rate of the MTBF case relative to the F case leads to smaller
particles (but larger than the HA-MTBF case) which have slower radial drift rates, but relatively large enhancements of material in the 
inner nebula, at least at this stage in the evolution because of the strong mass inflow of material from the outer disk.


Previous workers \citep[e.g.,][]{sv96,sv97,cc06,gar07,bra08,bir10} observed the transport of material from the outer disk to the inner regions, but generally did not see the same total enhancements as we do here. In all these studies one or more simplifications were made, but it is probably still true that the differential strength $Q_*$ is the most important single factor. In the case of \citet{sv96, sv97}, only icy material was treated, all solids vanished at the ice EF, no vapor tracking was included and a simplified particle growth algorithm was used, but some enhancement of solids in the ice-giant region was still seen due to inward drift from further out. While \citet{bir10} did study differential strengths, their models merely changed the particle strength at the iceline without removing the associated icy material, so it is unclear just how large an effect they would have predicted under more realistic assumptions. 

At the next level of detail, one sees discrete radial peaks and valleys in the abundance of solid material, some quite narrow and pronounced. These peaks generally coincide with locations just exterior to various evaporation fronts (EFs) due to water ice,  and ``organics", which dominate the condensible material in our model. We suspect that for our fiducial case, if the silicate EF were completely resolved within our radial grid, that a peak may appear there as well. Similar multiple features were pointed out by \citet{gar07}. The radial locations of these peaks vary with the radial location of the respective EFs, which are in turn dependent on the viscosity, temperature, and ultimately opacity of the disk. Thus, the system is coupled thermally and dynamically to the particle growth process. One can even discern fluctuations in the {\it gas surface density} associated with these structures (see figures \ref{fig:3pangasF}, \ref{fig:3pangasMTBF}, and \ref{fig:3pangasHA}, and section \ref{subsec:opac}). Enhanced local mass density and/or smaller particles lead to higher opacity (as long as their size exceeds the peak thermal wavelengths), which affects midplane temperature and the EF. In case F, loss of solids leads to strongly falling opacity everywhere, and a cooling nebula, with more rapidly evolving EF locations. In cases MTBF and HA-MTBF, EFs actually evolve very little over the period of these simulations, allowing greater solids buildup behind them. The radial structure of the post-EF solids enhancements varies considerably, from an extremely narrow spike outside the water ice EF in case F (and MTBF), to somewhat broader ``organic" buildups in cases MTBF and HA-MTBF (here a weak silicate buildup is
present), each with associated opacity peaks. 
In fact, a glance at the red curves for $\dot{M}_{\rm{d}}$ in the lower panels of Figs. \ref{fig:3pangasF}, \ref{fig:3pangasMTBF}, and \ref{fig:3pangasHA} shows that even at the last time depicted, convergence of solids mass into narrow zones 
is still ongoing. While \citet{gar07} suggested that the amplitude of these peaks reaches a steady state in the long run, longer runs will be very instructive. 

\subsubsection{Implications of mass redistribution and particle size for planetesimal formation:}
\label{subsubsec:planetes_imp} 
A subject of long-running interest is the still poorly-understood ``primary accretion" of asteroids (in the inner nebula) and TNOs/KBOs/comet nuclei (in the outer nebula). Recently, attention has focused primary accretion scenarios in which such objects ``form big", leapfrogging the various barriers to accretion of which only the first few are discussed here \citep[for the others, see][]{ida08,ng10,gre12,oo13}. Two somewhat disparate avenues to primary accretion may be distinguished: ``streaming instabilities" (SI) which operate nearly axisymmetrically in dense midplane layers \citep{gp00,yg05}, and ``turbulent concentration" (TC) which is a clumpier process, acting throughout the nebula but also leading most effectively to planetesimals near the midplane \citep{cuz01,cuz08,cuz10}. A full discussion of this topic is beyond the scope of this paper (see \citet{joh15} for a recent review).  The conditions needed for SI are easier to assess, and we defer a discussion of the implications for TC to a future paper.

For SI to operate requires a midplane particle-to-gas ratio $\rho_{\rm{d}}/\rho \gtrsim 1$ so the solids can drive the gas motions. This has traditionally led SI advocates to focus on some combination of large particles, low turbulent intensity, and global enhancement of solids, usually due to late-stage removal of the gas. Related suggestions involve a reliance on ``lucky particles" growing large enough to settle and lead to SI \citep[e.g.,][]{joh14}, maybe if trapped at ``pressure bumps" \citep{dra13}, but it seems to us that under nominal conditions, ``lucky" particles (those of radius $r_{\rm{L}}$ in our figures) are too rare and contain insufficient mass to drive collective effects such as SI (they might have other interesting implications as migrators, however). We believe it is more fruitful to explore the implications of the {\it mass-dominant} particles, represented by $r_{\rm M}$ and St$_{\rm M}$. It is possible to express the local midplane value of $\rho_{\rm{d}}/\rho$, simplifying Eq. (\ref{equ:subdisk}, also see Appendix B) as in \citet{cw06} to 
$h_{\rm{d}}/H = \sqrt{\alpha_{\rm t}/{\rm{St}}_{\rm{M}}}$:

\begin{equation}
\label{equ:SI}
\frac{\rho_{\rm{d}}}{\rho} = \frac{\Sigma_{\rm{d}}}{2h_{\rm{d}}}\frac{2H}{\Sigma}
                      = f\frac{H}{h_{\rm{d}}} = f \sqrt{{\rm St_{\rm{M}}}/\alpha_{\rm t}}.
\end{equation}

\begin{figure}
 \includegraphics[trim = 120 0 -120 0,clip,width=6.0in,height=6.0in,angle=0.0]{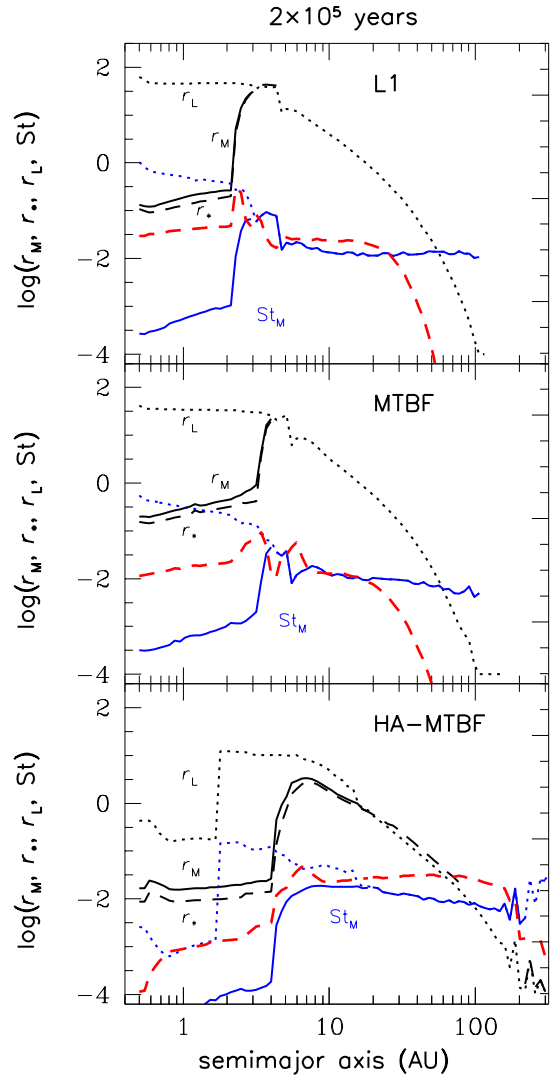}
\caption{Radial profile of $\rho_{\rm{d}}/\rho$ near the midplane is shown in the red dashed line. Streaming instabilities require $\rho_{\rm{d}}/\rho \gtrsim 1$. It appears that, for some cases (specifically the upper panel) SI may at some point be achievable over a narrow range of radii, right outside the ice EF - a location where the first giant planet has long been thought to arise. Of course, the absolute {\it location} where this circumstance occurs will depend on model parameters.}
\label{fig:SI3panel}
\end{figure}

\noindent
{\bf Figure \ref{fig:SI3panel}} shows $\rho_{\rm{d}}/\rho$ with other quantities as a function of $R$ (red dashed curve) for three representative models. In general, at this point in nebula evolution, 
$\rho_{\rm{d}}/\rho < (\ll)\, 0.1$, but there is one intriguing situation in which $\rho_{\rm{d}}/\rho$ may eventually approach a value where SI might be possible, and it is right where longstanding expectations are that the first giant planet would form - just outside the iceline (most prominently, upper panel for L1). Moreover, this situation occurs quite early 
in nebula evolution and perhaps a longer simulation may eventually achieve this condition without the need to await photoevaporation or some other process to deplete the nebula gas. At this time and place, 
our results are showing particle radii of just over 1 cm, but recall that icy aggregates may not be able to compact fully and the identical St$_{\rm M}$ could arise for porous particles perhaps $10$ or more times larger. Soon we will run our code for porous particles, and for longer times, to explore the possibilities for early planetesimal formation. While SI appears inoperative in the inner nebula at these timescales for $\alpha_{\rm t}$ values studied herein or larger, because the weak silicate aggregates do not grow large enough to settle into the required dense layers, the particle sizes and density enhancements at these times are in the range of interest for TC, and also agree with what is seen in meteorites. It is increasingly suspected that {\it some} planetesimals {\it did} form in the inner nebula even at these early times, and it will naturally be of great interest to continue these simulations to the Myr timescales on which most observed chondrites form.

\vspace{0.2in}
\subsubsection{Other implications of mass redistribution for observations of different kinds:}
\label{subsubsec:other imp}

The mass redistributions we see, fueled by the transport of material over large radial distances and the effects of Evaporation Fronts, have implications for a number of other currently puzzling observations. From the standpoint of mm-submm observations by ALMA and other radio observatories, the peaks and valleys in mass and opacity might provide a protoplanetary disk with a ringed appearance such as seen recently in HL Tau \citep{par15} and MWC 480 \citep{obe15}. Our simulations to date have not included EFs for the ``supervolatiles" thought responsible for the structure in MWC 480, but it would not be surprising to see a zoned or banded structure similar to that created by the ices, organics, and silicates we are currently modeling. 

From the standpoint of meteoritics, a number of implications are evident. The enhancement of vapor in the inner nebula, a combination of evaporated water ice and ``organics", is especially large in the L1 case (factor of $\sim 8$). Depending 
on the state of the evaporated ``organics" (see Sec. \ref{subsec:future}), whether CO, CH compounds, or higher hydrocarbons, the redox state of the inner nebula cannot help being influenced by this. 

Another current meteoritics puzzle involves the anomalously high $^{17,18}{\rm O}/^{16}{\rm O}$ ratios seen in most chondritic materials relative to the sun \citep{mck08}, an enrichment that seems to appear some $1-2\times 10^5$ years after the earliest ``CAI" minerals condense, and to grow with time \citep{cla03,sim12,ku12}. Current thinking is that the heavy isotopes are liberated by a UV self-shielding process in the upper layers of the outer nebula \citep{ly05}, become bound in icy particles, and drift inwards to evaporate in the inner nebula  \citep{yur05}. Photochemical models seem to be able to {\it produce} the needed material in a timely way \citep{lyo09} but there has never been a demonstration that it could be transported to the inner nebula on such short timescales. Now, it seems at least several plausible models indicate that substantial outer solar system material can indeed reach the inner solar system within $1-2\times 10^5$ years. More detailed models, coupling these transport models with $^{17,18}{\rm O}$ formation rates in the outer solar system, are in order. 

The peaks and valleys we see in the radial abundance of solids, even quite early on, may provide insight into the longstanding problem of the ``small Mars" relative to Earth and Venus \citep{cha01}; planetesimals formed from such material through the entire terrestrial planet region might have been quite nonuniform in initial surface mass density. Naturally, the absolute locations of peaks and valleys depend on the actual radial locations of the various EFs, as they evolve with time. It will be of interest to see what longer runs reveal, but the mere occurrence of mass peaks and valleys with up to order-of-magnitude density contrast in the terrestrial planet region on radial scales of an AU or so are surely of interest. Moreover, all of our models show large-scale inward drift of outer nebula solids, to varying degrees. This effect alone might explain the compact surface {\it solids} mass densities inferred by \citet{des07} from early, compact giant planet configurations, even while the enveloping gas disk, with a certain amount of dust, might extend to much larger sizes. It is also well-known that ``normal" ({\it i.e.} Kepler) planetary systems are much more packed with inner planets than is our solar system. The possibility of finding systematic causal factors for these important differences is intriguing.

\subsection{Evaporation fronts; local refineries}
\label{subsec:EFs}

Just outside EFs, we find significant peaks in the solids mass, created when drifting, vaporized material diffuses back outside the EFs, recondenses on small grains, and advects or diffuses some distance before becoming accumulated into a larger particle. These peaks and valleys in the fractional mass of solids, which move as the EFs evolve, may be observable in protoplanetary disks, as mentioned in Sec. \ref{subsec:mass}. Additionally, we see that EFs lead to a distillation effect whereby the {\it composition} of material residing some distance outside  each EF is enriched in the volatile that has just passed through that EF, and returned to recondense, leaving behind the more refractory material it had evaporated from, which continues to grow and drift radially inwards \citep[][previously referred to EFs as barriers to inward drift for the associated species]{gar07}. The red and magenta curves in the middle panels of Figs. \ref{fig:3panF}, \ref{fig:3panMTBF}, and \ref{fig:3panHA} track this compositional distillation through the internal {\it density} of local solids. For example, just outside the water ice EF the particle density drops from its ``cosmic abundance" value of roughly 1.5 to a nearly pure water ice value of 1, indicating that the solids there have become highly enriched in water ice relative to cosmic abundance.  Of course, the same effect would occur outside each EF and would occur for the various EFs we are not tracking here as well. It is a way of creating radial belts of solids with very different composition than material on either side. The radial width and shape of these refinery or distillation zones depends on $\alpha_{\rm t}$ primarily, but to some degree on the nature of the growth-by-sticking assumptions. Moreover, the radial zones evolve with time as the EFs and the mass flux through them vary, so longer simulations will be needed to what happens at later times. One can imagine interesting meteoritic and astrophysical chemistry associated with the EFs for silicates, troilite, and organics as well. This is a potentially rich area which also has implications for circumplanetary disks, and will be developed in future studies. 

One prediction can be made in this regard, which we should mention here. The ice-enriched nature of solids immediately outside the ice EF, combined with our preceding results mentioned above, that the most likely place for early planetesimal formation to occur by SI 
is just outside the water EF, leads us to predict that Jupiter, say, if that were the planet first formed just outside the snow line, wherever that arose due to energetics details, would have a {\it water-enhanced core} relative to cosmic abundance. Perhaps Juno can test this prediction. 

\subsection{Nebula opacity}
\label{subsec:opac}

We have followed the evolution of EFs using a self-consistent calculation for the photospheric and midplane temperatures, that employs a new model to compute the Rosseland mean opacity of aggregate particles \citep{cuz14a}. The opacity depends on particle size, number density and composition, so it depends on the evolving particle size 
distribution and abundance. We have found that the opacity in the outer disk becomes quite small if growth proceeds quickly (particles growing larger than a thermal wavelength have smaller cross-section per unit mass), and/or if the loss of material to inward drift is high (less material). This tendency is weaker for models with higher $\alpha_{\rm{t}}$ that keep particles smaller, allowing them
to be retained longer and be more easily carried outwards as the disk spreads. In the inner region, the opacity can become
larger with time, even though particles are growing to larger sizes (although slowly) if supply of material
from the outer regions can keep the mass densities high. A sustained high opacity can 
maintain higher temperatures in the inner regions for longer periods. We also note that the radially variable opacity can even, apparently, lead to ``lumps'' in the {\it gas surface density}, because the opacity
affects the local temperature and thus the viscosity in our $\alpha$-model. 

The evolution of opacity is important for two new hydrodynamic instabilities which apparently lead to sustained turbulence, without any need for magnetic fields or ionization, and are thus extremely important to the inner solar system which has been regarded as ``dead" or at best ``dull" from the standpoint of turbulence \citep{nel13,tur14,sk14,mar13,mar14}. The new instabilities (as currently understood) require a suitably rapid thermal equilibration, and thus prefer lower Rosseland mean opacities. The limiting opacities have been estimated by \citet{nel13} to be on the order of 1 cm$^2$ g$^{-1}$ (perhaps up to 5 cm$^2$ g$^{-1}$; Umurhan, personal communication 2015). In the outer nebula, our opacities remain at or below these nominal thresholds, but in the mass-enhanced inner nebula regions, opacities reach larger values. Whether the corresponding instabilities are damped or not under these conditions is thus a pressing topic for future study, perhaps even in a coupled fashion with particle growth. 

\subsection{Caveats and Future Work}
\label{subsec:future}

There are several areas in which the code can be improved, which will be introduced in later
papers. The initial conditions for the models in this paper are consistent with the earliest times in the nebula,
but we have not added ongoing infall from the parent cloud. Such effects can be included through relatively simplified approaches \citep[e.g.,][]{bir10,yc12}. 

None of the models presented here have reached a situation
in which the mass fraction of vapor in the enhanced inner nebula exceeds the local H$_2$-He gas density, 
but at least one has  come within a factor of $\sim 10$ (the L1 model, Sec. \ref{subsec:L1}) prompting us to consider
that this situation may arise for a given model or at longer simulation times. 
Since this added vapor affects the disk viscosity, it is important to include its
effects when this occurs. Furthermore, the local gas density influences the process of grain growth through the
relative and radial velocities, thus a more self-consistent treatment requires that the volatilized-condensible vapor component is included in any calculations that involve $\Sigma$.

There are several areas in which we can improve the treatment of particle growth and composition. Most pressing is to explore the role of porosity. The moments method we utilize for coagulation is ideally
suited for this \citep{ec08}. The fractal growth of particles by low velocity sticking
of small monomers will cause the particles to have a much lower density than the solid density of the monomers
themselves \citep[e.g.,][]{dom07,orm08,zso10}. Coupled with the stickiness and strength of icy particles, it has been argued
by some workers that these fluffy aggregates can grow large, but maintain low relative
velocities, allowing for continued growth without compaction \citep{oku12,oo13}. In the extreme case that a particle grows with a fractal dimension of exactly two, the stopping times remain the
same for a single monomer. Thus, a model like our Q107
case with higher porosity might allow particles to grow faster and overcome the radial drift barrier. Furthermore, as
we have noted in Sec. \ref{subsubsec:mig}, our assumption of a gaussian PDF for the relative velocities in our calculation
of the migrator destruction probabilities has been called into question from more recent work 
\citep[e.g., ][]{gal11,win12b}, thus we intend to explore the more rigorous model of \citet{gar13} in future modeling
efforts. The choice of PDF may very well affect the frequency of lucky particles which may modify some of our 
conclusions here.

Other improvements concern our treatment of EFs. Currently we make the simple assumption that inwardly migrating
particles lose all their volatiles at the corresponding EFs, while modeling the EF as a  radial range of nearly constant temperature, centered on the species evaporation temperature and buffered by evaporation of the associated source of opacity. Though this works well for particles of the size we are so far growing, 
it may fail to capture the case of a much larger particle where a particle's surface layers can insulate its material at depth while it drifts a considerable distance. Further sweepup of local refractories may effectively trap any volatiles
remaining deep within the particle and offer a way to deliver volatiles deeper into the inner regions of the
nebula \citep[e.g.,][]{sl00}. 

We have also assumed that evaporation and condensation are reversible
processes for simplicity, but this is certainly not always the case. In particular, the ``evaporation" of ``organics'' may instead be a decomposition of the original complex molecules into nearly incondensible species like CO or CH$_4$, which participate in gas-phase chemistry rather than recondensing outside of the ``organics" EF. This loss of refractory, carbon-rich organics must, at some level, be a one-way street, because none of the most primitive carbonaceous chondrites contains more than about one-tenth of the carbon found in comets (see Cuzzi et al 2003 for a discussion and references). 
This loss of refractory organics may effect particle strengths as well, because organics are often found to be ``sticky", perhaps extending the range of ``strong'' 
particles to locations somewhat further inwards than the water ice EF.

Our radiative transfer treatment (section \ref{subsec:therm}) assumes implicitly that the bulk of the  energy released by viscous angular momentum transport is dissipated near the midplane, or at least in a fashion proportional to gas density, and is transferred vertically upward and outward to space, making the midplane the highest temperature location. However, in cases where viscosity is MRI-related \citep{tur14}, or VSI-related \citep{nel13,sk14}, it may be that the dissipation maximizes at some altitude near one scale height, or even closer to the disk photosphere. Such a case can surely be modeled and probably parametrized for our evolutions. The magnitude of the effect can be estimated by considering that, in the worst case where the energy is dissipated right at the photosphere, 
half of it is lost to space immediately and half radiates downwards. So, the vertical temperature distribution (that we do not model) will become more nearly isothermal, but the midplane temperature at a given location may not change by more than a factor of $2^{1/4}$.  

We expect there will be other innovations as our understanding of the problem improves, but our code is at least well
suited and robust enough to readily accept all the new physics mentioned above. An additional layer of parallelization
in mass bins will greatly speed up our code which will allow for more complexity, and will be implemented
in the future. Finally, we note that this model can readily be applied to the circumplanetary disk environment.

\vspace{0.2in}
\noindent 
{\it Acknowledgements.} We wish to thank Fred Ciesla, Sandy Davis, Steve Desch, Pascale Garaud, Uma Gorti, Phil Marcus, and Orkan Umurhan for helpful conversations. We especially thank Chris Ormel for bringing our attention to a flaw in our diffusion model, in an initial version of the paper. 
We also wish to thank an anonymous reviewer for pointing out several aspects that will lead to the improvement of our models. We thank Cameron Wehrfritz for his help in the preparation of this manuscript. This work was supported by a grant from NASA's Origins of Solar Systems program and a large amount of cpu time awarded through NASA's HEC program, whose consultants also helped with parallelizing the code.

\renewcommand{\theequation}{A-\arabic{equation}}
\setcounter{equation}{0}  
\section*{Appendix A: Code Algorithms}

In Sections A.1-A.2, we describe how we separately solve the equations for the radial evolution
of the gas (Eq. A-[\ref{equ:sgas}]), and that of the vapor and dust components (Eq. A-[\ref{equ:diffadv}])
in the first two sections. In Sec. A.3., we then describe in more detail how the temperature and opacity 
are calculated. The calculations in Sections A.1-A.3 are all carried out at synchronization steps and thus 
are only done at regular intervals. In Sec. A.4. we further describe the details of how we solve for the relative velocities
which are required for both coagulative and incremental growth. Sections A.5 and A.6 describe the algorithm for growth beyond the fragmentation barrier, and radial drift of
migrators.
\vspace{0.2in}

\noindent
{\it A.1. Gas Evolution Equation}
\vspace{0.1in}

We use an implicit finite differencing scheme to solve Eq. (\ref{equ:sgas}). Radial bins $j$ are spaced
logarithmically over the radial range $R_{\rm{in}} \leq R_j \leq R_{\rm{out}}$ such that

\begin{equation}
\label{equ:radbin}
R_j = R_{\rm{in}}(R_{\rm{out}}/R_{\rm{in}})^{\frac{j}{n_{\rm{R}}-1}},
\end{equation}

\noindent
with $R_1=R_{\rm{out}}$ and $R_{n_{\rm{R}}}=R_{\rm{in}}$, and $n_{\rm{R}}$ is the total number of 
radial bins. We thus first rewrite Eq. (\ref{equ:sgas}) in logarithmic form, using equation [\ref{equ:sound}] and the relation $\nu = \alpha_{\rm{t}}cH = \alpha_{\rm{t}} c^2 /\Omega$: 

\begin{equation}
\label{equ:sgaslog}
\frac{\partial{\Sigma}}{\partial{t}} = \frac{3\Lambda}{R^2}\frac{\partial}{\partial{\ln{R}}}
\left\{R^{-1/2}\frac{\partial}{\partial{\ln{R}}}(R^2\alpha_{\rm{t}} T\Sigma)\right\},
\end{equation}

\noindent
where $\Lambda = \gamma k_{\rm{B}}(\mu^2_{\rm{H}}GM_\star)^{-1/2}$. Finite differencing of Eq.
(\ref{equ:sgaslog}) leads to the tridiagonal system $a_j\Sigma^{n+1}_{j-1} + b_j\Sigma^{n+1}_j + 
c_j\Sigma^{n+1}_{j+1} = e_j\Sigma^n_{j-1} + f_j\Sigma^n_j + g_j\Sigma^n_{j+1}$, where $n$ is the
time index for time $t$ and $n+1$ for time $t+\Delta t_{\rm{sync}}$. The coefficients are given by

\begin{equation}
\label{equ:trigas}
\begin{array}{lcl}
a_j & = & -\delta \Lambda R^{-1/2}_{j-1/2}(R_{j-1}/R_j)^2\alpha^{\rm{t}}_{j-1}T_{j-1}, \\
b_j & = & 1 + \delta \Lambda(R^{-1/2}_{j-1/2}+R^{-1/2}_{j+1/2})\alpha^{\rm{t}}_{j}T_j, \\
c_j & = & -\delta \Lambda R^{-1/2}_{j+1/2}(R_{j+1}/R_j)^2\alpha^{\rm{t}}_{j+1}T_{j+1}, \\
\end{array}
\end{equation}

\noindent
with $e_j=-a_j$, $g_j=-c_j$, $f_j = 2 - b_j$, and $\delta = (3/2)\Delta t_{\rm{sync}}/(\Delta \ln{R})^2$.
Since bin spacing in our code is logarithmic, the $j\pm 1/2$
refer to the logarithmic center between bins $j$ and $j\pm 1$.

\vspace{0.2in}
\noindent
{\it A.2. Advection-Diffusion Equation}
\vspace{0.1in}


The advection-diffusion equation can be differenced in a similar way to that of the gas evolution. We 
begin by rewriting the RHS  of Eq. (\ref{equ:diffadv}) in logarithmic form

\begin{equation}
\frac{\partial{\Sigma}^{\rm{v,d}}_i}{\partial{t}} = \frac{1}{R^2}\frac{\partial{}}{\partial{\,\ln{R}}}
\left\{D_{\rm{v,d}}\Sigma \frac{\partial{\alpha^{\rm{v,d}}_i}}{\partial{\,\ln{R}}} -
R V_{\rm{v,d}}\Sigma^{\rm{v,d}}_i\right\},
\end{equation}

\noindent
where $\Sigma^{\rm{v,d}}_i = \alpha^{\rm{v,d}}_i\Sigma$, 
the first term on the RHS side is the diffusion term with $D_{\rm{v,d}}$ the diffusion coefficient, 
and the second term on the RHS is the advective term. Dropping the specifics of whether we are treating
vapor or the dust for now, and also suppressing the species subscript $i$, finite differencing again 
leads to a tridiagonal system for each species $i$, where the coefficients are for the {\it diffusion}
term (1st term on RHS)

\begin{equation}
\label{equ:trisol}
\begin{array}{lcl}
a_j & = & -\delta \biggl(\Lambda^{\rm{d}}_jR^{-2}_j\alpha^{\rm{t}}_{j-1/2}\Sigma_{j-1/2}
T_{j-1/2}R^{3/2}_{j-1/2} - \\
  &   & \frac{1}{4}\alpha^{\rm{t}}_j\Sigma_j T_jR^{-1/2}_j\Delta \ln{R}(\Lambda^{\rm{d}}_{j+1}-
\Lambda^{\rm{d}}_{j-1})\biggr), \\
b_j & = & 1 + \delta \Lambda^{\rm{d}}_j R^{-2}_j\biggl(\alpha^{\rm{t}}_{j-1/2}\Sigma_{j-1/2}
T_{j-1/2}R^{3/2}_{j-1/2} + \\
  &  & \alpha^{\rm{t}}_{j+1/2}\Sigma_{j+1/2}T_{j+1/2}R^{3/2}_{j+1/2}\biggr), \\
c_j & = & -\delta \biggl(\Lambda^{\rm{d}}_jR^{-2}_j\alpha^{\rm{t}}_{j+1/2}\Sigma_{j+1/2}
T_{j+1/2}R^{3/2}_{j+1/2} + \\
 &  & \frac{1}{4}\alpha^{\rm{t}}_j\Sigma_j T_jR^{-1/2}_j\Delta \ln{R}(\Lambda^{\rm{d}}_{j+1}-
\Lambda^{\rm{d}}_{j-1})\biggr), \\
\end{array}
\end{equation}

\noindent
with $\delta = (1/2)\Delta t_{\rm{sync}}/(\Delta \ln{R})^2$, and now, recalling  
Eq. (\ref{equ:ddiff}) for $D_{\rm{d}}$,

\begin{equation}
\label{equ:ldiff}
\Lambda^{\rm{d}}_j = \sum_k \frac{\rho^{\rm{d}}_{k,j}}{\rho^{\rm{d}}_j}
\frac{\Lambda}{1 + {\rm{St}}^2_{k,j}},
\end{equation}

\noindent
for the dust component, whereas $\Lambda^{\rm{d}}_j = \Lambda$ for the vapor. The advection term (2nd
term on RHS) requires that we use upwind/downwind differencing for the derivative that depends on
the sign of the velocity $V_{\rm{v,d}}$. If $V_{\rm{v,d}} < 0$, we add terms to the $a_j$ and $b_j$:

\begin{equation}
\label{equ:vlt0}
\begin{array}{lcl}
a_j = a_j - \delta \Delta \ln{R} \,V^{n+1}_{j-1} R_{j-1}\Sigma_{j-1}/R^2_{j-1/2}, \\
b_j = b_j + \delta \Delta \ln{R} \,V^{n+1}_j R_j \Sigma_j/R^2_{j-1/2}, \\
\end{array}
\end{equation}

\noindent
whereas if $V_{\rm{v,d}} > 0$, we add terms to the $b_j$ and $c_j$:

\begin{equation}
\label{equ:vgt0}
\begin{array}{lcl}
b_j = b_j - \delta \Delta \ln{R} \,V^{n+1}_j R_j \Sigma_j/R^2_{j+1/2}, \\
c_j = c_j + \delta \Delta \ln{R} \,V^{n+1}_{j+1} R_{j+1} \Sigma_{j+1}/R^2_{j+1/2}. \\
\end{array}
\end{equation}

\noindent
Finally, the remaining coefficients are $e_j=-a_j$, $g_j=-c_j$ and $f_j = 2 - b_j$ with
velocities evaluated at current step $n$.

For the inner and outer edges of the disk, we impose a combination
of Dirichlet and Neumann boundary conditions (Robin boundary condition) through the mass flux

\begin{equation}
\label{equ:diffadvbc}
\frac{\partial{\alpha}^{\rm{v,d}}_i}{\partial{\,\ln{R}}} = -\frac{1}{D_{\rm{v,d}}\Sigma}
\left(\frac{\dot{M}^{\rm{v,d}}_i}{2\pi} - RV_{\rm{v,d}}\Sigma^{\rm{v,d}}_i\right),
\end{equation}

\noindent
where $\dot{M}^{\rm{v,d}}_i$ is the mass flux of vapor (or solids) of species $i$. The boundary conditions are 
also made implicit, and require the presence of ghost points just outside the boundaries. 
We extrapolate to find the values of the necessary quantities using polynomial methods. 

Application of the inner ($j=n_{\rm{R}}$) and outer ($j=1$) boundary conditions requires that we alter
the tridiagonal coefficients. We difference Eq. (\ref{equ:diffadvbc}) by taking the derivative centered
around the boundary points. In order to estimate the flux at time $t^{n+1}$ for species $i$, we define
$\dot{M}^{\rm{v,d}}_{i,j} = \dot{M}^{\rm{v,d}}_j\alpha^{\rm{v,d}}_{i,j}/f^{\rm{v,d}}_j$ where 
$\dot{M}^{\rm{v,d}}_j$ is the total mass of solids or vapor in a bin. The modified coefficients for
the boundaries, again dropping specifics of dust or vapor with the exception of the total fractional
masses $f^{\rm{v,d}}_j$ to avoid confusion, are then  

\begin{equation}
\label{equ:coefbc}
\begin{array}{lcl}
b^\prime_1 & = & b_1 - a_1\frac{\Delta \ln{R}}{\Lambda^{\rm{d}}_1 \alpha^{\rm{t}}_1 T_1 R^{1/2}_1}
\biggl(2V^{n+1}_1 + \\
 &  &  \dot{M}_1/\pi R_1f^{\rm{v,d}}_1\Sigma^{n+1}_1 \biggr); \\
f^\prime_1 & = & f_1 - e_1\frac{\Delta \ln{R}}{\Lambda^{\rm{d}}_1 \alpha^{\rm{t}}_1 T_1 R^{1/2}_1}\biggl(2V^n_1 + \\
 &  &  \dot{M}_1/\pi R_1f^{\rm{v,d}}_1\Sigma^n_1 \biggr); \\
b^\prime_{n_{\rm{R}}} & = & b_{n_{\rm{R}}} + c_{n_{\rm{R}}}\frac{\Delta \ln{R}}{\Lambda^{\rm{d}}_{n_{\rm{R}}} 
\alpha^{\rm{t}}_{n_{\rm{R}}} T_{n_{\rm{R}}} R^{1/2}_{n_{\rm{R}}}}\biggl(2V^{n+1}_{n_{\rm{R}}} + \\ 
  &   &  \dot{M}_{n_{\rm{R}}}/\pi R_{n_{\rm{R}}}f^{\rm{v,d}}_{n_{\rm{R}}}\Sigma^{n+1}_{n_{\rm{R}}} \biggr);  \\
f^\prime_{n_{\rm{R}}} & = & f_{n_{\rm{R}}} + g_{n_{\rm{R}}}\frac{\Delta \ln{R}}{\Lambda^{\rm{d}}_{n_{\rm{R}}} 
\alpha^{\rm{t}}_{n_{\rm{R}}} T_{n_{\rm{R}}} R^{1/2}_{n_{\rm{R}}}}\biggl(2V^n_{n_{\rm{R}}} + \\ 
  &  &  \dot{M}_{n_{\rm{R}}}/\pi R_{n_{\rm{R}}}f^{\rm{v,d}}_{n_{\rm{R}}}\Sigma^n_{n_{\rm{R}}} \biggr); 
\end{array}
\end{equation}

\noindent
with $c^\prime_1 = a_1 + c_1$, $g^\prime_1 = e_1 + g_1$, $a^\prime_{n_{\rm{R}}} = a_{n_{\rm{R}}} + 
c_{n_{\rm{R}}}$ and $e^\prime_{n_{\rm{R}}} = e_{n_{\rm{R}}} + g_{n_{\rm{R}}}$.

\vspace{0.2in}
\noindent
{\it A.3. Opacity}
\vspace{0.1in}

We employ the opacity model of \citet{cuz14a} in order to calculate the variable Rosseland mean
opacity $\kappa$ of the evolving particle size distribution. 
Calculation of $\kappa$ involves determining the wavelength-dependent, or monochromatic
opacities $\kappa_\lambda$ for a gas-particle mixture,
%
%
which depend on an extinction efficiency $Q_{\rm{e}}=Q_{\rm{a}}+Q_{\rm{s}}$ for grain size $r$ and
wavelength $\lambda$.
The extinction is due to a combination of pure absorption, which is re-radiated as 
heat, at an efficiency $Q_{\rm{a}}$, and scattering at an efficiency $Q_{\rm{s}}$. Following \citet{cuz14a},
we allow for nonisotropic scattering by adopting the common scaling \citep[e.g.,][]{vdh80}
\begin{equation}
\label{equ:extinc}
Q^\prime_{\rm{e}} = Q_{\rm{e}} - gQ_{\rm{s}} = Q_{\rm{a}} + (1-g)Q_{\rm{s}},
\end{equation}

\noindent
where the asymmetry parameter $g = \left<\cos \Theta\right>$ is the first moment of 
the distribution of the scattered component, or phase function $P(\Theta)$, which describes the degree of 
forward scattering ($g > 0$, see \citet{cuz14a}, their Eq. 3). Values of $g < 0$ are preferentially
back-scattering and $g = 0$ is isotropic.
\citet{cuz14a} give expressions for the absorption and scattering efficiencies in terms of the
electric polarizability \citep{vdh80}, which is a function of the particle size and complex particle 
dielectric constant.
%
%
%

{\it Aggregate particles and Effective Medium Theory (EMT)}. We assume in our model that particles are compositionally 
heterogeneous, being granular aggregates of much smaller grains of all compositions that condense at the local temperature, as is consistent with observations of
meteorites and interplanetary and cometary dust particles. Assuming our particles to be aggregates composed
of much smaller elements which are themselves smaller than any relevant wavelength allows us to use Garnett 
Effective Medium Theory to calculate average refractive indices of their ensemble 
\citep[see Appendix C,][]{cuz14a}. The Garnett model \citep[see, e.g.,][]{bh83} describes inclusions of 
some dielectric constant $\epsilon_0$ embedded in a homogeneous medium of dielectric constant 
$\epsilon_{\rm{m}}$ (which is vacuum here), and can be generalized to a multiple component, potentially 
porous medium composed of multiple species with 
different dielectric constants and volume fractions relative to the aggregate volume.

The monochromatic opacity $\kappa_\lambda$ is then calculated from our averaged indices (obtained using 
Garnett theory) with

\begin{equation}
\label{equ:kaplam}
\begin{split}
\kappa_\lambda = \frac{1}{\rho} \int \pi r^2 n_{\rm{d}}(r)
\left[Q^\prime_{\rm{e}}(r,\lambda)\right]\,dr \approx \\
\frac{\pi}{\rho} \sum_{k} w^{\rm{sim}}_k r^2_k b_k Q^\prime_{{\rm{e}},k}(\lambda)
\end{split},
\end{equation}

\noindent
where we have approximated the integral by a summation over all particle sizes $k$ weighted by their
discretized number densities $b_k$ (cm$^{-3}$), and the $w^{\rm{sim}}_k$ are Simpson's trapezoidal rule
coefficients. Note that $n_{\rm{d}}(r)\,dr$ has been replaced with the bulk number density $b_k$ which
applies to both the dust and migrator populations. The bulk number density for the dust population is 
given by

\begin{equation}
\label{equ:bk}
b^{\rm{d}}_k = \frac{2-q}{1-q}\, \rho_{\rm{d}} \left(\frac{m^{1-q}_k - m^{1-q}_{k-1}}
{m^{2-q}_L - m^{2-q}_{\rm{min}}}\right).
\end{equation}

\noindent
In Eq. (\ref{equ:bk}), $\rho_{\rm{d}}$ is the volume mass density of total solids in the column, 
$m_{\rm{min}}$ and $m_{\rm{L}}$ the minimum and maximum masses in the dust particle size distribution
(Sec. \ref{subsubsec:coag}), and subsequently $b_k = b^{\rm{m}}_k$ for migrator particles 
(Sec. \ref{subsubsec:mig}). The Rosseland mean opacity is then calculated from Eq. (\ref{equ:kappar}).

%
%
%
%

The algorithm for calculation of the midplane temperature for each radial bin $R_j$ is called at
every $t_{\rm{sync}}$ in conjunction with the gas evolution and solids and vapor advection and
diffusion. The calculation is iterative because the layer optical depth depends
on the opacity through the evolving size distribution, which depends on $T_j$. We use a root finding
technique to evaluate Eq. (\ref{equ:temp}) in which each temperature estimate requires that we recalculate
the dust fractions $\alpha^{\rm{d}}_{i,j}$ and particle densities $\rho^{\rm{p}}_j$, taking into account 
that the radial location can be an EF (Sec. \ref{subsubsec:drift}), to determine $\kappa$.
The new opacity estimate then determines a new $T_j$. Iterations continue until the change in $T_j$ on any 
subsequent iteration is less than some tolerance level. 

During the iteration process, there are as many as five possible circumstances regarding the previous
bin temperature $T_j$ and the new temperature $T^\prime_j$,

\begin{equation}
\label{equ:tcond}
\begin{array}{l}
1)\,\,T_j < T_i+\Delta T_{\rm{EF}}, \,\, T^\prime_j > T_i+\Delta T_{\rm{EF}}; \\
2)\,\,T_j > T_i+\Delta T_{\rm{EF}}, \,\, T^\prime_j < T_i+\Delta T_{\rm{EF}}; \\
3)\,\,T_j,T^\prime_j < T_i+\Delta T_{\rm{EF}}; \\
4)\,\,T_j > T_{i+1}-\Delta T_{\rm{EF}}, \,\, T^\prime_j < T_{i+1}-\Delta T_{\rm{EF}}; \\
5)\,\,T_i+\Delta T_{\rm{EF}} < T_j,T^\prime_j < T^\prime < T_{i+1}-
\Delta T_{\rm{EF}}; \\
\end{array}
\end{equation}

\noindent
which we consider when adjusting a radial bin's solids and vapor content. For example, in condition (1) and
also condition (3) for the case in which $T_j < T^\prime_j$, changes involve the conversion of solids of constituent $i$ into vapor:

\begin{equation}
\label{equ:cond1}
\begin{array}{l}
\alpha^{{\rm{d}}\prime}_{i,j} = \alpha^{\rm{d}}_{i,j}(T_i+\Delta T_{\rm{ER}} - 
T^\prime_j)(T_i+\Delta T_{\rm{ER}} - T_j)^{-1}; \\
\alpha^{{\rm{m}}\prime}_{i,j} = \alpha^{\rm{m}}_{i,j}(T_i+\Delta T_{\rm{ER}} - 
T^\prime_j)(T_i+\Delta T_{\rm{ER}} - T_j)^{-1}; \\
\alpha^{{\rm{v}}\prime}_{i,j} = \alpha^{{\rm{v}}}_{i,j} + \Delta \alpha^{\rm{d}}_{i,j} +
\Delta \alpha^{\rm{m}}_{i,j}, \\
\end{array}
\end{equation}

\noindent
where, e.g., $\Delta \alpha^{\rm{d}}_{i,j} = \alpha^{\rm{d}}_{i,j}-\alpha^{{\rm{d}}\prime}_{i,j}$.
For the case in which all remaining solids are converted to vapor (condition 1) then $\alpha^{\rm{d}}_{i,j}$
and $\alpha^{\rm{m}}_{i,j}$ are simply added to $\alpha^{\rm{v}}_{i,j}$. On the other hand, in a situation
where vapor is condensing, for example condition (2), or also (3) but with $T_j > T^\prime_j$ then we
partition vapor onto both dust and migrators in the appropriate {\it area} fractions

\begin{equation}
\label{equ:cond2}
\begin{array}{l}
\alpha^{{\rm{v}}\prime}_{i,j} = \alpha^{\rm{v}}_{i,j}(T^\prime_j - T_i +\Delta T_{\rm{ER}}) 
(T_j - T_i + \Delta T_{\rm{ER}})^{-1}; \\
\alpha^{{\rm{d}}\prime}_{i,j} = \alpha^{\rm{d}}_{i,j} + x^{\rm{s}}_j\Delta \alpha^{\rm{v}}_{i,j}; \\
\alpha^{{\rm{m}}\prime}_{i,j} = \alpha^{\rm{m}}_{i,j} + (1-x^{\rm{s}}_j)\Delta \alpha^{\rm{v}}_{i,j}, \\
x^{\rm{s}}_j = \sum_{k=1}^{n_*} \pi r^2_k\rho^{\rm{d}}_k/\left(\sum_{k=1}^{n_*} \pi r^2_k\rho^{\rm{d}}_k +
\sum_{k=n_*+1}^{n_{\rm{L}}} \pi r^2_k \rho^{\rm{m}}_k\right), \\
\end{array}
\end{equation}

\noindent
where $x_{\rm{s}}$ is the area fraction of dust in solids. The
iterated values of the mass fractions are then used to calculate the mean particle material densities and adjust
the particle mass distributions. The final values of the mass fractions are assigned when the
temperature has converged.

A special case in our code involves the conversion of troilite. Solid troilite is converted to vapor and
metallic iron in the fraction $f_{\rm{I}}=56/88$. For simplicity, we consider this reaction to be
completely reversible, where more care must be taken in the reverse process to assure the fractions
are correct.

\vspace{0.2in}
\noindent
{\it A.4. Relative Velocities}
\vspace{0.1in}


In our calculations, we assign a single particle-to-gas and particle-to-particle relative velocity for
the entire particle mass distribution (dust and migrators) within the subdisk ``dust" layer, which is defined
by the scale height $h_{\rm{D}}$ as described in section \ref{subsubsec:subdisk}.  This calculation uses a logarithmic spacing in particle radius from the minimum size $r_{\rm{min}}$ which is fixed in our code, 
to the largest size $r_{\rm{L}}$.

The stopping times of solid particles $t^{\rm{s}}_k$ depend on
the flow regime which depends on whether the radius of a particle is larger or smaller than the
gas molecular mean free path (Sec. \ref{subsubsec:Standts}). For any height $z$, the stopping time 
for particle of mass $m_k$ at radial location $R_j$ is

\begin{equation}
\label{equ:tsk}
t^{\rm{s}}_{k} = \frac{2r^2_k\rho^{\rm{p}}_j}{3c_j\rho_j\lambda^{\rm{mfp}}_j}
\left[D_{{\rm{t}},j} + \frac{\lambda^{\rm{mfp}}_j}{r_k}\left(A_{\rm{t}} + 
B_{\rm{t}}e^{-C_{\rm{t}}r_k/\lambda^{\rm{mfp}}_j}\right)\right],
\end{equation}

\noindent
where the gas density as a function of height is $\rho_j = \sqrt{2/\pi}(\Sigma_j/2H_j)
\exp{[-\frac{1}{2}(z/H_j)^2]}$
and the expression in brackets is the Epstein-to-Stokes regime transitional formula \citep{pod88},
with coefficients $A_{\rm{t}} =1.249$, $B_{\rm{t}} =0.42$, $C_{\rm{t}}=0.87$ and 

\begin{equation}
\label{equ:dtran}
D_{{\rm{t}},j} = \frac{\rho_j c_j \lambda^{\rm{mfp}}_j}{3\nu_{\rm{m}}} - (1/\varepsilon)\left[A_{\rm{t}} + 
B_{\rm{t}}e^{-C_{\rm{t}}\varepsilon}\right].
\end{equation}

\noindent
where $\nu_{\rm{m}} = \mu_{\rm{m}}/\rho$ is the molecular viscosity, and $\mu_{\rm{m}} = 1.3\times 10^{-4}$
g cm$^{-1}$ s$^{-1}$. The transition between the Epstein and Stokes regimes is defined
by $r_k \leq \varepsilon \lambda^{\rm{mfp}}_j$ where the factor $\varepsilon = 9/4$ is typically 
used. In our code we have adjusted this value to get a smoother transition and found $3/2$ works better. 
For $r_k \ll \lambda^{\rm{mfp}}_j$, the
stopping time formula reduces to the Epstein regime $t^{\rm{s}}_{k} = r_k\rho^{\rm{p}}_j/
c_j\rho_j$. For $r_k \gtrsim \lambda^{\rm{mfp}}_j$, we utilize the Stokes regime equation

\begin{equation}
t^{\rm{s}}_{k} = \frac{8}{3}\frac{\rho^{\rm{p}}_jr_k}{\rho_jC^{\rm{d}}_k\Delta 
V^{\rm{pg}}_{k}},
\end{equation}

\noindent
where $\Delta V^{\rm{pg}}_k$ is the particle-to-gas relative velocity. The drag coefficient $C^{\rm{d}}_k$
depends on the particle Reynolds number ${\rm{Re^p}}_k = 2r_k\Delta V^{\rm{pg}}_k/\nu_{\rm{m}}$ and is calculated 
using the prescription \citep{wei77}:

\begin{equation}
C^{\rm{d}}_k = 
\begin{cases}
24/{\rm{Re^p}}_k	&\,{\rm{for}}\,{\rm{Re^p}}_k< 1;\\
24/{\rm{Re^p}}^{0.6}_k	&\,{\rm{for}}\,{\rm{Re^p}}_k< 800;\\
0.44		&\,{\rm{for}}\,{\rm{Re^p}}_k\ge 800;\\
\end{cases}.
\end{equation}

\noindent
Since $t^{\rm{s}}_k$ depends on $\Delta V^{\rm{pg}}_k$ for ${\rm{Re^p}}_k > 1$, iterations will be
required to determine the correct stopping time, and thus relative velocities.

We begin with equations (\ref{equ:Ui}-\ref{equ:vg}) and solve for the gas radial and azimuthal 
velocities $u$ and $v$ at semimajor axis $R_j$ as

\begin{equation}
\label{equ:u}
u = \frac{\sum_k A_k\rho^{\rm{d}}_k\left[B U_k + 2\Omega V_k\right] + 2B\Omega \eta V_{\rm{K}}}{B^2 + \Omega^2},
\end{equation}

\begin{equation}
\label{equ:v}
v = \frac{-\sum_k A_k\rho^{\rm{d}}_k\left[(\Omega/2)U_k - B V_k\right] - \Omega^2\eta V_{\rm{K}}}{B^2 + \Omega^2},
\end{equation}

\noindent
where $B=\sum_l A_l\rho^{\rm{d}}_l$. These can then be inserted into the equations for the $U_k$ and 
$V_k$ and the
individual component can be removed from the sums. That is, we solve for the component $V_k$ (or $U_k$) 
by substituting the expression for $v$ into the component equation for $V_k$ (or $U_k$, e.g., Eq. 
[\ref{equ:Ui}] or [\ref{equ:Vi}]). This gives

\begin{equation}
V_k = \frac{(\Omega/2)F_kU_k + Y_k + \Omega^2A_k\rho \eta V_{\rm{K}}}{A_k\rho G_k},
\end{equation}

\noindent
where $G_k = BA_k\rho^{\rm{d}}_k - B^2 - \Omega^2$, $F_k = A^2_k\rho \rho^{\rm{d}}_k + B^2 + 
\Omega^2$, and

\begin{equation} 
Y_k = A_k\rho \sum_{l\neq k}A_l\rho^{\rm{d}}_l\left[(\Omega/2)U_l-BV_l\right].
\end{equation}

\noindent
Solving for the final expressions for the velocity components, we find

\begin{equation}
\begin{split}
U_k = -\biggl[A_k\rho G_k Z_k + 2\Omega F_k Y_k + 2\Omega A_k\rho (\Omega^2F_k+ \\
A_k\rho BG_k)\eta V_{\rm{K}}\biggr]\left[(A_k\rho G_k)^2 + (\Omega F_k)^2\right]^{-1}
\end{split},
\end{equation}

\begin{equation}
\begin{split}
V_k = -\biggl[(\Omega/2)F_k Z_k - A_k\rho G_k Y_k + A_k\rho \Omega^2(BF_k - \\
A_k\rho G_k)\eta V_{\rm{K}}\biggr]\left[(A_k\rho G_k)^2 + (\Omega F_k)^2\right]^{-1}
\end{split},
\end{equation}

\noindent
where $Z_k$ is given by

\begin{equation}
Z_k = A_k\rho \sum_{l\neq k}A_l\rho^{\rm{d}}_l\left[BU_l+2\Omega V_l\right].
\end{equation}

\noindent
The equations for $U_k$ and $V_k$ represent $2n$ equations that can be solved by matrix inversion using
standard methods. 
 
\vspace{0.2in}
\noindent
{\it A.5. Migrator Growth and Radial Drift}
\vspace{0.1in}

If we define the mass of the dust subdisk in a radial bin $R_j$ to be $M^{\rm{sd}}_j = 
2\rho^{\rm{d}}_j h^{\rm{d}}_j{\mathscr{A}}_j$ where the area of a radial bin is 
${\mathscr{A}}_j=\pi(R^2_j-R^2_{j+1})$ and the subdisk layer height $h_{\rm{d}}$ is defined in Sec. \ref{subsubsec:subdisk}),
then the total mass of new ``migrator'' material created in time $\Delta t_j$ is roughly (compare to Eq. 
[\ref{equ:mdrift}] of section \ref{subsubsec:drift}):

\begin{equation}
\label{equ:Mexc}
M^*_j\simeq M^{\rm{sd}}_j\left(\frac{m^{2-q}_{\rm{L}}-m^{2-q}_*}{m^{2-q}_{\rm{L}}-
m^{2-q}_{\rm{min}}}\right)_j,
\end{equation}

\noindent
Although the migrator population covers masses $m \geq m_*$, we consider the mass in this layer
$M^{\rm{sd}}_j$,
including the fraction of dust lying in this layer, to be available for sweepup by all migrators.
This mass does not reflect all the mass in radial bin $j$, as some fraction of ``dust'' occupies
the upper layers beyond $h_{\rm{d}}$. The migrator mass thus ``created'' in time $\Delta t_j$ is assigned to
the first total mass bin of the migrator population, $M^{\rm{m}}_{1,j} = M^*_j$, while already
existing total mass bins are shifted up by one in index.

{\it Drift mass.} Because we use an asynchronous time step scheme in our code, the timestep of
a radial bin $j$ is typically longer than that of bin $j-1$ which lies exterior to $j$. In order to 
account for the asynchronicity, we use a predictor/corrector method for the migrator mass drifting in.
There are three circumstances that arise within this scheme: (1) a time step for bin $j-1$ was called
at the same time as bin $j$; (2) no time step was called for bin $j-1$ when a step for $j$ was called;
and, (3) it is a synchronization step in which all bins are called, but with the appropriate time $\Delta t^{\rm{sync}}_j$ 
set for each radial bin such that all radial bins are brought to the same 
elapsed time. The synchronization step is
the step in which all drift mass is accounted for. The three cases can be summarized as

\begin{equation}
\label{equ:mdcases}
M_{\rm{pc}} = 
\begin{cases}
\begin{split}
M^{{\rm{drift}},N}_{j-1}\frac{N_j\Delta t_j -(N_{j-1}-1)\Delta t_{j-1}}{\Delta t_{j-1}} - \\
M^{{\rm{drift}},N-1}_{j-1}\frac{(N_j-1)\Delta t_j - (N_{j-1}-1)\Delta t_{j-1}}{\Delta t_{j-1}};\end{split} \\
M^{{\rm{drift}},N}_{j-1}(\Delta t_j/\Delta t_{j-1}); \\
\begin{split}
M^{{\rm{drift}},N}_{j-1} - M^{{\rm{drift}},N-1}_{j-1}\times \\
\frac{(N_j-1)\Delta t^{\rm{sync}}_j -
(N_{j-1}-1)\Delta t^{\rm{sync}}_{j-1}}{\Delta t^{\rm{sync}}_{j-1}},\end{split} \\
\end{cases}
\end{equation}

\noindent
where $\Delta t^{\rm{sync}}_j = N_{\rm{sync}}\Delta t_{\rm{min}} - N_j\Delta t_j$, and the superscript
$N$ refers to the most recent {\it stored} calculation, and $N-1$ the  calculation previous to $N$. 

Due to the nature of our parallelization of the code in radial bins, the most recent information for
the drift mass is not available until a synchronization step. An individual radial bin $j$ is assigned
to a single processor, but requires information from $j-1$ regarding the mass drifting in.
We store and use values from previous time steps for the mass that drifts out of bin $j-1$ into $j$
and we use those as the basis of our calculation which leads to some small ``out of phase'' transport
which is eventually corrected at a $t_{\rm{sync}}$. Given that the amount of mass drifting in compared
to the total mass of a bin is quite small, any temporary effect that may occur during regular time
steps is minimal.

$M_{\rm{pc}}$ represents a total mass of migrators of various sizes available to drift inwards into radial bin $j$ in time 
$\Delta t_{j-1}$, whose total
mass per mass bin $m_k$ is given by $M^{\rm{drift}}_{k,j-1}$, such that $M^{\rm{drift}}_{j-1} =
\sum_k M^{\rm{drift}}_{k,j-1}$.  The actual mass available in the time step $\Delta t_j$ (or in the case of a synchronization
step, $\Delta t^{\rm{sync}}_j$) is then scaled to $M^{{\rm{drift}}\prime}_{k,j-1} = (M_{\rm{pc}}/
M^{\rm{drift}}_{j-1})M^{\rm{drift}}_{k,j-1}$. The particle mass distribution for bin $j-1$ will be
different from bin $j$, thus we reinterpolate the drift mass distribution from bin $j-1$ onto the
same particle mass range as bin $j$ and add the contribution to the total migrator mass
distribution $M^{\rm{mig}}_{k,j}$. In the event that the particle mass range in bin $j-1$ does not
all lie within the mass range of bin $j$, we (1) return to the dust population of radial bin $j$  the mass
of all migrators that are in sizes smaller than the fragmentation barrier mass of bin $j$; and/or, (2) add new
particle mass and total migrator mass bins to radial bin $j$ for particles from bin $j-1$ that are larger than
the most massive pre-existing particle in bin $j$.

{\it Evaporation fronts.} If radial bin $j$ or $j-1$ (or both) lie within an EF, then inward drifting 
migrators will have their compositions altered through adjustment to the local nebula temperature. This
means that some or all of their volatile will be removed and added to the vapor content. For an EF
with evaporation temperature $T_i$, we adjust the mass fraction in species $i$ of migrators 
from bin $j-1$

\begin{equation}
\label{equ:alphaef}
\alpha_{\rm{EF}} = \alpha^{\rm{m}}_{i,j-1}\left(\frac{T_i+\Delta T_{\rm{EF}}-T_j}
{T_i+\Delta T_{\rm{EF}} - T_{j-1}}\right),
\end{equation}

\noindent
with the evaporated fraction added to the vapor content of bin $j$:

\begin{equation}
\label{equ:migef}
\alpha^{{\rm{v}}\prime}_{i,j} = \alpha^{\rm{v}}_{i,j} + (\alpha^{\rm{m}}_{i,j-1}-\alpha_{\rm{EF}})
\frac{\Sigma_{j-1}}{\Sigma_j}\frac{{\mathscr{A}}_{j-1}}{{\mathscr{A}}_j}
\frac{M^{{\rm{drift}}\prime}_{j-1}}{M^{\rm{m}}_{j-1}}.
\end{equation}

\noindent
In Eq. (\ref{equ:migef}) above, $M^{{\rm{drift}}\prime}_{j-1}$ is the adjusted total mass of migrators
drifting in to bin $j$ in time $\Delta t_j$, $M^{\rm{m}}_{j-1}$ is the total mass of migrators in bin
$j-1$, the ratio of which along with the ratios of gas density and bin area ensures we are adding the
correct mass fraction to bin $j$. In the special case of the troilite EF, the above equation will
have a $(1-f_I)$ factor multiplied to the second term on the RHS, and we add the remaining factor $f_I$
to the metallic iron dust fraction (see Appendix A.3).

{\it Migrator growth.} The incremental growth of migrators is described by Eq. (\ref{equ:dmkdt}). We 
assume that all migrator particles $m_{k,j}$ grow from both the dust content
in the subdisk layer, and potentially other migrators to some new size $m^\prime_{k,j}$ in time 
$\Delta t_j$. The fraction of dust and migrators accreted
can be determined from the individual terms on the RHS of Eq. (\ref{equ:dmkdt}). The change in the
total mass of migrators in mass bin $k$ is estimated then from

\begin{equation}
\label{equ:dmmig}
\Delta M^{\rm{m}}_{k,j} = \left(\frac{m^\prime_{k,j}}{m_{k,j}}-1\right) M^{\rm{m}}_{k,j},
\end{equation}

\noindent
which can be an overestimate for two reasons. 

The first reason is that there is a limited amount of dust mass
available in the subdisk that can be accreted. If it is found that more dust mass was accreted than lies
within the subdisk, the excess is subtracted out and the particle sizes, densities and accreted mass
are adjusted accordingly with the same factor taken from all sizes. Although the subdisk dust mass
becomes negligible, more dust remains at higher altitudes, which can be mixed downwards by settling
and diffusion. 

The second reason is that the initial calculation can allow more mass to be accreted from migrator mass bin $k$ by other
migrators than is available in mass bin $k$. Under these circumstances, the entire redistribution of the migrators
will need to be redone (perhaps multiple times) until only the mass available for accretion can be accreted. Under normal circumstances, after the growth calculation, we redistribute total migrator
mass according to the fraction of Eq. (\ref{equ:dmmig}) that was due to migrator accretion of
other migrators. For every migrator total mass bin $k$, we loop over other migrator total mass
bins $l\ge k$ (here $l$ and $l'$ represent dummy indices) and subtract out
what those bins accreted from mass bin $k$

\begin{equation}
\label{equ:migcor}
M^{{\rm{m}}\prime}_{k,j} = M^{\rm{m}}_{k,j} - (1-f_\rho)\sum_{l\ge k} \Delta M^{\rm{m}}_{l,j}\frac{S_{k,l}
\rho^{\rm{m}}_{k,j}}{\sum_{l^\prime}S_{l^\prime,l}\rho^{\rm{m}}_{l^\prime,j}},
\end{equation}

\noindent
where $S_{k,l}\rho^{\rm{m}}_{k,j}$ represents the density fraction of material of particle mass bin $k$
in the migrator population accreted by mass $l$, 
and $\sum_{l^\prime}S_{l^\prime,l}\rho^{\rm{m}}_{l^\prime,j}$ is the total density fraction over 
all masses accreted by $l$, and is thus related to the total mass of migrator material available in the 
subdisk for a migrator to accrete from. The factor $f_\rho$ is the fraction of dust accreted in 
$\Delta t_j$.

If it is found that more mass from a migrator bin is accreted than what is available ({\it i.e.},
$M^{{\rm{m}}\prime}_{k,j} < 0$), we define a scaling coefficient for each bin $k$ found to have more mass
accreted from it than available:

\begin{equation}
\label{equ:migscale}
{\mathcal{A}}_{k,j} = M^{\rm{m}}_{k,j}\left[(1-f_\rho)\sum_{l} \Delta M^{\rm{m}}_{l,j}\frac{S_{k,l}
\rho^{\rm{m}}_{l,j}}{\sum_{l^\prime}S_{l^\prime,l}\rho^{\rm{m}}_{l^\prime,j}}\right]^{-1},
\end{equation}

\noindent
which weights the amount of material that is subtracted from a bin due to accretion from other bins, and
becomes the coefficient for the sum on the RHS of Eq. (\ref{equ:migcor}). The overestimated mass is
then $1-{\mathcal{A}}_{k,j}$ which is subtracted out of $\Delta M^{\rm{m}}_{k,j}$ once iterations
that ensure the modified condition for Eq. (\ref{equ:migcor}) are satisfied for all $k$.
 
Once all growth calculations are done, as a final step we perform a reinterpolation of the mass histogram
of migrators (including all related quantities) to a logarithmic mass bin spacing, and make
sure that the criterion for the maximum number of bins per mass or radius decade is adhered to (see Drazkowska et al. 2013, 2014).

\vspace{0.2in}
\noindent
{\it A.6. Particle Destruction Probability}
\vspace{0.1in}

Rather than adding another time consuming calculation in our migrator growth subroutine that
would involve integration over multiple integrals to properly calculate the destruction probabilities,
we discretize the destruction rates of a migrator particle of mass $m_k$ due to dust and to other migrators as

\begin{equation}
\label{equ:dusdest}
\dot{{\mathscr{P}}}^{\rm{d}}_k = \sum_{l=1}^{l_*} {\mathscr{R}}^{\rm{coll}}_{l,k}\zeta_{l,k},
\end{equation}

\begin{equation}
\label{equ:migdest}
\dot{{\mathscr{P}}}^{\rm{m}}_k = \left(1 - {\mathscr{P}}^{\rm{m}}_k\right)\sum_{l=l_*+1}^{k} 
{\mathscr{R}}^{\rm{coll}}_{l,k}\zeta_{l,k},
\end{equation}

\noindent
where summation to $l=l_*$ is over all dust masses up to the fragmentation barrier $m_*$, and the summation
from $l_*+1$ to $k$ is over all migrators of mass $m_k$ and smaller. Equation (\ref{equ:migdest}) for 
migrators contains an additional factor $\zeta_{l,k}$ that depends on the probability of {\it destruction} of 
migrators during the current time step.
We account for the second integral (over time of growth) by using a discrete number of points over 
the time interval as it grows from $m_k$ to $m^\prime_k$. Different mass values between $m_k$ and
$m^\prime_k$, along with their associated relative velocities, determine the destruction probability
over the timestep. Because growth is relatively small in time $\Delta t_j$ over a wide range of values of turbulent intensity 
$\alpha_{\rm{t}}$, a few points is generally enough, but more points are used when growth is fast,  
such as it might be the case for low $\alpha_{\rm{t}}$. The total probability of destruction of a migrator in mass
bin $k$ is

\begin{equation}
\label{equ:dprobd}
\begin{split}
{\mathscr{P}}_k = {\mathscr{P}}^{\rm{d}}_k + {\mathscr{P}}^{\rm{m}}_k = 
\frac{\Delta t_j}{N_p}\sum_{p=1}^{N_p}\left(\sum_{l=1}^{l_*} {\mathscr{R}}^{\rm{coll}}_{l,k}
\zeta_{l,k}\right)_p + \,\,\,\,\,\,\,\,\,\,\, \\
\,\,\,\,\,\,\,\,\,\,\,1 - \exp{\left[-\frac{\Delta t_j}{N_p}\sum_{p=1}^{N_p}
\left(\sum_{l=l_*+1}^{k} {\mathscr{R}}^{\rm{coll}}_{l,k}\zeta_{l,k}\right)_p\right]},
\end{split}
\end{equation}

\noindent
where the index $p$ and $N_p$ refer to the number of points used in the estimate.
Once ${\mathscr{P}}_k$ is determined, we can calculate the total
mass of particles of size $k$ destroyed as $M^{\rm{dest}}_k = {\mathscr{P}}_kM^{\rm{m}}_{k,j}$; this
mass is returned to the dust population. Bins that suffer total destruction are removed {\it if} they are 
at the tail end of the distribution only. Bins with zero mass in the middle of the distribution are 
allowed.

\vspace{0.2in}

\renewcommand{\theequation}{B-\arabic{equation}}


\setcounter{equation}{0}  

\section*{Appendix B: Particle Diffusivity and Schmidt Number}

A quick survey of the fluid dynamics literature reveals that the term ``Schmidt Number" (Sc) is widely used as the ratio of gas viscosity $\nu$ to particle diffusivity $D_{\rm{d}}$ for particles of arbitrary $t_{\rm{s}}$. Minor complications arise if ${\rm Sc_g} =  \nu/D_{\rm g}$, for the gas itself, is significantly different from unity {\citep{ste90,ha10,ha12} because many derivations of $D_{\rm d}$ relate particle diffusivity to gas diffusivity {\citep{vol80,cuz93,ch03,sh04,yl07}; we will assume ${\rm Sc_g}=1$, but deviations are easily allowed for. Here we have adopted the formulation of $D_{\rm d}$ and Sc in \citet[][henceforth YL07]{yl07}, who incorporated orbital dynamics effects. For discussion and validation of these formulations see \citet{car11}, who also showed that the resulting particle layer thickness $h_{\rm d}$ varied as St$^{-1/2}$ for St = 0.1 to 100. This innocent-looking St-dependence, familiar from the results of \citet[][henceforth D95]{dub95}, hides several subtleties, and is not {\it quite} the final answer, as we sketch below. 

A closed-form expression for the {\it particle} vertical scale height $h_{\rm d}$ can be obtained by integrating the vertical mass balance equation for the concentration {\it ratio} $\rho_{\rm d}/\rho$ (D95 Eq. 32; Garaud et al. 2007, henceforth G07, her Eq. 21), or treating mass balance using scaling arguments \citep[][henceforth CW06 (their Eqns. 4-7)]{cw06}. Written in terms of ${\rm Sc}$, without worrying about its proper form (see however \citet{sh04} for a pre-orbit-dynamics assessment), Eq. (22) of G07 for $h_{\rm d}/H$ is exactly reproduced (ignoring constants of order unity) by combining Eqns. (37) and (39) of D95 (but note that a factor of ${\kappa}_{\rm t0} = 1/{\rm Sc}$ has been lost in D95 Eq. (37), because Eq. (36) requires the normalized diffusivity $\overline{\kappa}_{\rm t}={\kappa}_{\rm t0} \alpha_{\rm{t}}/{\rm St}$). Solving the diffusion equation for a particle/gas {\it ratio} is a preferred approach; because neither CW06 nor YL07 accounted for vertical variation of gas density; their expressions give $h_{\rm D} > H$  when ${\rm St} < \alpha_{\rm{t}}$, which is unphysical. 


CW06 assumed a constant gas density and so their result is directly comparable to the scale height $h/H$ for the ratio $\rho_{\rm d}/\rho$ derived by D95 (her Eq. 36); in fact, it is identical if both are expressed in terms of Sc generically. Thus, it can also be transformed to a proper particle layer scale height $h_{\rm d}/H$ using Eq. (39) of D95, also becoming identical (in terms of Sc, for a single particle) to Eq. (22) of G07 (again neglecting constants of order unity):
\begin{equation}
h_{\rm d}/H = (1 + {\rm St\, Sc} / \alpha_{\rm{t}})^{-1/2}.
\end{equation} 
However, all these previous results from mass balance equations assumed a particle ``settling time" $t_{\rm sett}$ given by the terminal velocity of a single particle under vertical solar gravity $=1/{\rm St}\,\Omega $). This is not appropriate for St$\gg 1$ particles which are underdamped, as pointed out by \citet{wei84,wei88} who suggested using the inclination damping time of a layer of large particles ($=t_{\rm{s}}$) as the ``settling time" when St$\gg 1$. This idea was incorporated into the numerical models of \citet{cuz93} and used in a bridging expression constructed by YL07 (their Eq. 2) to derive their Eq. (6) for $h/H$, which ironically reduced exactly to the small-St (${\rm Sc}=1$) expression in CW06 or Eq. (37) of D95. This is because (relative to the earlier expressions), the larger ${\rm Sc}= 1+{\rm St}^2$ of YL07 (using their physically motivated Eq. 5) is exactly canceled by allowing for the slower ``settling" times of very large particles. Equations (4-7) of CW06, substituting ${\rm Sc} = 1+{\rm St}^2$ and $t_{{\rm sett}} = t_{\rm{s}}+1/{\rm St}\,\Omega$ from YL07, naturally lead to the same result: $h/H = (\alpha_{\rm{t}}/{\rm St})^{1/2}$. While this has the proper behavior to account for the particle layers in the simulations of \citet{car11}, which never exceed 0.1$H$ in thickness, it remains nonphysical in the small-St regime for reasons noted above, so must be converted using D95 Eq. (39) into a true particle scale height:
\begin{equation}
h_{\rm d}/H = (1 + {\rm St} / \alpha_{\rm{t}})^{-1/2},
\end{equation} 
which is valid for all combinations of St and $\alpha_{\rm{t}}$ in global turbulence, where the energetic eddy frequencies are comparable to the orbit frequency. In nonturbulent nebulae with densely settled midplane layers, the near-midplane turbulence is shear driven and eddy times $\tau_{\rm{e}}$ are faster, requiring modification to these expressions; for discussions see \citet{cuz93} or YL07 (unfortunately Fig. 6 of YL07 does not correctly represent the model of Cuzzi et al. 1993, which also includes a dependence on $\tau_{\rm{e}}$, leading to the ambiguity of YL07 Fig. 7 discussed by \citet{car11}, but this does not affect the overall results of YL07). 

\vspace{0.2in}
\renewcommand{\theequation}{C-\arabic{equation}}


\setcounter{equation}{0}  

\section*{Appendix C: Code Tests}
 
\noindent

{\it C.1. Coagulation}

\vspace{0.1in}

Our numerical code uses the moments method of \citet{ec08} to solve the Smoluchowski equation 
(Sec.\ref{subsubsec:coag}) under the assumption that the dust distribution can be treated as a powerlaw
$f(m) \propto m^{-q}$ up to some fragmentation mass $m_*$. The algorithm for the moments has been tested
for both simple and realistic collisional kernels against the brute force solution to the coagulation
equation for the cases in which the relative velocities between particles is systematic, or in the case 
where they are driven by turbulence. We specifically utilize the ``modified explicit'' approach to follow the 
growth of the largest particle $r_{\rm{L}}$ \citep[Eq. (28),][]{ec08}. This approach was also compared
to the alternative growth formalism of \citet{gar07} and good agreement was found. Thus we do not repeat
any tests of its accuracy here.

\vspace{0.2in}

\noindent

{\it C.2. Mass and Radius Binning}

\vspace{0.1in}

\begin{figure}
 \resizebox{0.9\linewidth}{!}{%
 \includegraphics{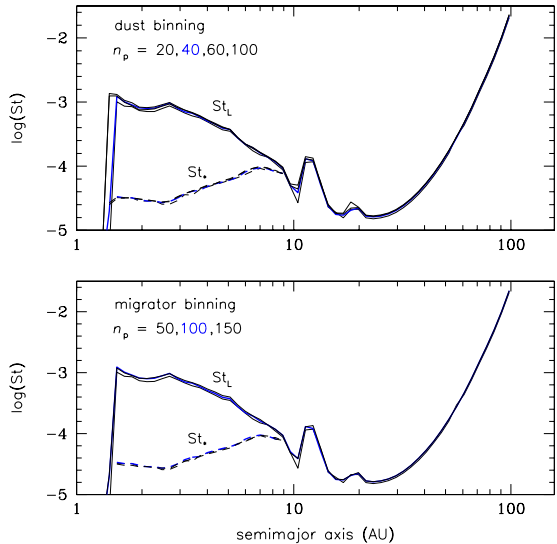}}
\caption{\underline{Upper panel:} Comparison of the particle Stokes numbers after $10^3$ years for
four different choices for the number of bins per radius decade $n_{\rm{p}}$ for the dust population. The
blue curve is our nominal choice for this paper. \underline{Lower panel:} As above, but for the migrator population.}
\label{fig:bintest}
\end{figure}

In this section we check the influence of the choice of number $n_{\rm{p}}$ of logarithmically spaced 
bins per decade radius on the growth
of particles in our code, both for the dust and migrator distributions. The moments method only requires
a defined mass or radius histogram when calculating the collisional kernel, for example when solving for
the particle-to-particle relative velocities, which must be integrated over in order to calculate the growth 
rate of $m_{\rm{L}}$. This histogram is always well defined because of the strict adherence to
a powerlaw in mass or radius. Nevertheless, because reasonably accurate growth requires sufficient sampling
of the kernel quantities, especially for the migrators, the number of bins chosen per 
decade can have an effect on the results.
 
In the upper panel of {\bf Figure \ref{fig:bintest}}, we compare the Stokes numbers St$_{\rm{L}}$ 
and St$_*$ (where defined) for a run of $10^3$ years for the fragmentation only case and an assumed 
$\alpha_{\rm{t}} = 5\times 10^{-3}$, for four different choices of $n_{\rm{p}}$. The blue curve represents the
value we have chosen for this paper. Even given the variation of St$_{\rm{L}}$ over orders of magnitude, 
all of the choices produce very similar results with the lowest value $n_{\rm{p}}=20$ producing slightly more
variation than higher values, but certainly within acceptable levels. The greatest difference occurs in the innermost
bin, which also happens to coincide with the edge of an EF. This is not surprising since the temperature variation
is quite sensitive to the particle size distribution. From these tests, it would appear that a choice of 
$n_{\rm{p}}=60$ (whose curve appears most similar to the value of 100) may be a best choice going forward. It is
also possible to vary $n_{\rm{p}}$ with $R$.

In the bottom panel of Fig. \ref{fig:bintest}, we have instead varied the number of migrator radius bins per decade
while using our nominal value for the dust distribution. We compare our nominal value for the migrator bins 
$n_{\rm{p}}=100$ (blue curve) to smaller and larger values and find that this choice is more than sufficient as the
$n_{\rm{p}} = 100$ and 150 lie right on top of each other. We note that our treatment of the migrator distribution
binning is different from that of the dust distribution which is effectively Eulerian as are all the usual solutions
of the Smoluchowski equation, which lead to the well-known concerns about mass resolution \citep{dra14}. The migrator bins are, in fact,
Lagrangian in that the migrator mass bins grow at different rates, and thus even a logarithmically equally spaced
grid would become unequal in its spacing with time. However, in order to maintain a consistency
of spacing, we rebin the migrator distribution, strictly conserving mass, after every iteration. The generally wide agreement
 between the variety of choices for $n_{\rm{p}}$ give us confidence that growth is being treated well at 
all particle sizes we achieve in our models.

\vspace{0.2in}

\noindent

{\it C.3. Variation of $N_{\rm{sync}}$}

\vspace{0.1in}

\begin{figure}
 \resizebox{0.9\linewidth}{!}{%
 \includegraphics{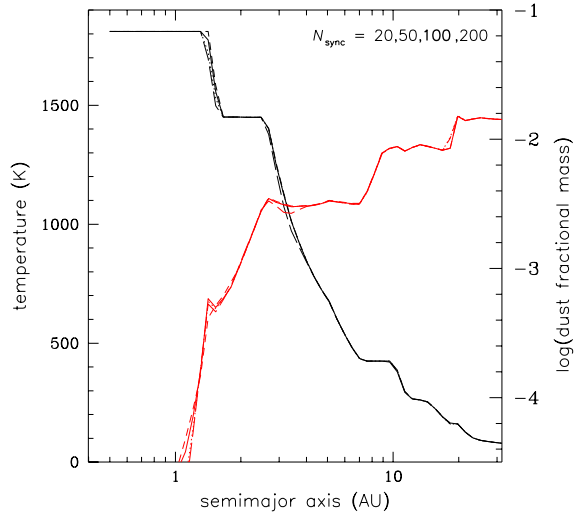}}
\caption{Temperature (black) and dust fraction (red) curves for different choices of the synchronization
step. Our nominal choice is given by the solid curves. The relatively little variation indicates that the global
processes such as the gas evolution and temperature calculation need not be done so frequently.}
\label{fig:tsync}
\end{figure}

Global calculations in our code, such as the gas evolution, temperature and dust diffusion and advection do not 
need to be done every time step. This is the basis for our use of an asynchronous time stepping scheme. In this 
section we compare different choices of the number of steps $N_{\rm{sync}}$ our code executes before calling a
synchronization step in which all properties of the nebula are brought to the same time.

In {\bf Figure \ref{fig:tsync}} we compare four different choices of $N_{\rm{sync}}$ with our nominal choice of
100 shown by the solid curves. We plot both the temperature (black curves) and the dust fractional mass (red curves)
for the same model as above after $10^3$ years. Again, we find that the agreement across the different choices is
more than adequate indicating that these processes are not changing quickly over a time scale of 
$N_{\rm{sync}}\times \Delta t_{\rm{min}}$. Most of the variation is due to the sensitivity of the temperature to
changes in the particle size distribution. Even these appear to be minor, especially in this case of high 
$\alpha_{\rm{t}}$ in which growth can be rapid.
 


\end{document}